\documentclass[vecphys]{svmult}

\usepackage{makeidx}         
\usepackage{graphicx}        
\usepackage{multicol}        
\usepackage[bottom]{footmisc}

\makeindex             

\def\la{\mathrel{\mathpalette\fun <}}
\def\ga{\mathrel{\mathpalette\fun >}}
\def\fun#1#2{\lower3.6pt\vbox{\baselineskip0pt\lineskip.9pt
  \ialign{$\mathsurround=0pt#1\hfil##\hfil$\crcr#2\crcr\sim\crcr}}}
\newcommand{\lfrac}[2]{{{#1}{/}{#2}}}
\def\ee #1 {\times 10^{#1}}

\begin{document}

\title*{The Effects of Large-Scale Magnetic Fields\\
on Disk Formation and Evolution}
\titlerunning{Chapter VI: The Effects of Large-Scale Magnetic Fields}
\author{Arieh K\"onigl\inst{1}\and Raquel Salmeron\inst{2}} 
\institute{Department of Astronomy \& Astrophysics, The University of
Chicago, Chicago  IL 60637, USA \texttt{akonigl@uchicago.edu}
\and Research School of Astronomy \& Astrophysics and Research School of
Earth Sciences, The Australian National University, Canberra ACT 0200,
Australia \texttt{raquel@mso.anu.edu.au}} 
%
%
\maketitle

\section{Introduction}
\label{sec:intro}

In Chap.~V [Magnetohydrodynamics of Protostellar Disks] it was shown that a
sufficiently highly conducting Keplerian disk that is threaded by a weak
magnetic field will be subject to the magnetorotational instability
(MRI) and may evolve into a turbulent state in which the field is
strongly amplified and has a {\em small-scale}, {\em disordered}
configuration. This turbulence has been proposed as the origin of the
effective viscosity that enables matter to accrete by transferring
angular momentum {\em radially} out along the disk plane. In this
chapter we focus on an alternative mode of angular momentum transport
that can play an important role in protostellar disks, namely, {\em
vertical} transport through the disk surfaces effected by the stresses
of a {\em large-scale}, {\em ordered} magnetic field that threads the
disk.

The possible existence of a comparatively strong, ``open'' magnetic
field over much of the extent of at least some circumstellar disks
around low- and intermediate-mass protostars is indicated by far-IR and
submillimeter polarization measurements, which have discovered an
ordered, hourglass-shaped field morphology on sub-parsec scales in
several molecular clouds (e.g., Schleuning 1998; Girart et al. 2006;
Kirby 2009). The polarized radiation is attributed to thermal emission
by spinning dust grains whose short axes are aligned along the magnetic
field (e.g., Lazarian 2007). The detected hourglass morphology arises
naturally in molecular cloud cores in which a large-scale magnetic field
provides dynamical support against the core's self-gravity (see
Sect.~\ref{subsec:forces}). In this picture the field is {\em
interstellar} in origin and is part of the Galactic magnetic field
distribution. As discussed in Sect.~\ref{sec:formation}, the inward
gravitational force can become dominant and the core then undergoes
dynamical collapse to form a central protostar and a circumstellar
disk. The magnetic field is dragged in by the infalling matter and could
in principle lead to a large-scale ``open'' field configuration in the
disk.

An ordered magnetic field that threads a disk can exert a magnetic
torque that removes angular momentum from the interior gas. This angular
momentum can be carried away along the field lines either by {\em
torsional Alfv\'en waves} in a process known as magnetic braking (see
Sect.~\ref{subsec:braking}) or through a rotating outflow in what is
known as a {\em centrifugally driven wind} (CDW; see
Sect.~\ref{subsec:wind}). These mechanisms could supplement or even
entirely supplant the radial transport along the disk plane invoked in
traditional disk models: by turbulent stresses as in the MRI scenario
mentioned above or through gravitational torques in a self-gravitating
disk as described in Chap.~IV [Disk Hydrodynamics]. In the case of
radial transport, the angular momentum removed from the bulk of the
matter is deposited into a small amount of gas at the outer edge of the
disk. In the case of vertical transport, this angular momentum is
deposited into a small fraction of the disk mass (the tenuous surface
layers of the disk) that is removed as a CDW or else (when magnetic
braking operates) into the ambient medium through which the torsional
Alfv\'en waves propagate. The introduction of this new transport channel
has profound implications to the structure and properties of disks in
which it is a major contributor to the angular momentum budget and
potentially also to the strong connection that has been found between
accretion and outflow phenomena in young stellar objects. This is
discussed in Sect.~\ref{sec:vertical}, where we also consider how to
determine which of the two possible angular momentum transport modes
(radial or vertical) operates at any given location in a magnetically
threaded disk and whether these two modes can coexist.

Of course, large-scale, ordered magnetic fields can also be produced
{\em in situ} by a {\em dynamo} process; we consider this alternative
possibility for the origin of the disk field in
Sect.~\ref{subsec:driving}. In the case of the Sun, high-resolution
observations made at extreme ultraviolet and soft X-ray wavelengths
and transformed into spectacular false-color images have revealed a
complex web of organized structures that appear as loops and
prominences near the stellar surface but simplify to a more uniform
distribution further out (e.g., Balogh et al. 1995). There is growing
evidence that Sun-like stars are already magnetically active in the
protostellar phase and, in fact, generate fields that are a thousand
times stronger than that of the present-day Sun. The dynamical
interaction between such a field and a surrounding accretion disk
through which mass is being fed to the nascent star could have
important evolutionary and observational consequences. We consider
this in Sect.~\ref{sec:disk-star}. We conclude with a summary and a
discussion of future research directions in Sect.~\ref{sec:conclude}.

\section{MHD of Magnetically Threaded Disks}
\label{sec:MHD}

Before getting into the specifics of the various topics outlined in
Sect.~\ref{sec:intro}, we present a general discussion of the
dynamical properties of ordered magnetic fields in relation to
protostellar disks and of the main methods that have been applied to
their study.

\subsection{Magnetic Forces}
\label{subsec:forces}

The magnetic force per unit volume on a magnetized fluid element is
given by
\begin{equation}
\label{eq:MHD1}
\frac{\vec{J}\times\vec{B}}{c} = - \nabla \left ( \frac{B^2}{8\pi}
\right ) + \frac{\vec{B}\cdot\nabla\vec{B}}{4\pi} \; ,
\end{equation}
where $\vec{J}$ is the current density, $\vec{B}$ is the
magnetic field vector, $c$ is the speed of light, and we substituted
\begin{equation}
\label{eq:MHD2}
\vec{J} = \frac{c}{4\pi}\nabla \times \vec{B}
\end{equation}
(neglecting the displacement current in Amp\`ere's Law on the assumption
that all speeds are $\ll c$) and used a vector identity to obtain the
two terms on the right-hand side of Eq.~(\ref{eq:MHD1}). As was already
noted in Chap.~V, the first term represents the magnetic pressure force
and the second one the force due to magnetic tension. To build intuition
it is useful to consider the magnetic {\em field lines}. Because of the
solenoidal (i.e., absence of monopoles) condition on the magnetic field,
\begin{equation}
\label{eq:MHD3}
\nabla \cdot \vec{B} = 0\; ,
\end{equation}
{\em the flux $\Psi$ of ``open'' field lines through the disk} (the
integral $\int \vec{B} \cdot \D\vec{S}$, where $\D\vec{S}$ is a disk
surface-area element) {\em is conserved}. This is a noteworthy result:
it says that, even if the disk is resistive in its interior, the flux of
magnetic field lines that thread it (also referred to as the {\em
poloidal flux}) will not be destroyed (although poloidal flux could be
added to, or removed from, the disk through its outer and inner
boundaries).  In the MHD picture, the field lines can be thought of as
rubber bands that exert {\em tension} of magnitude $B^2/4\pi$ directed
{\em along} the field and {\em pressure} of magnitude $B^2/8\pi$ {\em
normal} to their local direction (e.g., Parker 1979).

The nonmagnetic forces acting on a cloud core or a disk are the force of
gravity (mostly self-gravity in the case of a core and central-mass
gravity in the case of a nearly Keplerian disk), $-\varrho \nabla \Phi$
(where $\varrho$ is the mass density and $\Phi$ is the gravitational
potential), and the thermal pressure force $-\nabla P$.  There may be an
additional force, associated with the momentum flux of MHD waves, in a
turbulent system.  Such turbulence could give rise to an effective
viscosity, but, as is generally the case in astrophysical systems, the
effects of microphysical shear viscosity would remain negligible (e.g.,
Frank et al. 2002). Protostellar disks typically have sufficiently low
temperatures that the dominant forces in the disk plane are gravity and
magnetic tension, whereas in the vertical direction the thermal pressure
force is always important, balancing the downward force of gravity or of
the magnetic pressure gradient.

These notions can be illustrated by considering the equilibrium
configurations of cloud cores (e.g., Mouschovias 1976).  Within the
interstellar gas that ultimately ends up in the core the magnetic
field lines are initially nearly parallel to each other and the field
is well coupled to the matter (with the magnetic flux being, in the
jargon of MHD, ``frozen'' into the matter). Since a magnetic field
exerts no force parallel to itself, matter can slide relatively easily
along the field to form a flattened mass distribution. The oblate
structure predicted by this scenario is consistent with the observed
shapes of starless cores in the Orion giant molecular cloud (e.g.,
Tassis 2007).
\footnote{Note, however, that many cores do not exhibit a clear oblate
structure and that, in fact, many of the cores in Taurus are apparently
prolate and may have formed from the fragmentation of the filamentary
clouds in which they are embedded (e.g., Di Francesco et al. 2007).}
It is convenient to describe this configuration with the
help of a cylindrical coordinate system $\{r, \phi, z\}$ centered on
the forming star, with the $z$ axis aligned with the initial magnetic
field direction. A vertical hydrostatic equilibrium in which the
(upward) thermal (and possibly turbulent) pressure force balances the
(downward) gravitational force is established on a dynamical
(free-fall) time scale and is by and large maintained throughout the
subsequent evolution of the core, as verified by numerical simulations
(e.g., Fiedler \& Mouschovias 1993; Galli \& Shu 1993), including in
cases of filamentary clouds that are initially elongated in the field
direction (e.g., Nakamura et al. 1995; Tomisaka 1996).  Self-gravity
acting in the radial direction tends to pull the field lines
inward. The hourglass shape revealed by the polarization measurements
is produced because these field lines remain anchored in the cloud
envelope. This results in magnetic tension that is associated mainly
with the term $(B_z/4\pi)\D{B_r}/\D{z}$ in Eq.~(\ref{eq:MHD1}) and is
in complete analogy with the force exerted by a stretched rubber
band. The field morphologies revealed by the polarization measurements
are interpreted as arising from the approximate balance between this
force and radial gravity. This interpretation is supported by H I and
OH Zeeman measurements of the line-of-sight field amplitude and by
estimates of the plane-of-sky field strength using the measured
dispersion in the field orientation (the Chandrasekhar-Fermi method),
both of which typically imply roughly virialized cores (with ordered and
turbulent magnetic fields contributing approximately equally to the
overall support of the cloud against self-gravity; e.g., Ward-Thompson
et al. 2007).

In applying these ideas to the disks that form from the gravitational
collapse of cores, one should bear in mind the following two
points. First, the flux threading the disk is sufficiently strongly
concentrated that the bending of the field lines between the midplane
(where $B_r=B_\phi=0$ by reflection symmetry) and the disk surface can
be large enough to make $B_{r,{\rm s}}$ (where the subscript `s' denotes
the surface) comparable to $B_z$ (which, in turn, changes little between
the midplane and the surface if the disk is thin). Consequently, {\em
magnetic squeezing} by the $z$-gradient of the magnetic pressure
associated with the $B_r$ (and possibly also $B_\phi$) field components
can become comparable to, or even exceed, the downward force of
gravity. This is indeed a key property of the wind-driving disk models
discussed in Sect.~\ref{subsec:disk}. Second, the density scale height
$h$ in the disk typically satisfies $h\ll r$, implying a rapid decrease
of the density with $z$ even as the magnetic field amplitude changes
little for $z \la r$. Therefore, magnetic forces generally dominate all
other forces on scales $\la r$ above the disk surface, and the field
there assumes a so-called ``force-free'' field configuration
($\vec{J}\times\vec{B}\approx 0$). According to Eq.~(\ref{eq:MHD1}), in
this case the magnetic tension force has to balance the magnetic
pressure force, which points outward (as $B_z^2$ increases toward the
center). This implies that the field lines in ths region assume a
vase-like, ``concave in'' morphology (i.e., bending toward the vertical
axis), in contradistinction to the hourglass-like, ``convex out'' shape
that they have inside the disk. This behavior accounts for the initial
collimation of disk-driven MHD winds (see Sect.~\ref{subsec:wind}).

\subsection{Magnetic Braking}
\label{subsec:braking}

Plucked rubber bands (or strings) carry waves whose phase velocity is
the square root of the ratio of the tension to the mass density. Using
again the analogy to magnetic field lines, one immediately obtains the
phase speed of (transverse) Alfv\'en waves, $v_{{\rm A}z}= B_z/\sqrt{4\pi
\varrho}$ (taking the background field to point in the $\mathbf{\hat z}$
direction). When the transverse magnetic field of the wave points in the
azimuthal direction, the
corresponding (torsional) Alfv\'en wave carries angular momentum. A
rotating molecular cloud core or disk will twist the magnetic ``rubber
bands'' that thread it. The degree of twisting fixes the pitch
$|B_{\phi,{\rm s}}/B_z|$ of the field lines and is determined, in turn,
by the ``load'' on the other end of the band (i.e., by the inertia of
the external matter to which the field lines are coupled). This twisting
represents a transfer of angular momentum from the core or disk to the
external medium (subscript `ext'). One can thus determine $B_{\phi,{\rm
s}}$ by equating the torque per unit area (normal to $\mathbf{\hat z}$)
exerted on each of the two surfaces of the flattened core or disk, $r
B_z B_{\phi,{\rm s}}/4\pi$, to minus the rate per unit area of angular
momentum carried by the waves from each surface and deposited in the
(initially nonrotating) ambient medium, $-\varrho_{\rm ext} r_{\rm
ext}^2 \Omega_B v_{\rm A ext}$, where $\Omega_B$ is the angular
velocity of the field line (which is conserved along the field under
ideal-MHD conditions; see Sect.~\ref{subsec:wind}). Poloidal flux
conservation along the field makes it possible to relate the radius $r$
and field $B_z$ in the core or disk to the corresponding quantities in
the external medium, $B_z r \D{r} = B_{z,{\rm ext}} r_{\rm ext}
\D{r_{\rm ext}}$, where the ambient field (assumed to be uniform on
large scales) can be expressed in terms of the poloidal flux $\Psi$ by
$B_{z,{\rm ext}} = \Psi/\pi r_{\rm ext}^2$. One then obtains (cf. Basu
\& Mouschovias 1994)
\begin{equation}
\label{eq:MHD4}
B_{\phi,{\rm s}}=-\, \frac{\Psi}{\pi r^2}\, \frac{v_{B\phi}}{v_{\rm
A ext}}\; ,
\end{equation}
where $v_{B\phi} = r \Omega_B$. The general meaning of
$v_{B \phi}$ is discussed in Sect.~\ref{subsec:nonideal}; unless
Ohm diffusivity dominates in the core/disk, it can be identified
with the midplane angular velocity of the particles into which the field
lines are frozen (see Eq.~(\ref{eq:wind8})).

\subsection{Centrifugal Wind Driving}
\label{subsec:centrifugal}

As noted in Sect.~\ref{subsec:forces}, the dynamics just above the disk
surface is magnetically dominated, i.e., the magnetic energy density
there is larger than the thermal, gravitational, and kinetic energy
densities of the gas. The comparatively large electrical conductivity in
this region implies that the poloidal ($r,\, z$) gas
velocity is parallel to the poloidal magnetic field (see
Sect.~\ref{subsec:wind}) and the bulk particle motions can be
approximated as those of beads along rotating, rigid, massless wires
(Henriksen \& Rayburn 1971).  This mechanical analogy is useful for
deriving the criterion for the centrifugal launching of disk winds
(Blandford \& Payne 1982). We neglect thermal effects in this derivation
and, correspondingly, regard the disk as being infinitely thin. Since
the field geometry only varies on a scale $\sim r$ in the force-free
zone, very close to the disk surface the field lines can be regarded as
being nearly straight. Considering thus a straight wire that intersects
a Keplerian disk at a distance $r_0$ from the center and makes an angle
$\theta$ to the disk normal, the balance of gravitational and
centrifugal forces {\em along} the wire implies that, in equilibrium,
the effective potential $\Phi_{\rm eff} (y)$ satisfies
\begin{equation}
\label{eq:MHD5}
\frac{\partial \Phi_{\rm eff}}{\partial y} = \frac{y + \sin{\theta}}{(1 + 2 y
\sin{\theta} + y^2 )^{3/2}} - y \sin^2{\theta} - \sin{\theta} = 0\; ,
\end{equation}
where the dimensionless variable $y \equiv s/r_0$ measures the distance
$s$ along the wire. The equilibrium is unstable when $\partial ^2
\Phi_{\rm eff} /\partial y^2<0$ (corresponding to a local maximum of
$\Phi_{\rm eff}$) , which at $y =0$ occurs for $\theta >30^{\circ}$.
Hence, centrifugal driving sets in if
\begin{equation}
\label{eq:MHD6}
\frac{B_{r,{\rm s}}}{B_z} > \frac{1}{\sqrt{3}}\; .
\end{equation}
This wind-launching criterion plays a key role in wind-driving disk
models. It is worth noting that, in contrast to typical stellar winds,
in which the outflowing gas must ``climb out'' of basically the entire
gravitational potential well at the stellar surface, in the case of
outflows from a rotationally supported, infinitely thin disk the depth
of the (gravitational + centrifugal) potential well is lower by a factor
of 2 on account of the rotation, and gas can in principle ``escape to
infinity'' without any added thermal push if there is a sufficiently
strongly inclined, rigid channel (the magnetic field lines).  Real
disks, however, are not completely cold and therefore have a finite
thickness. When this is taken into account, the effective potential
attains a maximum at some height above the disk surface, and thermal
pressure forces are required to lift gas from the disk up to that point
(e.g., Ogilvie 1997). This and some other properties of CDWs are
considered in Sect.~\ref{subsec:wind}.

\subsection{Nonideal MHD}
\label{subsec:nonideal}

Unlike the approach of classical electrodynamics, in which it is common
to consider the magnetic field as being generated by current flows
according to the Biot-Savart law, in MHD practice it is often more
illuminating to focus on the magnetic field and to regard the current
density as a subordinate quantity that is determined, through
Eq.~(\ref{eq:MHD2}), by how the field is shaped by its interaction with
matter (e.g., Parker 2007). The current, in turn, helps to determine the
neutral-fluid--frame (denoted by a prime) electric field
$\vec{E^\prime}$ according to Ohm's law
\begin{equation}
\label{eq:MHD7}
\vec{J} = \tens{\sigma}\cdot \vec{E^\prime} = \sigma_{\rm O}
\vec{E^\prime}_\parallel + \sigma_{\rm H} \vec{\hat B} \times
\vec{E^\prime}_\perp + \sigma_{\rm P} \vec{E^\prime}_\perp \; ,
\end{equation}
where the conductivity has been expressed as a tensor to take account of
the inherent anisotropy that an ordered magnetic field induces in the
motions of charged particles. Here the subscripts $\parallel$ and
$\perp$ denote vector components that are, respectively, parallel and
perpendicular to the unit vector $\vec{\hat B}$, whereas $\sigma_{\rm
O}$, $\sigma_{\rm H}$, and $\sigma_{\rm P}$ are, respectively, the
Ohm, Hall, and Pedersen conductivity terms.

Under ideal-MHD conditions, the conductivity is effectively infinite and
the comoving electric field vanishes. This is an adequate approximation
for describing the dynamics of a disk wind or of the medium surrounding
the core/disk. However, within the core or disk themselves the degree
of ionization is generally low, so finite conductivity effects must be
taken into account in the dynamical modeling.
When the conductivity is low, each charged particle
species (denoted by a subscript `j') develops a drift velocity
$\vec{v}_{\rm d,j}\equiv \vec{v}_{\rm j} - \vec{v}$ with respect to the
neutral fluid velocity (which we approximate by the average fluid
velocity $\vec{v}$, as appropriate for a weakly ionized medium). The
drift velocities can be calculated from the equations of motion of these
species, each of which is well approximated by a steady-state balance
between the Lorentz force and the drag force $\vec{F}_{\rm
nj}$ exerted by collisions with the neutrals,
\begin{equation}
\label{eq:MHD8}
Z_{\rm j} e\left(\vec{E^\prime} + \frac{\vec{v}_{\rm d,j}}{c} \times
\vec{B}\right) = - \vec{F}_{\rm nj} = m_{\rm j} \gamma_{\rm j} \varrho
\vec{v}_{\rm d,j} \;,
\end{equation}
where $Z_{\rm j}$ is the (signed) particle charge in units of the
electronic charge $e$, $\gamma_{\rm j} \equiv <\sigma v>_{\rm j}/(m_{\rm
j} + m)$, and $<\sigma v>_j$ is the rate coefficient for collisional
momentum transfer between particles of mass $m_{\rm j}$ and neutrals (of
mass $m$ and mass density $\varrho$). This collisional interaction in
turn exerts a force $\vec{F}_{\rm jn} = - \vec{F}_{\rm nj}$ on the
neutrals. The degree of coupling between a given charged species and the
magnetic field is measured by the {\em Hall parameter}, defined as the
ratio of the particle's gyrofrequency to its collision frequency
$\nu_{\rm j n} = \gamma_{\rm j} \varrho$ with the neutrals:
\begin{equation}
\label{eq:MHD9}
\beta_{\rm j} \equiv \frac{|Z_{\rm j}|eB}{m_{\rm j} c}\frac{1}{\gamma_{\rm
j}\varrho} \; .
\end{equation}
In this expression, $B \equiv |\mathbf{B}|\, sgn\{B_z\}$ is the {\em
signed} magnetic field amplitude, with the sign
introduced to keep the dependence of the Hall conductivity on the
magnetic field polarity (see Eq.~(\ref{eq:MHD13}) below).
When $|\beta_{\rm j}| \gg 1$ the coupling is good and the collision term
can be neglected in comparison with the magnetic term; in this case the
Lorentz force is approximately zero, corresponding to the near vanishing
of the electric field in the charged particle's frame. On the other
hand, when $|\beta_{\rm j}| \ll 1$ the coupling is poor; in this case the
magnetic term in Eq.~(\ref{eq:MHD8}) can be neglected.

By summing up Eq.~(\ref{eq:MHD8}) over the particle species one obtains
\begin{equation}
\label{eq:MHD10}
\sum_{\rm j} n_{\rm j} \vec{F}_{\rm jn} = \frac{\vec{J}\times\vec{B}}{c}\; .
\end{equation}
This shows explicitly that, in a weakly ionized medium, the Lorentz
force (which acts only on the charged particles) is transmitted to the
bulk of the matter only through a collisional drag, which involves a
relative motion between the charged and neutral components. Therefore,
if magnetic forces are important in such a medium, its structure is {\em
inherently not static}. This is exemplified by the behavior of a
magnetically supported molecular cloud core (e.g., Shu et al. 1987).
The magnetic field that threads the core is anchored in
the comparatively well-ionized cloud envelope, but in the core's
interior the degree of ionization is low and the magnetic tension force
is transmitted to the predominantly neutral gas through ion--neutral
drag. Since the ions (taken in what follows to constitute a single
species denoted by the subscript `i') are well coupled to the magnetic
field and thus remain nearly fixed in space, the associated {\em
ambipolar diffusion} drift entails an inward motion of the neutral
particles toward the center of mass. If the evolution lasts longer than
the ambipolar diffusion time ($\sim (R/v_{\rm A})^2\gamma_{\rm
i}\varrho_{\rm i}$, which can be inferred from Eqs.~(\ref{eq:MHD14})
and~(\ref{eq:MHD18}) below), the central concentration will become large
enough to cause the core to become gravitationally unstable, and
dynamical collapse will ensue.

Expressing the current density in terms of the charged particles'
drifts,
\begin{equation}
\label{eq:MHD11}
\vec{J}= \sum_{\rm j} e n_{\rm j} Z_{\rm j} \vec{v}_{\rm d,j}
\end{equation}
(where the charged particles have number densities $n_{\rm j}$ and satisfy
charge neutrality, $\sum_{\rm j} n_{\rm j} Z_{\rm j} = 0$), and using
Eq.~(\ref{eq:MHD8}), one can solve for the conductivity tensor
components in Eq.~(\ref{eq:MHD7}):
\begin{equation}
\label{eq:MHD12}
\sigma_{\rm O} = \frac{ec}{B}\sum_{\rm j} n_{\rm j}|Z_{\rm j}|\beta_{\rm
j}\;,
\end{equation}
\begin{equation}
\label{eq:MHD13}
\sigma_{\rm H} = \frac{ec}{B}\sum_{\rm j} \frac{n_{\rm j}|Z_{\rm j}|}
{
1+\beta_{\rm j}^2}\; ,
\end{equation}
and
\begin{equation}
\label{eq:MHD14}
\sigma_{\rm P} = \frac{ec}{B}\sum_{\rm j} \frac{n_{\rm j}|Z_{\rm
j}|\beta_{\rm j}}{ 1+\beta_j^2}
\end{equation}
(e.g., Cowling 1976; Wardle \& Ng 1999). Note that $\sigma_{\rm H}$ (and
correspondingly the Hall term in Ohm's law) depends on an odd power of
the magnetic field amplitude and can therefore have either a positive or
a negative sign (reflecting the magnetic field polarity). This leads to
qualitative differences in the behavior of the disk solutions in the
Hall regime for positive and negative values of $B_z$.

The conductivity tensor formalism is useful for constructing realistic
disk models in which the relative magnitudes of the different
conductivities can change as a function of height even at a single
radial location as a result of the variation in the density and in the
dominant ionization mechanism as one moves between the disk surface and
the midplane (see Sect.~\ref{subsec:disk}). It is nevertheless
instructive to relate this formalism to the classical diffusivity
regimes considered in Chap.~V. This is best done by solving for the
fluid-frame electric field,
\begin{equation}
\label{eq:MHD15} c
\vec{E^\prime} = \eta_{\rm O} \nabla \times \vec{B} + \eta_{\rm H} (
\nabla \times \vec{B}) \times \vec{\hat B} + \eta_{\rm A} (\nabla \times
\vec{B})_{\perp}\; ,
\end{equation}
where the Ohm, Hall, and ambipolar diffusivities are given,
respectively, by
\begin{equation}
\label{eq:MHD16}
\eta_{\rm O} = \frac{c^2}{4 \pi \sigma_{\rm O}} \; ,
\end{equation}
\begin{equation}
\label{eq:MHD17}
\eta_{\rm H} = \frac{c^2}{4 \pi \sigma_{\perp}} \frac{\sigma_{\rm
H}}{\sigma_{\perp}}\; ,
\end{equation}
and
\begin{equation}
\label{eq:MHD18}
\eta_{\rm A} = \frac{c^2}{4 \pi \sigma_{\perp}} \frac{\sigma_{\rm
P}}{\sigma_{\perp}} - \eta_{\rm O}\; ,
\end{equation}
with $\sigma_\perp \equiv (\sigma_{\rm H}^2 + \sigma_{\rm P}^2)^{1/2}$
(e.g., Mitchner \& Kruger 1973;
Nakano et
al. 2002). The three distinct regimes are delineated by
\begin{eqnarray}
\label{eq:MHD19}
\sigma_{\rm O} \approx \sigma_{\rm P} \gg |\sigma_{\rm H}| && \quad\quad
\rm{Ohm} \nonumber\\
\sigma_{\rm O} \gg |\sigma_{\rm H}| \gg \sigma_{\rm P} && \quad\quad
\rm{Hall} \\
\sigma_{\rm O} \gg \sigma_{\rm P} \gg |\sigma_{\rm H}| && \quad\quad
\rm{ambipolar} \nonumber \; .
\end{eqnarray}
When ions and electrons (subscript `e') are the only charged species,
one has $\eta_{\rm H} = \beta_{\rm e} \, \eta_{\rm O}$ and $\eta_{\rm A} =
\beta_{\rm i}\, \beta_{\rm e}\, \eta_{\rm O}$, with the respective Hall
parameters given numerically by $\beta_{\rm i} = q \beta_{\rm e} \approx
0.46 \, (B/ n_{13})$ (e.g., Draine et al. 1983; where $q \approx 1.3 \ee
-3 \, T_2^{1/2}$, $n_{13} = n_{\rm H}/(10^{13}\, {\rm cm}^{-3})$, $B$ is
measured in Gauss, $T_2 = T/(10^2\, {\rm K}$), and the mean ion mass is
$m_{\rm i} = 30 \, m_{\rm H}$). In this case the three classical regimes
correspond, respectively, to the Hall parameter ranges $|\beta_{\rm i}|
\ll |\beta_{\rm e}| \ll 1$ (Ohm), $|\beta_{\rm i}| \ll 1 \ll |\beta_{\rm e}|$
(Hall), and $1 \ll |\beta_{\rm i}| \ll |\beta_{\rm e}|$ (ambipolar).
\begin{figure}
\centering
\includegraphics[height=8cm]{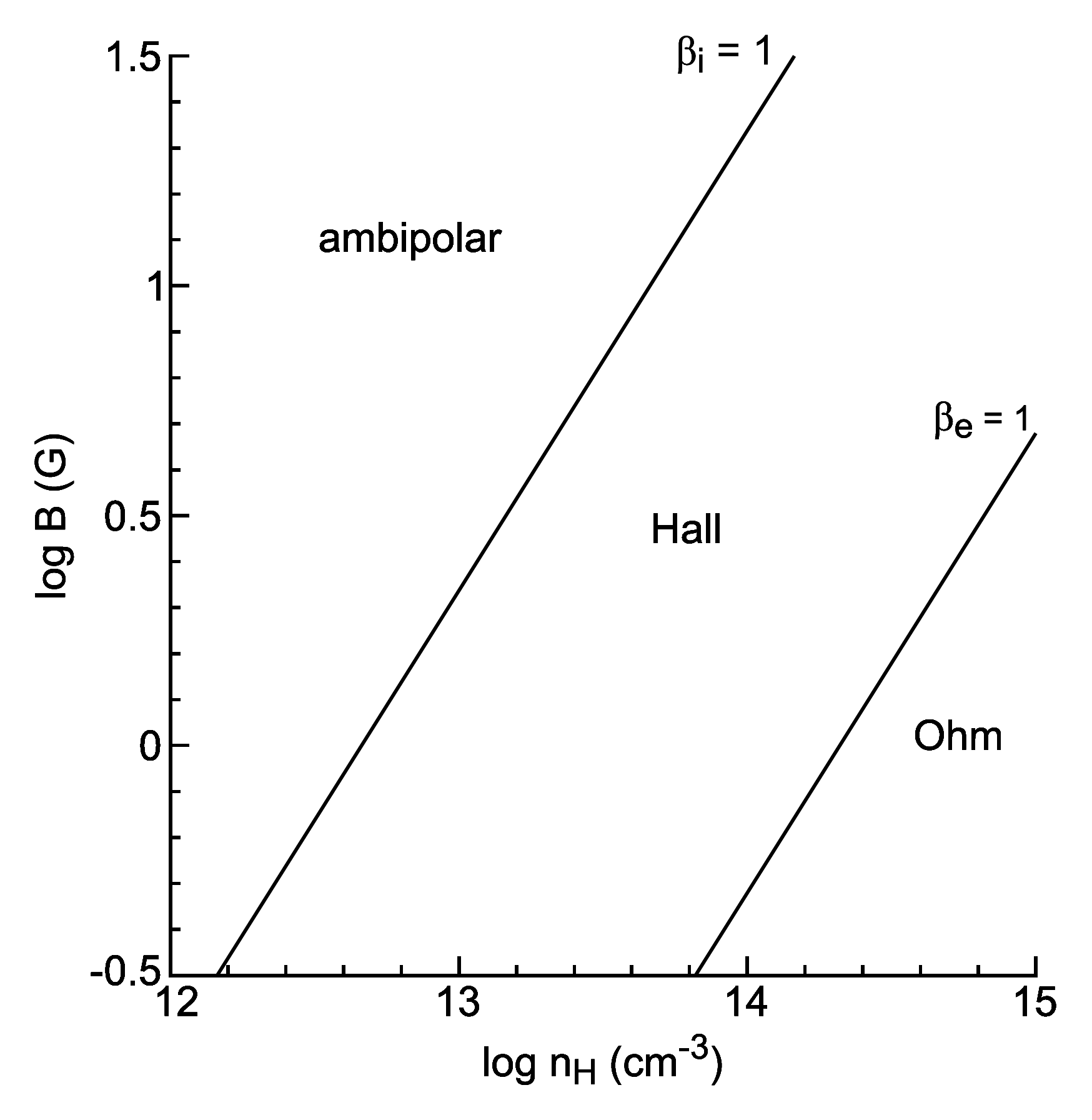}
\caption{Magnetic diffusivity regimes in the $\log{n_{\rm
H}}-\log{B}$ plane for $T = 280\,$K.} \label{fig6_1}
\end{figure}

Figure~\ref{fig6_1} shows the regions of ambipolar, Hall and Ohm
dominance in the $\log{n_{\rm H}}-\log{B}$ plane.\footnote{For a
complementary figure in the $\log{n_{\rm H}}-\log{T}$ plane, see Fig.~2
of Chap.~V.} Molecular cloud cores and the outer regions ($\ga 10\, {\rm
AU}$) of protostellar disks typically correspond to the ambipolar
regime, as do the disk surface regions at smaller radii; disk midplanes
on scales $\sim 1-10\, {\rm AU}$ are often dominated by Hall
diffusivity, whereas Ohm diffusivity characterizes the midplanes of
disks on scales $\sim 0.1-1\, {\rm AU}$. In the innermost ($\la
0.1\,$AU) regions of the disk, where the temperature increases above
$\sim 10^3\,$K and the gas becomes collisionally ionized (e.g., Gammie
1996; Li 1996a), anomalous Ohm diffusivity (the enhanced drag between
positive and negative charge carriers due to scattering off
electromagnetic waves generated by current-driven plasma instabilities)
might play a role. Note that the precise extent of the different regimes
depends on the radial profile of the disk column density, since the
latter determines the degree of penetration of the ionizing radiation or
cosmic rays (see Sect.~\ref{subsubsec:ionize}). For a given mass
accretion rate, the column density depends on the nature of the angular
momentum transport mechanism. In particular, transport by a large-scale,
ordered magnetic field is generally more efficient than transport by a
small-scale, disordered field,\footnote{This can be seen by representing
the $r\phi$ turbulent stress component as $\alpha P$ (where the constant
$\alpha$ is typically $< 1$; cf. Balbus \& Papaloizou 1999), so the
ratio of the torques exerted by the ordered field and by the turbulent
stress is $(-B_z B_{\phi,{\rm s}}/4\pi \langle P \rangle )(r/\alpha H)$
(where $\langle\rangle$ denotes vertical averaging over the disk
half-thickness $H$), which is typically $\gg 1$.} resulting in higher
inflow speeds and correspondingly lower column densities and higher
degrees of ionization in CDW-mediated accretion than in MRI-based
turbulent disks.

The behavior of the electric field is governed by Faraday's Law,
\begin{equation}
\label{eq:MHD20}
\frac{\partial \vec{B}}{\partial t} = - c \nabla
\times \vec{E}\; ,
\end{equation}
where $\vec{E} = \vec{E^\prime}-\vec{v} \times \vec{B}/c$ is the
lab-frame electric field. In view of Eq.~(\ref{eq:MHD15}), when the
resistive term in the expression for $\vec{E^\prime}$ can be neglected
and the only charge carriers are ions and electrons, one can express the
ambipolar and Hall contributions to $c\vec{E}$ as $(\vec{v}-\vec{v}_{\rm
i})\times \vec{B}$ and $(\vec{v}_{\rm i}-\vec{v}_{\rm e})\times
\vec{B}$, respectively (see K\"onigl 1989 and Chap.~V), yielding
$\partial \vec{B}/\partial t = \nabla \times (\vec{v}_{\rm e} \times
\vec{B})$, which indicates that in this case the field lines are frozen
into the electrons (the particle species with the highest mobility
$|Z_{\rm j}|e/m_{\rm j}$). It is also seen that the ideal-MHD limit
$\vec{E} = - \vec{v}\times \vec{B}/c$ is approached when the
ion--neutral and ion--electron drift speeds are small in comparison with
the bulk speed. It further follows that, in the ambipolar regime, the
ions and electrons drift together relative to the neutrals (i.e.,
$|\vec{v}_{\rm i} - \vec{v}_{\rm e}| \ll |\vec{v}_{\rm i} - \vec{v}|$),
so the field is effectively frozen also into the ions, whereas in the
Hall regime the ions and neutrals essentially move together and the
electrons drift relative to them ($|\vec{v}_{\rm i} - \vec{v}_{\rm e}|
\gg |\vec{v}_{\rm i} - \vec{v}|$). More generally, one can define an
effective velocity $\vec{v}_B$ for the poloidal flux surfaces that
applies also in the Ohm regime, when these surfaces are no longer frozen
into any particle species (although, as noted in
Sect.~\ref{subsec:forces}, they continue to maintain their
identity). Focusing on the midplane, where $B=B_z$, this is done through
the relation $c\vec{E}= - \vec{v}_B \times \vec{B}$ (cf. Umebayashi \&
Nakano 1986). As we just observed, in the absence of Ohm
diffusivity (and still considering only ion and electron charges)
this relation is satisfied if one substitutes $\vec{v}_{\rm e}$ for
the velocity, which verifies that $\vec{v}_B= \vec{v}_{\rm e}$ in this
case. However, in the Ohm regime the azimuthal field-line velocity
\begin{equation}
\label{eq:MHD21}
v_{B \phi, 0} = - c E_{r,0}/B_0
\end{equation}
can differ from that of the most mobile charged particle component, and
similarly for the radial flux-surface velocity
\begin{equation}
\label{eq:MHD22}
v_{Br,0} = cE_{\phi,0}/B_0\; ,
\end{equation}
where the subscript `0' denotes the midplane.

A departure from ideal MHD may lead to energy dissipation at a rate (per
unit volume) $\vec{J} \cdot \vec{E}^\prime$. As can be inferred from
Eqs.~(\ref{eq:MHD7}) and~(\ref{eq:MHD15}), both the Ohm and ambipolar
terms in Ohm's law contribute to Joule heating, but not the Hall term.
It may perhaps seem puzzling that energy is dissipated in the ambipolar
regime even though the field lines remain frozen to the charged
particles, but one can directly associate the heating in this case with
the ion--neutral collisional drag: $\vec{J} \cdot \vec{E}^\prime \approx
n_{\rm i} \vec{v}_{\rm d,i} \cdot \vec{F}_{\rm in}$ (assuming $Z_{\rm i}
\gg 1$; see Eqs.~(\ref{eq:MHD8}) and~(\ref{eq:MHD11})). Joule dissipation
would be the main internal heating mechanism in disk regions where
angular momentum transport is dominated by a large-scale, ordered
field. Ambipolar heating, in particular, could also play an important
role in the thermal structure of disk-driven winds (see
Sect.~\ref{subsec:wind}).

Just as we arrived at a combination of physical variables that measures
the degree of coupling between a charged particle and the magnetic field
(namely, the Hall parameter; Eq.~(\ref{eq:MHD9})), we will find it
useful to identify an analogous measure for the neutrals. It turns out
that an appropriate parameter combination of this type for
Keplerian disks, which plays a similar role in MRI-based systems
(where the neutrals couple to a small-scale, disorderd field) and in
wind-driving disks (where the coupling is to a large-scale, ordered
field), is the {\em Elsasser number}
\begin{equation}
\label{eq:MHD23}
\Lambda \equiv \frac{v_{\rm A}^2}{\Omega_{\rm K}\eta_{\perp}}\; ,
\end{equation}
where
\begin{equation}
\label{eq:MHD24}
\eta_\perp =  \frac{c^2}{4\pi \sigma_\perp}
\end{equation}
is the ``perpendicular'' magnetic diffusivity and $\Omega_{\rm K}$ is
the Keplerian angular velocity. This parameter is $\gg 1$ and $\ll 1$
for strong and weak neutral--field coupling, respectively.\footnote{The
parameter $\Lambda$ is distinct from the Lundquist number $S\equiv
v_{\rm A} L/\eta_{\rm O})$ and from the magnetic Reynolds number
Re$_{\rm M}\equiv V L/\eta_{\rm O}$ (where $V$ and $L$ are
characteristic speed and length-scale, respectively), which have been
used in similar contexts in the literature. It was first introduced
into the protostellar disk literature in this form (but using a
different symbol) by Wardle (1999).}  The Elsasser number is used in
planetary dynamo theory to measure the ratio of the Lorentz to Coriolis
forces. Since both the MRI mechanism and the field-mediated vertical
angular momentum transport involve magnetic coupling to Keplerian
rotation, it is not surprising that this parameter also arises in the
disk context. In the three main conductivity regimes it reduces to
\begin{equation}
\label{eq:MHD25}
\Lambda ~ \Rightarrow ~ \left \{ \matrix{ ~ \frac{\gamma_{\rm i} \varrho_{\rm
i}}{\Omega_{\rm K}} ~ \equiv ~ \Upsilon \quad\quad \rm{ambipolar} \cr\cr
\beta_{\rm i}\, \Upsilon \quad\quad \rm{Hall} \cr\cr
\beta_{\rm e}\, \beta_{\rm i}\, \Upsilon \quad\quad \rm{Ohm} \; ,}\right.
\end{equation}
where we again assumed, for simplicity, that the only charged
particles are ions and electrons. The ambipolar limit has a clear
physical meaning: it represents the ratio $\Upsilon$ of the
Keplerian rotation time to the neutral--ion momentum-exchange
time. This parameter has emerged as the natural measure of the
field--matter coupling in the wind-driving disk models of Wardle \&
K\"onigl (1993; see Sect.~\ref{subsec:disk}) as well as in studies of
the linear (e.g., Blaes \& Balbus 1994) and nonlinear (e.g., Mac Low
et al. 1995; Brandenburg et al. 1995; Hawley \& Stone 1998) evolution
of the MRI in such disks. Indeed, since the ions are well coupled to
the field in this limit ($|\beta_{\rm i}|\gg 1$), the neutrals will be
well coupled to the field if their momentum exchange with the charged
particles (which is dominated by their interaction with the
comparatively massive ions) occurs on a time scale that is short in
comparison with the dynamical time $\Omega_{\rm K}^{-1}$
(corresponding to $\Upsilon > 1$). In the Hall regime $\Lambda$
is equal to this ratio of time scales multiplied by $|\beta_{\rm
i}|$. This product has figured prominently in the classification of
wind-driving disk solutions (K\"onigl et al. 2010)
as well as in linear studies of disk MRI in this regime (e.g., Wardle
1999; Balbus \& Terquem 2001). In this case, too, it has a clear physical
meaning. In contradistinction to the ambipolar regime, the {\em ions
are not well coupled} to the field in the Hall limit ($|\beta_{\rm
i}|\ll 1$). In order for the neutrals to be well coupled to the field
it is, therefore, not sufficient for them to be well coupled to the
ions ($\Upsilon > 1$); rather, the product $|\beta_{\rm
i}|\Upsilon$ must be $>1$ in this case. In the Ohm regime, even
{\em the electrons are not well coupled} to the field ($|\beta_{\rm e}|
\ll 1$), so now the product $\beta_{\rm e} \beta_{\rm i} \Upsilon$
has to exceed 1 to ensure an adequate coupling of the neutrals to the
field. The condition $\Lambda \ga 1$ in fact characterizes both the
linear (e.g., Jin 1996) and the nonlinear (e.g., Sano et al. 2002)
behavior of the MRI in this regime, and the Elsasser number is again a
natural parameter for classifying viable wind-driving disk solutions
in this case (with each sub-regime corresponding to a distinct lower
bound on $\Lambda$; see K\"onigl et al. 2010).

\subsection{Similarity Solutions}
\label{subsec:similar}

Even the most simplified models of magnetically threaded protostellar
disks lead to systems of nonlinear partial differential equations
(PDEs). In some cases, such as the disk--star interaction considered in
Sect.~\ref{sec:disk-star}, it has proven necessary to carry out
full-fledged numerical simulations to get definitive answers about the
expected behavior. However, for other aspects of the problem, such as
the disk formation (Sect.~\ref{sec:formation}) and the wind-driving disk
structure (Sect.~\ref{sec:vertical}), it has been possible to make
analytic (or, rather, semianalytic) progress by looking for {\em
similarity solutions}, an approach that converts the PDEs into ordinary
differential equations (ODEs; e.g., Landau \& Lifshitz 1987).
This type of solution does not incorporate either inner or outer
radial boundaries, so it can be justified only at radii that
are sufficiently far removed from the disk's actual edges. However, the
basic properties revealed by these solutions have invariably been
confirmed by simulations, and the analytic approach has the advantage of
making the qualitative behavior more transparent and of making it possible
to better investigate the parameter dependence of the solutions and to
incorporate more physics and dynamic range than is yet possible
numerically. The semianalytic and numerical approaches have thus played
complementary roles in shedding light on these questions.

In the case of the disk-formation problem, the separate
dependences on the spatial scale $r$ and the time $t$ are subsumed into
a single dependence on the dimensionless combination $r/Ct$, taking
advantage of the fact that the isothermal sound speed $C$ is nearly
uniform (on account of the efficient cooling by dust grains) in cloud
cores and in the outer regions of circumstellar disks. The
self-similarity of the derived solutions is expressed by the fact that
they depend solely on the ratio $r/t$ and not on the separate values of
$r$ and $t$. In the case of the stationary, axisymmetric wind-driving disk
models, the separate dependences on the spatial coordinates $z$ and $r$
are subsumed into a single dependence on the dimensionless combination
$z/r$ (or, equivalently, on $z(r,r_0)/r_0$, where $z(r,r_0)$ describes the
shape of a magnetic field line that intersects the midplane at
$r_0$). In these {\em radially self-similar} models all physical
variables scale as power laws of the {\em spherical} radius
$R=(z^2+r^2)^{1/2}$ and the differential equations depend only on the
polar angle $\theta = {\rm arc}\cot{(z/r)}$. Further details are given in
Secs.~\ref{sec:formation} and~\ref{sec:vertical}, respectively.

\section{Disk Formation and Early Evolution}
\label{sec:formation}

Rotationally supported circumstellar disks evidently originate in the
collapse of self-gravitating, rotating, molecular cloud cores. Molecular
line observations (e.g., Goodman et al. 1993; Kane \& Clemens 1997) have
established that a majority of dense ($\ga 10^4\, {\rm cm}^{-3}$) cloud
cores show evidence of rotation, with angular velocities $\sim
3{\times}10^{-15}-10^{-13}\, {\rm s}^{-1}$ that tend to be uniform on
scales of $\sim 0.1\, {\rm pc}$, and with specific angular momenta in
the range $\sim 4{\times}10^{20}-3{\times}10^{22}\, {\rm cm}^2\, {\rm
s}^{-1}$.  The cores can transfer angular momentum to the ambient gas by
magnetic braking (Sect.~\ref{subsec:braking}), and this mechanism also
acts to align their angular momentum vectors with the local large-scale
magnetic field (e.g., Machida et al. 2006). This alignment occurs
on a dynamical time scale and hence can be achieved even in cores whose
lifetimes are not much longer than that (as in certain models of the
turbulent ISM; e.g., Elmegreen 2000).  The dynamical collapse might
occur as a result of mass rearrangement in the core on the ambipolar
diffusion time (e.g., Mouschovias et al. 2006; see
Sect.~\ref{subsec:nonideal}) or sooner if the core is close to the
critical mass for collapse (the effective Jeans mass) from the start
(e.g., Elmegreen 2007).  In this section we consider the core collapse
problem in the context of angular momentum transport by a large-scale,
ordered magnetic field; an alternative scenario involving gravitational
torques is discussed in Chap.~IV.

Once dynamical collapse is initiated and a core goes into a
near-free-fall state, the specific angular momentum is expected to be
approximately conserved, resulting in a progressive increase in the
centrifugal force that eventually halts the collapse and gives rise to a
rotationally supported disk on scales $\sim 10^2\, {\rm AU}$. These
expectations are consistent with the results of molecular-line
interferometric observations of contracting cloud cores (e.g., Ohashi et
al. 1997; Belloche et al. 2002). In this picture,
the disk rotation axis should be aligned with the direction of the
large-scale magnetic field that threads the cloud. Observations have not
yet yielded a clear-cut answer to whether this is indeed the case in
reality (e.g., M\'enard \& Duch\^ene 2004; Vink et al. 2005), and it is
conceivable that the field in some cases is too weak to align the core's
rotation axis or even control its contraction, or that additional
processes (such as fragmentation, disk warping, etc.) play a role. In
what follows we nevertheless continue to pursue the implications of the
basic magnetically supported cloud picture.

\subsection{Modeling Framework}
\label{subsec:framework}

Since the gravitational collapse time is much shorter than the local
ambipolar diffusion time in the core, the magnetic field lines at first
move in with the infalling matter. However, once the central mass begins
to grow, ambipolar diffusion becomes important within the gravitational
``sphere of influence'' of the central mass (Ciolek \& K\"onigl 1998;
Contopoulos et al. 1998).  When the incoming matter enters this region,
it decouples from the field and continues moving inward. The decoupling
front, in turn, moves outward and steepens into an {\em ambipolar
diffusion shock}. The existence of this C-type MHD shock was first
predicted by Li \& McKee (1996).  The transition from a nearly freely
falling, collapsing core to a quasi-stationary, rotationally supported
disk involves a strong deceleration in a {\em centrifugal shock}. This
shock typically occurs at a smaller radius than the ambipolar-diffusion
shock and is hydrodynamic, rather than hydromagnetic, in nature.

Based on the arguments presented in Sect.~\ref{subsec:forces} it should
be a good approximation to assume that the gas rapidly establishes force
equilibrium along the field and therefore to consider only motions in
the radial direction.\footnote{In reality, mass can also be added to
the system from the polar directions. Numerical simulations of
axisymmetric collapse in which new mass is added {\em only} through
vertical infall have tended to produce disk-to-star mass ratios $\sim
1$, much higher than typically observed. A better agreement with
observations may, however, be obtained when nonaxisymmetric density
perturbations and resultant gravitational torques are included in the
calculations (e.g., Vorobyov \& Basu 2007).} One can obtain semianalytic
solutions for this effectively 1D time-dependent problem by postulating
$r-t$ self-similarity, with a similarity variable
\begin{equation}
\label{eq:form1}
x ~ \equiv ~ \frac{r}{C\, t}
\end{equation}
(see Sect.~\ref{subsec:similar}). This modeling approach is motivated by
the fact that core collapse is a {\em multiscale} problem, which is
expected to assume a self-similar form away from the outer and inner
boundaries and not too close to the onset time (e.g., Penston 1969;
Larson 1969; Hunter 1977; Shu 1977). This has been verified by numerical
and semianalytic treatments of restricted core-collapse problems --
with/without rotation and with/without magnetic fields. Although the
constancy of the isothermal sound speed $C$ that underlies the
ansatz~(\ref{eq:form1}) is not strictly maintained throughout the
solution domain (in particular, $C$ scales roughly as $r^{-1/2}$ in
large portions of the disk that forms around the central star), this is
of little consequence to the results since thermal stresses do not play
a major role in the dynamics of the collapsing core. For a typical sound
speed $C= 0.19\ {\rm km\, s^{-1}}$, $x=1 \Leftrightarrow \{400,\
4000\}\, {\rm AU}$ at $t=\{10^4,\, 10^5\}\, {\rm yr}$.

Krasnopolsky \& K\"onigl (2002, hereafter KK02) constructed self-similar
solutions of rotating magnetic molecular cloud cores that are subject to
ambipolar diffusion. These solutions reveal many of the basic features
of star and disk formation in the core-collapse scenario and are
discussed in the remainder of this section. To incorporate ambipolar
diffusion into the self-similarity formulation it is necessary to
assume that the ion density scales as the square root of the neutral
density: $\varrho_{\rm i} = {\mathcal{K}} \varrho^{1/2}$. As discussed in
KK02, this should be a good approximation for the core-collapse problem:
it applies on both ends of a density range spanning $\sim 8$ orders of
magnitude, which corresponds roughly to radial scales $\sim 10-10^4\,
{\rm AU}$, with ${\mathcal{K}}$ varying by only $\sim 1$ order of magnitude
across this interval.

To allow mass to accumulate at the center in a 1D rotational collapse,
an angular momentum transport mechanism must be present. KK02 assumed
that {\em vertical} transport through magnetic braking continues to
operate also during the collapse phase of the core evolution. To
incorporate this mechanism into the self-similar model it is necessary
to assume that $v_{\rm A ext}$, the Alfv\'en speed in the external
medium, is a constant.\footnote{A nearly constant value $v_{\rm A ext}
\approx 1\ {\rm km\, s^{-1}}$ is, in fact, indicated in molecular clouds
in the density range $\sim 10^3-10^7\, {\rm cm}^{-3}$ (e.g., Crutcher
1999).} KK02 verified that, in their derived solutions, magnetic braking
indeed dominates the most likely alternative angular-momentum transport
mechanisms --- MRI-induced turbulence and gravitational
torques. However, they also found that angular momentum transport by a
CDW arises naturally (and may dominate) in the Keplerian disk that forms
in their fiducial solution (see Sect.~\ref{subsec:s_s_soln}).

\subsection{Basic Equations}
\label{subsec:core_eqns}

The mass and momentum conservation relations in their {\em differential}
form are given by:

\noindent Mass

\begin{equation}
\label{eq:form2}
\frac{\partial\varrho}{\partial t} + \frac{1}{r}\frac{\partial}{\partial
r}\left(r\varrho v_r\right) = -\frac{\partial}{\partial z}\left(\varrho
v_z\right)\; ,
\end{equation}

\noindent Radial Momentum

\begin{eqnarray}
\label{eq:form3}
\varrho\frac{\partial v_r}{\partial t} +\varrho v_r \frac{\partial
v_r}{\partial r}
&=&
\varrho g_r - C^2\frac{\partial \varrho}{\partial r}
+\varrho\frac{v_\phi^2}{r} +\frac{B_z}{4\pi}\frac{\partial B_r}{\partial z}\cr
& & \mbox{}
-\frac{\partial}{\partial r}\left(\frac{B_z^2}{8\pi}\right)
-\frac{1}{8\pi r^2}\frac{\partial}{\partial r}\left(rB_\phi\right)^2
-\varrho v_z\frac{\partial v_r}{\partial z}\; ,
\end{eqnarray}

\noindent Angular Momentum

\begin{equation}
\label{eq:form4}
\frac{\varrho}{r}\frac{\partial}{\partial t}\left(r v_\phi\right) +
\frac{\varrho v_r}{r}\frac{\partial}{\partial r}\left(r v_\phi\right)=
\frac{B_z}{4\pi}\frac{\partial B_\phi}{\partial z} + \frac{B_r}{4\pi
r}\frac{\partial}{\partial r}\left(rB_\phi\right)
-\varrho v_z\frac{\partial}{\partial z}(rv_\phi)\; ,
\end{equation}

\noindent Vertical Hydrostatic Equilibrium

\begin{equation}
\label{eq:form5}
C^2\frac{\partial\varrho}{\partial z} = \varrho g_z -
\frac{\partial}{\partial z}
\left(\frac{B_\phi^2}{8\pi}+\frac{B_r^2}{8\pi}\right) +
\frac{B_r}{4\pi}\frac{\partial B_z}{\partial r} \; ,
\end{equation}
where $g_r$ and $g_z$ are, respectively, the radial and vertical
components of the gravitational acceleration. We have not yet dropped
the terms that involve the vertical velocity component (except in the
vertical momentum equation, which we assume takes on its hydrostatic form).

We now integrate these equations over the core/disk thickness $2H$. We
use the {\em thin-disk approximation} ($H(r) \ll r$) and assume that the
density, radial velocity, azimuthal velocity, and radial gravity are
constant with height, and that so is also $B_z$ (except when it
is explicitly differentiated with respect to $z$, in which case we
substitute $\lfrac{\partial B_z}{\partial
z}=-r^{-1}({\partial}/{\partial r})(rB_r)$ from $\nabla \cdot
\vec{B}=0$). We also assume $B_r(r,z)=B_{r,{\rm s}}(r)\, [z/H(r)]$ (and
similarly for $B_\phi$), and, after deriving expressions that are valid
to order $(H/r)^2$, further simplify by dropping all terms
$\mathcal{O}(H/r)$ except in the combination $[B_{r{\rm
s}}-H(\lfrac{\partial B_z}{\partial r})]$ (see KK02 for details). We
thus obtain the {\em vertically integrated} conservation equations:

\noindent Mass

\begin{equation}
\label{eq:form6}
\frac{\partial\Sigma}{\partial t} +\frac{1}{r} \frac{\partial}{\partial r}
\left(r\Sigma v_r\right) = -\frac{1}{2\pi r}\frac{\partial{\dot{M}}_{\rm
w}}{\partial r}\; ,
\end{equation}

\noindent Radial Momentum

\begin{equation}
\label{eq:form7}
\frac{\partial v_r}{\partial t} + v_r\frac{\partial v_r}{\partial r}=
g_r -\frac{C^2}{\Sigma}\frac{\partial\Sigma}{\partial r}
+\frac{B_z}{2\pi\Sigma}\left(B_{r,{\rm s}}-H\frac{\partial B_z}{\partial
r}\right) +\frac{J^2}{r^3}\; ,
\end{equation}

\noindent Angular Momentum

\begin{equation}
\label{eq:form8}
\frac{\partial J}{\partial t} + v_r\frac{\partial J}{\partial r}=
\frac{rB_zB_{\rm \phi s}}{2\pi\Sigma}\; ,
\end{equation}

\noindent Vertical Hydrostatic Equilibrium

\begin{equation}
\label{eq:form9}
\frac{\Sigma C^2}{2 H} = \frac{\pi}{2}G\Sigma^2 + \frac{GM_*\varrho
H^2}{2r^3} + \frac{1}{8\pi}\left( B_{\phi,{\rm s}}^2
+B_{r,{\rm s}}^2-B_{r,{\rm s}}\, H\frac{\partial B_z}{\partial r}
\right)\, ,
\end{equation}
where $\Sigma = 2 \varrho H$ is the surface mass density, $J = r v_\phi$
is the specific angular momentum of the matter, $G$ is the gravitational
constant, and $M_*$ is the mass of the central protostar. In the
integrated equations we have implemented the 1D flow approximation by
setting $v_z =0$, but we retained the term on
the right-hand side of Eq.~(\ref{eq:form6}) to allow for mass loss through
a disk wind (at a rate ${\dot{M}}_{\rm w}$). Such a wind could carry
angular momentum, but we did not include vertical particle angular
momentum transport in Eq.~(\ref{eq:form8}) since most of a disk wind's
angular momentum initially resides in the magnetic field (see
Sect.~\ref{subsec:wind}). In any case, we proceed to solve the
equations by assuming at first that no wind is present and that the only
mechanism of angular momentum transport is magnetic braking.

The dominant charge carriers in the pre-collapse core are ions and
electrons. Adopting this composition, we approximate the {\em ion
equation of motion} in the ambipolar-diffusion limit by
\begin{equation}
\label{eq:form10}
\varrho \nu_{\rm ni} \vec{v}_{\rm d}= \frac{1}{4\pi}
(\nabla\times\vec{B})\times\vec{B}\; ,
\end{equation}
where $\nu_{\rm ni}$ is the neutral--ion collision frequency (see
Eqs.~(\ref{eq:MHD8}) and~(\ref{eq:MHD10})). This relation yields the
components of the drift velocity:
\begin{equation}
\label{eq:form11}
v_{\rm d,\phi}=\frac{B_zB_{\phi,{\rm s}}}{2\pi\nu_{\rm ni}\Sigma}\; ,
\end{equation}
\begin{equation}
\label{eq:form12}
v_{{\rm d},r}=\frac{B_z}{2\pi\nu_{\rm ni}\Sigma}\left(B_{r,{\rm
s}}-H\frac{\partial B_z}{\partial r}\right)\; .
\end{equation}

Magnetic braking is incorporated via Eq.~(\ref{eq:MHD4}), in which we
identify $v_{b\phi}$ with $v_{\rm i, \phi}$ (see
Sect.~\ref{subsec:nonideal}). Expressing $v_{\rm i,\phi}$ in terms of
$v_\phi$ and $v_{\rm d,\phi}$ (Eq.~(\ref{eq:form11})) and imposing a cap
($\delta \la 1$) on $|B_{\phi,{\rm s}}/B_z|$ (to account for the possible
development of a kink instability above the disk surface), one gets
\begin{equation}
\label{eq:form13}
B_{\phi,{\rm s}}=-\min\left[\frac{\Psi}{\pi r^2} \frac{v_\phi}{v_{\rm A ext}}
\left( 1+ \frac{\Psi B_z}{2\pi^2r^2\nu_{\rm ni}\Sigma v_{\rm A ext}}
\right)^{-1}\, ,\, \delta B_z\right]\; .
\end{equation}

The flux conservation relation $\lfrac{\partial\Psi}{\partial t}=
-2\pi r v_{Br}B_z$ (obtained from Eqs.~(\ref{eq:MHD20})
and~(\ref{eq:MHD22})), which describes the advection of poloidal flux
by the infalling matter, can be written in this limit as
\begin{equation} \label{eq:form14} \frac{\partial\Psi}{\partial t}=
-2\pi r v_{{\rm i},r}B_z =-2\pi r\left(v_r+v_{{\rm d},r}\right)B_z\; ,
\end{equation} with $v_{{\rm d},r}$ given by
Eq.~(\ref{eq:form12}). The surface value of $B_r$ that appears in the
latter equation can be determined in the limit of a potential field
(${\bf \nabla \times B} = 0$) outside an infinitely thin disk from an
$r$ integral of the midplane value of $B_z$: 
\begin{equation}
\label{eq:form15} 
B_{r,{\rm s}} = \int_0^\infty \D{k}\, J_1(kr) \int_0^\infty \D{r^\prime}
r^\prime [B_z(r^\prime) - B_{\rm ref}]J_0(kr^\prime) 
\end{equation} 
(e.g., Ciolek \& Mouschovias 1993), where $J_0$ and $J_1$ are Bessel
functions of order 0 and 1, respectively, and $B_{\rm ref}$ is the
uniform ambient field at ``infinity'' (which is henceforth neglected).
Although the current-free limit of the force-free medium outside the
core/disk is not exact, as the presence of a $B_\phi$ component, which
typically involves a distributed poloidal electric current, is required
for angular momentum transport above and below the core/disk surfaces,
the magnitude of $|B_{\rm \phi,{\rm s}}/B_z|$ (as deduced from
Eq.~(\ref{eq:form13})) is typically small at the start of the collapse
(e.g., Basu \& Mouschovias 1994). Furthermore, even after this ratio
increases to $\sim 1$ as matter and field become concentrated near the
center, the magnitude of the enclosed (poloidal) current ($I =
(c/2)r|B_\phi|$) increases relatively slowly with decreasing $r$
(scaling as $r^{-1/4}$ in the circumstellar disk; see
Eq.~(\ref{eq:form30}) below), so overall we expect the
expression~(\ref{eq:form15}) to be a fair approximation.  A similar
integral (with $\Sigma(r)$ replacing $B_z$) can be written for $g_r$
(Toomre 1963). These integrals can be approximated by their {\em
monopole} terms
\begin{equation}
\label{eq:form16}
B_{r,{\rm s}}=\frac{\Psi(r,t)}{2\pi r^2}\; ,
\end{equation}
\begin{equation}
\label{eq:form17}
g_r=-\frac{GM(r,t)}{r^2}
\end{equation}
(cf. Li \& Shu 1997), where $M(r,t)$ is the mass enclosed within a
radius $r$ at time $t$. When $B_z(r)$ (or $\Sigma(r)$) scales as
$r^{-q}$, one still obtains the monopole expression for $B_{r,{\rm
s}}$ (or $g_r$), but with a $q$-dependent coefficient ${\cal{O}}(1)$
(Ciolek \& K\"onigl 1998).

\subsection{Self-Similarity Formulation}
\label{subsec:s_s_formulation}

The various physical quantities of the problem can be expressed as
dimensionless functions of the similarity variable $x$
(Eq.~(\ref{eq:form1})) in the following fashion:
\begin{equation}
\label{eq:form18}
H(r,t)=Ct\,h(x)
\, ,\ \ \Sigma(r,t)=(C/2\pi Gt)\,\sigma(x)\; ,
\end{equation}
\begin{equation}
\label{eq:form19}
v_r(r,t)=C\,u(x)
\, ,\ \ v_\phi(r,t)=C\,w(x)\; ,
\end{equation}
\begin{equation}
\label{eq:form20}
g_r(r,t)=(C/t)\,g(x)
\, ,\ \ J(r,t)=C^2t\,j(x)\; ,
\end{equation}
\begin{equation}
\label{eq:form21}
M(r,t)=(C^3t/G)\,m(x)\, ,
\ \ \dot{M}_{\rm a}(r,t)=(C^3/G)\,\dot{m}(x)\; ,
\end{equation}
\begin{equation}
\label{eq:form22}
\vec{B}(r,t)=(C/G^{1/2}t)\,{\vec{b}}(x)\, ,
\Psi(r,t)=(2\pi C^3t/G^{1/2})\psi(x)\;
\end{equation}
where $\dot{M}_{\rm a}$ is the mass accretion rate.

The vertical force-balance relation~(\ref{eq:form9}) yields a quadratic
equation for the normalized disk half-thickness $h$,
\begin{equation}
\label{eq:form23}
\left(\frac{\sigma m_*}{x^3} - b_{r,{\rm s}}\frac{d b_z}{d
x}\right)h^2+\left(b_{r,{\rm s}}^2+b_{\phi,{\rm s}}^2+\sigma^2\right)h -
2\sigma= 0\; ,
\end{equation}
whose solution is
\begin{equation}
\label{eq:form24}
h=\frac{\hat \sigma x^3}{2\hat m_*} \left[-1 + \left(1+\frac{8\hat
m_*}{x^3\hat \sigma^2}\right)^{1/2}\right]\; ,
\end{equation}
where $\hat m_*\equiv m_*-x^3b_{r,{\rm s}}(db_z/dx)/\sigma$ and $\hat \sigma
\equiv \sigma + (b_{r,{\rm s}}^2+b_{\phi,{\rm s}}^2)/\sigma$.

The initial ($t=0$) conditions, which in the self-similar model also
represent the {\em outer asymptotic} ($r\rightarrow \infty$) values,
correspond to a collapsing core just before it forms a central point
mass. Based on previous analytic and numerical work, KK02 adopted
\begin{equation}
\label{eq:form25}
\sigma\rightarrow\frac{A}{x},
\ b_z\rightarrow\frac{\sigma}{\mu_\infty},
\ u\rightarrow u_\infty,\ w\rightarrow w_\infty\ \ {\rm{as}}\ x\rightarrow
\infty\; ,
\end{equation}
with $A=3$, $\mu_\infty=2.9$, $u_\infty=-1$. From the constituent equations one
can also derive the {\em inner asymptotic} behavior (corresponding to $r
\rightarrow 0$ at a fixed $t$) in the Keplerian disk:
\begin{eqnarray}
\label{eq:form26}
\dot m=m&=&m_*\; , \\
\label{eq:form27}
j&=&m_*^{1/2}x^{1/2}\; ,\\
\label{eq:form28}
-u&=&(m_*/\sigma_1)x^{1/2}\; ,\\
\label{eq:form29}
\sigma&=&
\frac{(2\eta/3\delta)(2m_*)^{1/2}}{[1+(2\tau/3\delta)^{-2}]^{1/2}}
x^{-3/2}\nonumber\\
&\equiv& \sigma_1x^{-3/2}\; ,\\
\label{eq:form30}
b_z=-b_{\phi,{\rm s}}/\delta&=&
[{m_*}^{3/4}/(2\delta)^{1/2}]x^{-5/4}\; ,\\
\label{eq:form31}
b_{r,{\rm s}}=\psi/x^2&=&(4/3)b_z\; ,\\
\label{eq:form32}
h&=&\{2/[1+(2\tau/3\delta)^2]m_*\}^{1/2}x^{3/2}\; ,
\end{eqnarray}
where $\tau \equiv (4\pi G \varrho)^{1/2}/\nu_{\rm n i}$.

The equations are solved as a boundary value problem using the above
asymptotic relations. A solution is determined by the values of the four
model parameters: $\tau$, $\delta=|B_{\rm \phi s} /B_z|$, $w_\infty =
v_{\phi}(t=0)/C$, and $\alpha \equiv C/v_{\rm A ext}$. The scaling
parameter $m_*$ for the central mass and the mass accretion rate is
obtained as an eigenvalue of the problem.

\subsection{Self-Similar Collapse Solutions}
\label{subsec:s_s_soln}

Figure~\ref{fig6_2} shows a fiducial solution corresponding to
the parameter combination $\tau = 1$, $\delta = 1$, $w(0)=0.73$, and
$\alpha = 0.8$ (which yield $m_* = 4.7$). In this case the initial
rotation is not very fast and the braking is moderate, leading to the
formation of a disk (with outer boundary at the centrifugal shock radius
$x_{\rm c}=1.3\times 10^{-2}$) within the ambipolar-diffusion (AD) region
(delimited by the AD shock radius $x_{\rm a}=0.41 \approx 30\, x_{\rm
c}$). One can distinguish the following main flow regimes:
\begin{figure}
\centering
\includegraphics[height=8cm]{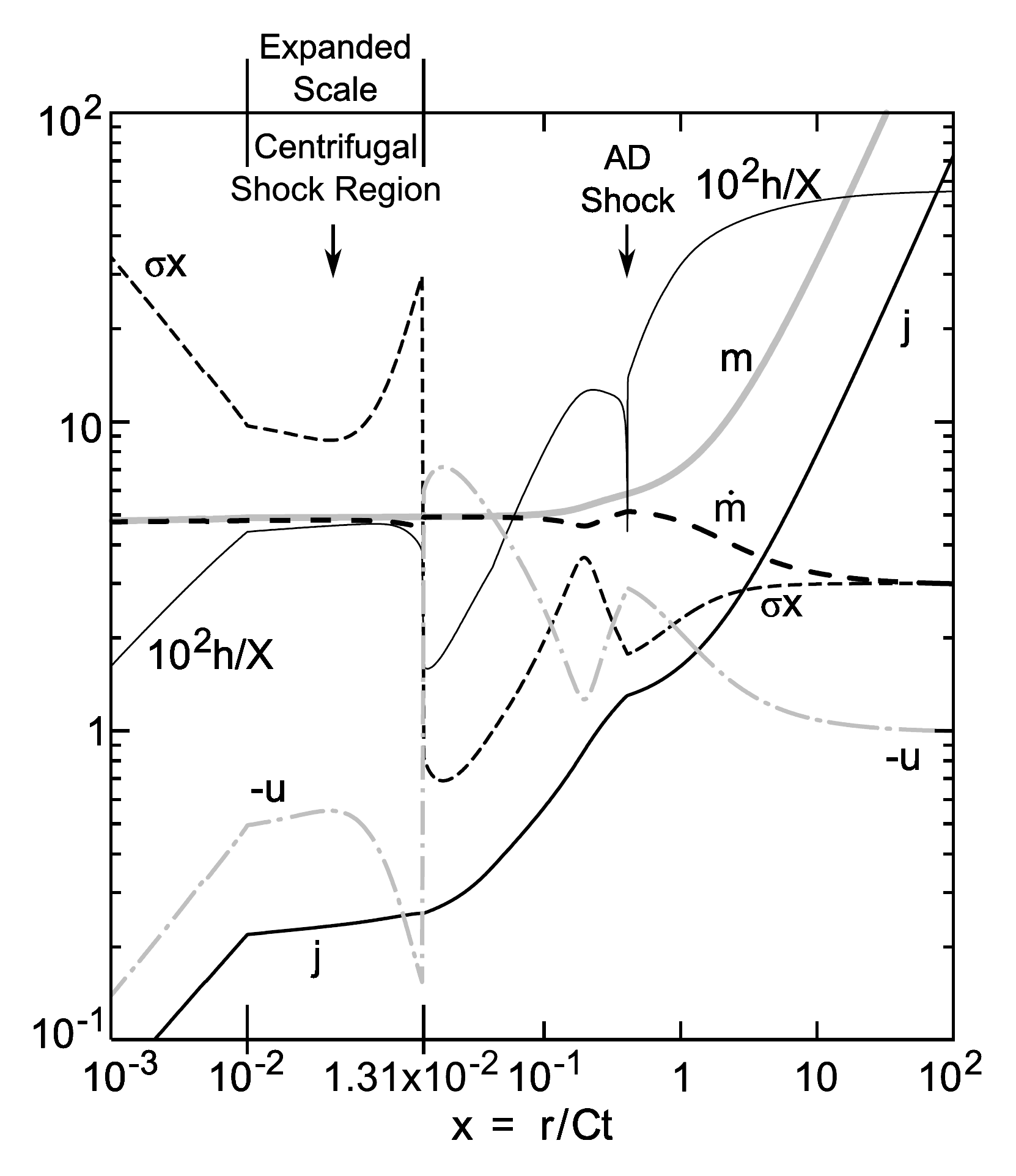}
\caption{Behavior of key normalized variables in the fiducial
self-similar disk formation solution as a function of the similarity
variable $x$ (see Eqs.~(\ref{eq:form18})--(\ref{eq:form21})).}
\label{fig6_2}
\end{figure}
\begin{itemize}
\item
Outer region ($x>x_{\rm a}$): ideal-MHD infall.
\item
AD shock: resolved as a continuous transition (but
may in some cases contain a viscous subshock); KK02 estimated
$x_{\rm a}\approx \sqrt{2}\tau/\mu_\infty$.
\item
Ambipolar diffusion-dominated infall ($x_{\rm c} < x <
x_{\rm a}$): near free fall controlled by the central star's gravity.
\item
Centrifugal shock: its location depends sensitively on the diffusivity
parameter $\tau$, which affects the amount of magnetic braking for
$x<x_{\rm a}$; KK02 estimated $x_{\rm c}\approx (m_*w_\infty^2/A^2)
{\rm exp}{\{-(2^{3/2}m_*/\mu_\infty)^{1/2}\tau^{-3/2}\}}$.
\item
Keplerian disk ($x<x_{\rm c}$): asymptotic behavior
(Eqs.~(\ref{eq:form26})--(\ref{eq:form32})) is approached after a
transition zone representing a massive ring (of width $\sim 0.1\, x_{\rm
c}$ and mass $\sim 8\%$ of the disk mass within $x_c$, which itself is
$\la 5\%$ of $m_*$).
\end{itemize}

The inner asymptotic solution implies that, at any given time, the disk
satisfies $\dot M_{\rm a}(r) = {\rm const}$, $\Sigma(r)\propto r^{-3/2}$,
$B\propto r^{-5/4}$, and $B_{r,{\rm s}}/B_z = 4/3$.\footnote{The surface
density scaling is the same as that inferred for the ``minimum mass''
solar nebula (e.g., Weidenschilling 1977). Note, however, that this
scaling is also predicted for a self-similar Keplerian disk with an
``$\alpha$ viscosity'' (Tsuribe 1999) as well as in certain models in which
gravitational torques dominate the angular momentum transport (e.g., Lin
\& Pringle 1987; Voroboyov \& Basu 2007).}  If the derived power-law
dependence of $B_z$ on $r$ is used in Eq.~(\ref{eq:form15}) one obtains
$B_{r,{\rm s}}/B_z = 1.428$, slightly less than the result found with
the monopole approximation~(\ref{eq:form16}), but still representing a
strong bending of the field lines (by $\sim 55^{\circ}$ from the normal)
that is significantly larger than the minimum of $30^{\circ}$ required
for launching a centrifugally driven wind from a ``cold'' Keplerian disk
(Eq.~(\ref{eq:MHD6})). This implies that protostellar disks formed in
this fashion are likely to drive disk outflows over much of their radial
extents, at least during early times when they still accumulate mass
from the collapsing core. Interestingly, the {\em steady-state, radially
self-similar} CDW solution of Blandford \& Payne (1982) also yields a
radial magnetic field scaling $\propto r^{-5/4}$. As discussed by KK02,
this suggests that angular momentum transport by a CDW can be formally
incorporated into the disk-formation solution, modifying some of the
details (such as the mass accretion rate onto the star, which could be
reduced by up to a factor of $\sim 3$) although not its basic
properties.  When interpreted as a wind-driving disk model, the
asymptotic solution corresponds to a weakly coupled disk
configuration (using the nomenclature introduced in
Sect.~\ref{subsec:disk}).

Despite being based on a number of simplifying assumptions, the fiducial
solution demonstrates that vertical angular momentum transport along
interstellar magnetic field lines can be sufficiently efficient to allow
most of the mass of a collapsing molecular cloud core to end up (with
effectively no angular momentum) at the center, with the central mass
dominating the dynamics well beyond the outer edge of the disk even as
the inflow is still in progress. The vertical transport is thus seen to
resolve the so-called angular momentum problem in star formation
(although the exact value of the protostar's angular momentum is
determined by processes near the stellar surface that are not included
in this model; see Sect.~\ref{sec:disk-star}). To the extent that
self-similarity is a good approximation to the situation in such cores,
it is conceivable that T Tauri (Class II) protostellar systems, whose
disk masses are typically inferred to be $\la 10\%$ of the central mass,
have had a similarly low disk-to-star mass ratio also during their
earlier (Class 0 and Class I) evolutionary phases. This possibility
remains to be tested by observations.

The fiducial solution also reveals that the AD shock,
even though it is located well outside the region where
the centrifugal force becomes important, helps to enhance
the efficiency of angular momentum transport through
the magnetic field amplification that it induces. The
revitalization of ambipolar diffusion behind the AD shock in turn
goes a long way toward resolving the magnetic flux problem
in star formation (the several-orders-of-magnitude
discrepancy between the empirical upper limit on the magnetic
flux of a protostar and the flux associated with the
corresponding mass in the pre-collapse core), although it is conceivable
that Ohm diffusivity in the innermost regions of the disk also plays an
important role in this process (e.g., Shu et al. 2006;
Tassis \& Mouschovias 2007). The details of the flux
detachment outside the Ohm regime can be modified if one takes account
of the fact that the flux is strictly frozen into the electrons but not
necessarily into the ions (i.e., if one includes also the Hall term in
Ohm's Law; see Sect.~\ref{subsec:nonideal}). In particular, Tassis \&
Mouschovias (2005) found that a quasi-periodic series of
outward-propagating shocks may develop in this case.

The shocks that form in the collapsing core, and in particular the
centrifugal shock (which is the strongest), will process the incoming
material and may have implications to the composition of protoplanetary
disks (e.g., the annealing of silicate dust; see Harker \& Desch 2002
and Chap.~VIII [Dust Processing and Mineralogy in Protoplanetary
Accretion Disks]).  Note in this connection that the shock velocity
$v_{\rm sh}$ relative to the intercepted matter has roughly the
free-fall magnitude $\propto (M/r)^{1/2}$, which is constant for any
given value (including, in particular, $x_{\rm c}$) of the similarity
variable. The postshock temperature (which scales as $v_{\rm sh}^{1/2}$)
will thus not vary with time if the inflow remains self-similar; the
postshock density, on the other hand, will decrease with time as the
shock moves to larger radii.

\begin{figure}
\includegraphics[width=0.45\linewidth]{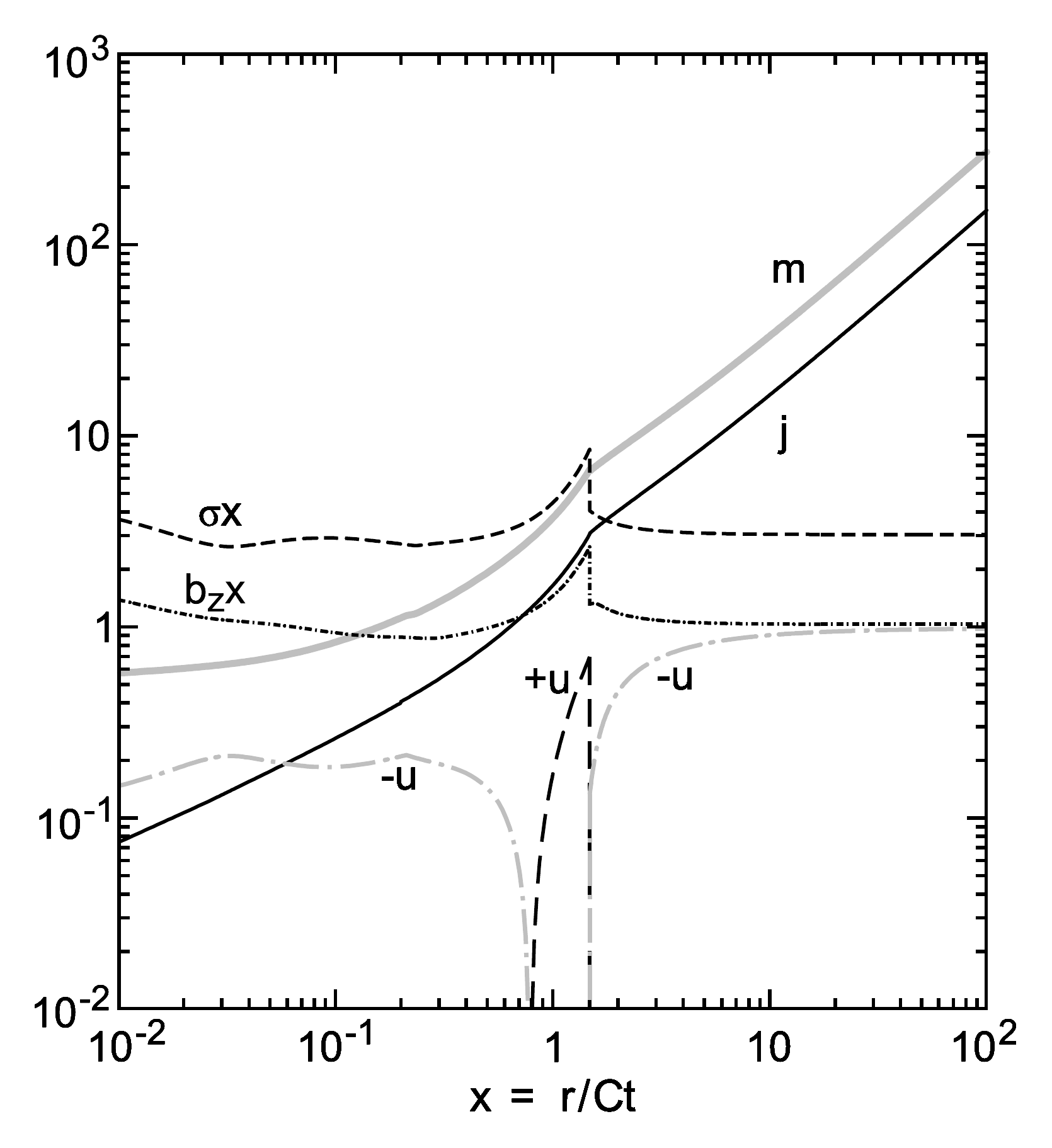}
\hfill
\includegraphics[width=0.45\linewidth]{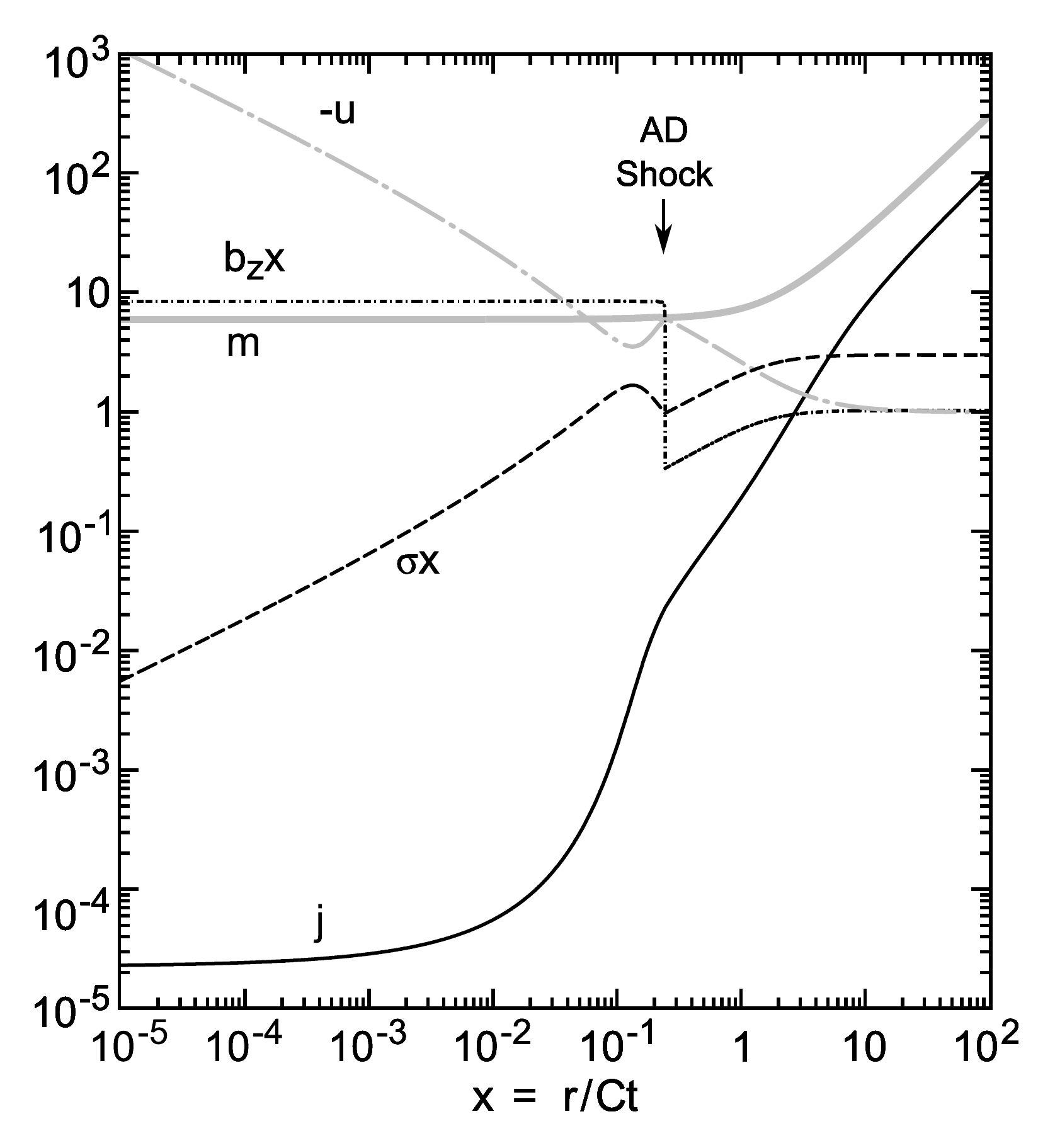} 
\caption{Behavior of key normalized variables in representative
fast-rotation ({\bf left}) and strong-braking ({\bf right}) solutions
of the collapse of a rotating, magnetic, molecular cloud core as a
function of the similarity variable $x$ (see
Eqs.~(\ref{eq:form18})--(\ref{eq:form22})).} \label{fig6_3}
\end{figure}

By modifying the model parameters, one can study the range of
possible behaviors in real systems. Figure~\ref{fig6_3}
shows two limiting cases, which bracket the fiducial solution
(cf. Fig.~\ref{fig6_2}).

\smallskip\noindent
The {\em fast rotation} case differs from the fiducial solution
primarily in having a large initial-rotation parameter
($w(0)=1.5$). It has the following distinguishing features:
\begin{itemize}
\item  The centrifugal shock is located within the
self-gravity--dominated (and ideal-MHD) region; a back-flowing
region is present just behind the shock.
\item The central mass is comparatively small ($m_* = 0.5$),
giving rise to a non-Keplerian outer disk region.
\item The ideal-MHD/ambipolar-diffusion transition
occurs behind the centrifugal shock and is gradual rather than sharp.
\end{itemize}
The {\em strong braking} case ($\tau=0.5$, $\delta=10$, $w(0)=1$,
$\alpha = 10$, yielding $m_*=5.9$) is characterized by
large values of the braking parameters $\delta$ and $\alpha$. It
is distinguished by having
\begin{itemize}
\item no centrifugal shock (or circumstellar disk); essentially all the
angular momentum is removed well before the inflowing gas reaches the
center, and the $x\rightarrow\ 0$  behavior resembles that of the nonrotating
collapse solution of Contopoulos et al. (1998).
\end{itemize}
This case may be relevant to the interpretation of slowly rotating young
stars that show no evidence of a circumstellar disk (e.g., Stassun et
al. 1999, 2001), which would seem puzzling if all protostars were born
with significant rotation that could only be reduced as a result of
their interaction with a disk (see Sect.~\ref{sec:disk-star}). In the
strong-braking interpretation, both the slow rotation and the absence of
a disk have the same cause. It is also worth noting in this connection
that the physical conditions that underlie the fiducial solution
presented above --- namely, a highly flattened, magnetized core whose
braking is controlled by the inertia of a comparatively low-density
ambient gas --- may not be generally applicable. In particular, it has
been argued (see Mellon \& Li 2009 and references therein) that
magnetized molecular clouds may retain a significant amount of mass
outside the collapsing core that would dominate the braking process and
could place the core in the strong-braking regime where disk formation
would be suppressed. Although there may be ways to alleviate or even
circumvent this problem (e.g., Hennebelle \& Ciardi 2009; Duffin \&
Pudritz 2009), there is obviously a need for additional observational
and theoretical work to clarify the physical conditions that
characterize core collapse in real clouds.

While the semianalytic self-similarity model described in this section
is useful for identifying important features of disk formation from the
collapse of rotating cores, the intricacies of this process can
only be studied numerically. Newly developed 3D, nonideal-MHD,
nested-grid codes (e.g., Machida et al. 2007) are already
starting to provide new insights into this problem.

\section{Wind-Driving Protostellar Disks}
\label{sec:vertical}

In this section we discuss wind-driving disk models. We begin by
highlighting the observational evidence for a close link between
outflows and accretion disks in protostellar systems and the
interpretation of this connection in terms of a large-scale magnetic
field that mediates the accretion and outflow processes. We next outline
the theory of centrifugally driven winds, describe the incorporation of
CDWs into equilibrium disk models, and conclude by addressing the
stability of such disk/wind configurations.

\subsection{The Disk--Wind Connection}
\label{subsec:connection}

We give a brief survey of the observational findings on protostellar
outflows and their connection to accretion disks. Besides the explicit
references cited, the reader may consult some of the many review
articles written on this topic over the years (e.g., Cabrit 2007; Calvet
et al. 2000; Dutrey et al. 2007; K\"onigl \& Pudritz 2000; Millan-Gabet
et al. 2007; Najita et al. 2007; Ray et al. 2007; Watson et al. 2007).

\subsubsection{Bipolar Outflows and Jets}
\label{subsubsec:outflows}

Bipolar molecular outflows and narrow atomic (but sometimes also
molecular) jets are ubiquitous phenomena in protostars, with nearly 1000
collimated outflows of all sorts already known (e.g., Bally et
al. 2007). The bipolar lobes are usually understood to represent ambient
molecular material that has been swept up by the much faster, highly
supersonic jets and winds that emanate from the central star/disk system
(e.g., Arce et al. 2007).  Jets associated with
low-bolometric-luminosity ($L_{\rm bol}<10^3 L_\odot$) protostars have
velocities in the range $\sim 150-400\, {\rm km\, s}^{-1}$, large
($>20$) Mach numbers, and inferred mass outflow rates $\sim
10^{-9}-10^{-7}\; M_\odot \, {\rm yr}^{-1}$. They are collimated on
scales of a few 10s of AU and exhibit opening angles as small as $\sim
3-5^\circ$ on scales of $10^3-10^4\, {\rm AU}$. High-resolution
observations of optically visible jets from classical T Tauri stars
(CTTSs) reveal an onion-like morphology, with the regions closer to the
axis having higher velocities and excitations and appearing to be more
collimated. Detailed optical/near-IR spectral diagnostic techniques have
been developed and applied to classical (Herbig-Haro) protostellar jets,
making it possible to directly estimate the neutral densities in the
forbidden-line emission regions. Recent results (Podio et al. 2006)
yield an average mass outflow rate of $5\times 10^{-8}\, M_\odot \, {\rm
yr}^{-1}$, markedly higher than previous estimates. Outflows have also
been detected by optical observations of intermediate-mass ($\sim 2-10\,
M_\odot$) Herbig Ae/Be stars and other high-luminosity sources. The jet
speeds and mass outflow rates in these protostars are larger by a factor
$\sim 2-3$ and $\sim 10-100$, respectively, than in the low-$L_{\rm
bol}$ objects.

\subsubsection{The Link to Accretion Disks}
\label{subsubsec:link}

{\em Strong correlations} are found between the presence of {\em outflow
signatures} (P-Cygni line profiles, forbidden line emission, thermal
radio radiation, well-developed molecular lobes) and {\em accretion
diagnostics} (UV, IR, and millimeter emission excesses, inverse P-Cygni
line profiles) in T Tauri stars (e.g., Hartigan et al. 1995). Such
correlations evidently extend smoothly to protostars with masses of
$\sim 10\ M_\odot$.  A related finding is that the apparent decline in
outflow activity with stellar age follows a similar trend exhibited by
disk frequency and inferred mass accretion rate. In addition,
correlations of the type $\dot M \propto L_{\rm bol}^q$ (with $q \sim
0.6-0.7$) have been found in both low-$L_{\rm bol}$ and high-$L_{\rm
bol}$ protostars for mass {\em accretion} rates and for mass {\em
outflow} rates in ionized jets as well as in bipolar molecular
lobes. Furthermore, CTTS-like accretion and outflow phenomena have now
been detected also in very-low-mass stars and brown dwarfs (e.g.,
Mohanty et al. 2005).

These findings strongly suggest that outflows are powered by accretion
and that the same basic physical mechanism operates in both low- (down
to nearly the planetary mass limit) and intermediate-mass protostars,
and possibly also in some higher-mass objects (e.g., Shepherd 2005). The
accretion proceeds through circumstellar disks, which can be directly
probed by means of high spectral- and spatial-resolution (in particular,
interferometric) observations at submillimeter, millimeter, mid-IR,
near-IR, and optical wavelengths. The disks appear to be rotationally
supported (for $r \la 100\, {\rm AU}$), and when the rotation law can be
determined, it is usually consistent with being Keplerian ($v_\phi
\propto r^{-1/2}$). Recent data indicate that $M_{\rm disk} \la M_*$ at
least up to $M_* \sim 20\, M_\odot$.

Strong evidence for a disk origin for the observed outflows is available
for {\em FU Orionis outbursts} in rapidly accreting young protostars
(e.g., Hartmann \& Kenyon 1996). It is inferred that the emission
during an outburst (of typical duration $\sim 10^2\, {\rm yr}$)
originates in a rotating disk and that the outflow represents a wind
that accelerates from the disk surface (with $\lfrac{\dot M_{\rm
w}}{\dot M_{\rm a}} \sim 0.1$; $\dot M_{\rm a} \sim 10^{-4}\, M_\odot \,
{\rm yr}^{-1}$). It has been suggested (Hartmann 1997) that most of the
mass accumulation and ejection in low-mass protostars occurs during
recurrent outbursts of this type. Interestingly, estimates in CTTSs
(e.g., Kurosawa et al. 2006; Ray et al. 2007) indicate that $\lfrac{\dot
M_{\rm w}}{\dot M_{\rm a}}$ has a similar value (i.e., $\sim 0.1$) also
during the quiescent phases of these protostars.

\subsection{Magnetic Driving of Protostellar Outflows}
\label{subsec:driving}

The most widely accepted explanation of protostellar outflows is that
they tap the rotational kinetic energy of the disk (and/or central
object) and are accelerated centrifugally from the disk (or stellar)
surface by the stress of a large-scale, ordered magnetic field that threads the
source. In this subsection we review the arguments that have led to this
picture and list a few alternatives for the origin of the field.

\subsubsection{Outflow Driving Forces}
\label{subsubsec:driving_forces}

The momentum discharges inferred from observations of protostellar jets
are compatible with the values deduced for the bipolar molecular
outflows but are typically a factor $\sim 10^2-10^3$ higher than the
radiation-pressure thrust $L_{\rm bol}/c$ produced by the stellar
luminosity, ruling out {\em radiative acceleration} as a dominant
driving mechanism in low/intermediate-mass protostars (e.g., Richer et
al.  2000). Radiative effects could however be important in driving
photoevaporative disk outflows, particularly in high-$L_{\rm bol}$
systems (e.g., Hollenbach et al. 1994; see Chap.~VII [The Dispersal of
Disks around Young Stars]). Disk heating by a luminous protostar (such
as a Herbig Be star) could potentially also induce a line-driven wind
from the inner disk (Drew et al. 1998). Another effect is radiation
pressure on the dusty outer regions of a disk wind, which could act to
decollimate the streamlines even in comparatively low-$L_{\rm bol}$
systems (see Sect.~\ref{subsubsec:wind_observe} and cf. K\"onigl \&
Kartje 1994).

{\em Thermal pressure} acceleration is commonly discounted as a dominant
mechanism since the requisite high (effectively virial) temperatures are
not generally observed at the base of the flow. It has, however, been
suggested that, under suitable conditions, the thermal energy released
in shocks at the boundary layer between the disk and the star could be
efficiently converted into outflow kinetic energy (Torbett 1984; Soker
\& Regev 2003). Even if this is not a dominant mechanism, thermal
pressure effects are nevertheless important in the mass loading of
hydromagnetic winds (see Sect.~\ref{subsubsec:critical}) and could
potentially also play a significant role in the initial acceleration of
disk outflows of this type (e.g., Pesenti et al. 2004).

A general result for protostellar jets can be obtained by combining
\begin{enumerate}
\item $\dot M_{\rm jet} v_{\rm jet} \sim 10^{2}-10^3\, L_{\rm bol}/c$
\item $L_{\rm jet} \sim \dot M_{\rm jet} v_{\rm jet}^2$
\item $v_{\rm jet} \sim 10^{-3}\, c$
\item $L_{\rm bol} \sim L_{\rm acc}$ ($\sim G M_* \dot M_{\rm a}/R_*)$
\end{enumerate}
(where $R_*$ is the stellar radius), which implies that, on average, the
jet kinetic luminosity is related to the released accretion luminosity
by $L_{\rm jet} \sim 0.1-1\, L_{\rm acc}$.\footnote{Note that in
protostars that are beyond the main accretion phase $L_{\rm bol}$ can be
expected to exceed $L_{\rm acc}$ (e.g., Tilling et al. 2008), which would increase the inferred value of $\lfrac{L_{\rm
jet}}{L_{\rm acc}}$ in these sources.} Such a high ejection efficiency
is most naturally understood if the jets are driven {\em
hydromagnetically} (see Eqs.~(\ref{eq:wind11}) and~(\ref{eq:wind13})
below). This mechanism also provides a natural explanation of the strong
collimation exhibited by protostellar jets (see
Sect.~\ref{subsubsec:exact_wind}).

\subsubsection{Origin of Large-Scale Disk Magnetic Field}
\label{susubsec:origin}

Perhaps the most likely origin of a large-scale, open magnetic field
that can launch a CDW from a protostellar disk is the interstellar field
that supports the natal molecular cloud core. We have already considered
the observational evidence for such a core-threading field
(Sect.~\ref{sec:intro}), its dynamical effect on the core
(Sect.~\ref{subsec:forces}), and how it is advected inward by the
inflowing matter during gravitational collapse
(Sect.~\ref{sec:formation}). However, to complete the argument it is
still necessary to demonstrate that, once a rotationally supported disk
is established, magnetic flux that is brought in to its outer edge
remains sufficiently well coupled to the inflowing matter to be
distributed along the disk plane. A potential problem arises, however,
in turbulent disks in which only radial angular momentum transport is
present. In particular, in a disk of characteristic radius $R$ and
half-thickness $H\ll R$ in which angular momentum transport is due to an
effective turbulent viscosity $\nu_{\rm turb}$, inward dragging of
magnetic field lines is possible only if the effective turbulent
magnetic diffusivity $\eta_{\rm turb}$ satisfies $\eta \la (H/R)\,
\nu$,\footnote{This follows from a comparison of the mass inflow speed,
$\sim \nu_{\rm turb}/R$, as inferred from the angular momentum
conservation equation, with the field diffusion speed relative to the
gas, $\sim \eta_{\rm turb}/H$, as inferred from the flux conservation
equation.} which may not occur naturally in such a system (Lubow et
al. 1994a).\footnote{Recent numerical simulations have indicated that
the turbulent $\eta$ is, in fact, of the order of the turbulent $\nu$ in
disks where the turbulence derives from the magnetorotational
instability (Guan \& Gammie 2009; Lesur \& Longaretti 2009). Note,
however, that efficient inward advection of magnetic flux may
nevertheless be possible in this case on account of the expected
suppression of the instability near the disk surfaces (e.g., Rothstein
\& Lovelace 2008).} The introduction of vertical angular momentum
transport along a large-scale magnetic field offers a straightforward
way out of this potential dilemma by decoupling these two processes:
magnetic diffusivity is determined by the ionization state and
composition of the disk (or else by the local turbulence, if one is
present) and no longer has the same underlying physical mechanism as the
angular momentum transport. The disk formation models presented in
Sect.~\ref{sec:formation} provide concrete examples of how open magnetic
field lines can be self-consistently incorporated into a disk in which
they constitute the dominant angular momentum transport channel (see
also Spruit \& Uzdensky 2005).

An alternative possibility is that the outflow is driven by the {\em
stellar dynamo-generated field}. This could happen either along field
lines that have been effectively severed after they penetrated the
disk, as in the X-wind scenario (e.g., Shu et al. 2000), or along
field lines that are still attached to the star. However, even in the
latter case the outflow is envisioned to result from an interaction
between the stellar magnetic field and the disk (see
Sect.~\ref{sec:disk-star}). It has also been proposed that a
large-scale, open field could be generated by a {\em disk dynamo}
(e.g., Tout \& Pringle 1996; von Rekowski et al. 2003; Blackman \& Tan
2004; Pudritz et al. 2007; Uzdensky \& Goodman 2008). In this
connection it is worth noting that, even if the disk dynamo generates
small-scale, closed magnetic loops, it is conceivable, extrapolating
from the situation on the Sun, that some of these loops could be
dynamically extended to sufficiently long distances (in particular,
beyond the respective Alfv\'en surfaces) for them to become
effectively open (Blandford \& Payne 1982).

Direct evidence for the presence of magnetic fields in either the
outflow or the disk has been scant, but this may not be surprising in
view of the considerable observational challenges. Among the few actual
measurements are the strong circular polarization that was detected on
scales of $\sim 20\, {\rm AU}$ in T Tau S (Ray et al. 1997), which was
interpreted as a field of at least several gauss that was advected from
the origin by the associated outflow,\footnote{Circular polarization on
much larger physical scales was detected along the HH 135-136 outflow,
where it was similarly interpreted as evidence for a helical magnetic
field (Chrysostomou et al. 2008).} and the meteoritic evidence that
points to the presence in the protosolar nebula of a $\sim 1\, {\rm G}$
field at $r \sim 3\; {\rm AU}$ (Levy \& Sonett 1978). In view of the
strong indications for disk-launched outflows in FU Orionis systems, it
is also worth mentioning the Zeeman-signature least-square deconvolution
measurement in FU Ori (Donati et al. 2005), which was interpreted as a
$\sim 1\, {\rm kG}$ field (with $|B_\phi| \sim B_z/2$) originating on
scales of $\sim 0.05\, {\rm AU}$ in the associated disk (with the
direction of $B_\phi$ being consistent with its origin in shearing of
the vertical field by the disk differential rotation). Further support
for a magnetic disk--outflow connection has been inferred from
measurements of apparent rotations in a number of protostellar jets,
although, as noted in Sect.~\ref{subsubsec:generic} below, this evidence
is still controversial.

\subsection{Centrifugally Driven Winds}
\label{subsec:wind}

Our focus here is on the role of CDWs as an angular momentum transport
mechanism. We survey their main characteristics and concentrate on those
aspects that are relevant to the construction of ``combined'' disk/wind
models, the topic of the next subsection.

\subsubsection{Wind Equations}
\label{subsubsec:wind_eq}

The qualitative basis for such outflows was considered in
Sect.~\ref{subsec:centrifugal}. In a formal treatment, the winds are
analyzed using the equations of time-independent, ideal
MHD.\footnote{The ideal-MHD assumption is justified by the fact that the
charged particles' drift speeds rapidly become much lower than the bulk
speed as the gas accelerates away from the disk surface; see
Sect.~\ref{subsec:nonideal}.} These equations are:

\noindent
Mass conservation (continuity equation)

\begin{equation}
\label{eq:wind1}
\nabla \cdot ( \varrho \vec{v}) = 0\; ,
\end{equation}

\noindent
Momentum conservation (force equation)
\begin{equation}
\label{eq:wind2}
\varrho \vec{v} \cdot \nabla \vec{v} = - \nabla P - \varrho \nabla \Phi
+ \frac{1}{4 \pi} \left ( \nabla \times \vec{B}\right ) \times \vec{B}\; ,
\end{equation}
where we take the thermal pressure to be given by the perfect gas law
\begin{equation}
\label{eq:wind3}
P = \frac{k_{\rm B} T \rho}{\mu m_{\rm H}}
\end{equation}
(where $T$ is the temperature, $k_{\rm B}$ is Boltzmann's constant,
$\mu$ is the molecular weight and $m_{\rm H}$ is the hydrogen atom mass),

\smallskip\noindent
Induction equation (magnetic field evolution)

\begin{equation}
\label{eq:wind4}
\nabla \times (\vec{v} \times \vec{B} ) = 0
\end{equation}
(cf. Eq.~(\ref{eq:MHD20})), and the solenoidal condition
(Eq.~(\ref{eq:MHD3})). In general, one also needs to specify an
{\em entropy conservation} equation (balance of heating
and cooling). For simplicity, however, we specialize to
{\em isothermal} flows (spatially uniform isothermal sound speed
$C=\sqrt{P/\varrho}$).

\subsubsection{Generic Properties}
\label{subsubsec:generic}

Concentrating on {\em axisymmetric} flows, it is convenient to decompose
the magnetic field into poloidal and azimuthal components,
$\vec{B}=\vec{B}_{\rm p} + \vec{B}_{\phi}$, with $\vec{B}_{\rm p} =
(\nabla A \times \vec{\hat \phi})/r$, where the poloidal flux function
$A(r, z)$ can be expressed in terms of the poloidal magnetic flux by $A
= \Psi/2\pi$. The flux function is constant along $\vec{B}$ ($\vec{B}
\cdot \nabla A = 0$) and thus can be used to label the field lines. The
solution of the induction equation is given by
\begin{equation}
\label{eq:wind5}
\vec{v} \times \vec{B} = \nabla \chi \; ,
\end{equation}
which shows that the electric field $\vec{E} = - (\vec{v} \times \vec{B})/c$
is derivable from an electrostatic potential ($\chi$). Since $\lfrac{\partial
\chi}{\partial \phi} = 0$ on account of the axisymmetry, it follows that
$\vec{v}_{\rm p} \parallel \vec{B}_{\rm p}$. Equivalently,
\begin{equation}
\label{eq:wind6}
\varrho \vec{v}_{\rm p} = k \vec{B}_{\rm p}\; ,
\end{equation}
where $k$ is a flux-surface constant ($\vec{B} \cdot \nabla k = 0$, using
$\nabla \cdot (\varrho \vec{v})=0$ and $\nabla \cdot \vec{B} =
0$). Physically, $k$ is the {\em wind mass load function},
\begin{equation}
\label{eq:wind7}
k = \frac{\varrho {v}_{\rm p}}{{B}_{\rm p}} = \frac{\D{\dot M_{\rm
w}}}{\D{\Psi}}\; ,
\end{equation}
whose value is determined by the physical conditions near the top of the
disk (more precisely, at the sonic, or slow-magnetosonic, critical
surface; see discussion below).
By taking the vector dot product of $\vec{B}_{\rm p}$ with
Eq.~(\ref{eq:wind5}) and using Eq.~(\ref{eq:wind6}) one finds that the
{\em field-line angular velocity} $\Omega_B = \Omega -
(\lfrac{kB_{\phi}}{\varrho r})$, where $\Omega$ is the matter angular
velocity, is also a flux-surface constant. By writing this relation as
\begin{equation}
\label{eq:wind8}
B_{\phi} = \frac{\varrho r}{k} ( \Omega - \Omega_B)
\end{equation}
it is seen that $\Omega_B$ can be identified with the angular velocity
of the matter at the point where $B_\phi =0$ (the disk midplane,
assuming reflection
symmetry).\footnote{\label{fn:Omega_B}Eqs.~(\ref{eq:wind5})--(\ref{eq:wind8})
also hold in a nonideal fluid so long as the field remains frozen into a
certain particle species (such as the electrons); in this case $\vec{v}$
and $\Omega$ are replaced by the velocity and angular velocity,
respectively, of that species (e.g., K\"onigl 1989).}

Using the $\phi$ component of the momentum conservation
equation~(\ref{eq:wind2}) and applying Eq.~(\ref{eq:wind6}) and the
field-line constancy of $k$, one finds that
\begin{equation}
\label{eq:wind9}
l = r v_{\phi} - \frac{r B_{\phi}}{4 \pi k}
\end{equation}
is a flux-surface constant as well, representing the conserved {\em
total specific angular momentum} (matter + electromagnetic
contributions) along a poloidal field line (or streamline). In a CDW the
magnetic component of $l$ dominates over the matter component near the
disk surface, $\frac{1}{\varrho v_{\rm p}} \frac{r |B_{\rm p} B_\phi|}{4\pi}
\gg r v_\phi\,$, whereas at large distances this inequality is
reversed. The transfer of angular momentum from the field to the matter
is the essence of the {\em centrifugal acceleration process}. This
transfer also embodies the capacity of such a wind to act as an
efficient angular momentum transport mechanism. In fact, as already
noted in Sect.~\ref{subsec:braking}, $r B_{z,{\rm s}}
B_{\phi,{\rm s}}/4\pi$ represents the magnetic torque per unit area that
is exerted on each surface of the disk. As was also already discussed in
connection with magnetic braking, the value of $B_{\phi,{\rm s}}$ is
determined by the conditions outside the disk, essentially by the
inertia of the matter that absorbs the transported angular momentum and
exerts a back torque on the disk. In the case of a CDW, $B_{\phi{\rm
s}}$ is effectively fixed by the regularity condition at the {\em
Alfv\'en critical surface}, which is the largest cylindrical radius from
which information (carried by Alfv\'en waves) can propagate back to the
disk. This condition yields the value of the ``Alfv\'en lever arm''
$r_{\rm A}$, which satisfies
\begin{equation}
\label{eq:wind10}
l = \Omega_B\, r_{\rm A}^2 \, .
\end{equation}
The rate of angular momentum transport by the wind is thus $\sim \dot
M_{\rm w} \Omega_0 r_{\rm A}^2$ (where we replaced $\Omega_B$ by the
midplane value of $\Omega$), whereas the rate at which angular momentum
is advected inward by the accretion disk is $\sim \dot M_{\rm a}
\Omega_0 r_0^2$. Hence, wind transport can enable accretion at a rate
\begin{equation}
\label{eq:wind11}
\dot M_{\rm a} \simeq (r_{\rm A}/ r_0)^2 \, \dot M_{\rm w}\; .
\end{equation}
CDW solutions can have $r_{\rm A} / r_0 \sim 3$ for reasonable
parameters, indicating that such outflows could transport the bulk of
the disk angular momentum in protostellar systems if $\dot M_{\rm w}
\simeq 0.1\, \dot M_{\rm a}$, which is consistent with the observationally
inferred mass outflow rates.

By taking the vector dot product of $\vec{B}_{\rm p}$ with
Eq.~(\ref{eq:wind2}), one finds that the specific energy
\begin{equation}
\label{eq:wind12}
{\cal{E}} = \frac{1}{2} v^2 -\frac{r \Omega_B B_\phi B_{\rm p}}{4\pi
\varrho v_{\rm p}} + w + \Phi \; ,
\end{equation}
where $w = C^2\, \ln{(\varrho/\varrho_{\rm A})}$ (with $\varrho_{\rm A}
= 4\pi k^2$ being the density at the Alfv\'en surface) is the specific
enthalpy, is constant along flux surfaces. This is the {\em generalized
Bernoulli equation}, in which the magnetic term arises from the poloidal
component of the Poynting flux $c\, \vec{E} \times \vec{B}/4 \pi$. Using
Eqs.~(\ref{eq:wind9}) and~(\ref{eq:wind10}), this term can be rewritten
as $\Omega_0\, (\Omega_0 r_{\rm A}^2 - \Omega r^2)$, and by approximating
${\cal{E}} \approx v_{{\rm p},\infty}^2/2$ as $r \rightarrow \infty$ and assuming
that $(r_{\rm A}/r_0)^2 \gg 1$, one can estimate the value of the
asymptotic poloidal speed as
\begin{equation}
\label{eq:wind13}
v_{{\rm p},\infty} \simeq 2^{1/2} \Omega_0 r_{\rm A} \; ,
\end{equation}
or $v_{{\rm p},\infty} / v_{\rm K} \approx 2^{1/2} r_{\rm A} / r_0$. This shows
that such outflows are capable of attaining speeds that exceed (by a
factor of up to a few) the Keplerian speed $v_{\rm K} = r_0 \Omega_0$ at
their base and thus could in principle account for the measured
velocities of protostellar jets.

By combining ${\cal{E}}$, $\Omega_B$, and $l$, one can form the
field-line constant
\begin{equation}
\label{eq:wind14}
{\cal{H}} \equiv {\cal{E}} - \Omega_B l = \frac{1}{2} v^2 - r^2 \Omega_B
\Omega + \Phi \; ,
\end{equation}
where we omitted $w$ from the explicit expression. Evaluating at $r_0$
gives ${\cal{H}}=-\frac{3}{2}v_{\rm K}^2= -\frac{3}{2}(G M_*
\Omega_0)^{2/3}$, and when this is combined with the form of ${\cal{H}}$
at large distances ($r_\infty$) one gets
\begin{equation}
\label{eq:wind15}
r_\infty v_{\phi}(r_\infty) \Omega_0 - \frac{3}{2} (G
M_*)^{2/3}\Omega_0^{2/3} - \frac{1}{2} \left [v_{\rm p}^2(r_\infty) +
  v_{\phi}^2(r_\infty)\right ] \approx 0\; .
\end{equation}
If one could measure the poloidal and azimuthal speeds at a location
$r_\infty$ in a protostellar jet and estimate the central mass $M_*$,
one would be able to use Eq.~(\ref{eq:wind15}) (regarded as a cubic in
$\Omega_0^{1/3}$) to infer the jet launching radius $r_0 = (G
M_*/\Omega_0^2)^{1/3}$ (Anderson et al. 2003). One could
then also use Eqs.~(\ref{eq:wind8}) and~(\ref{eq:wind10}) to estimate
$|B_{\phi \infty}/B_{\rm p \infty}|$ and $r_{\rm A}/r_0$, respectively,
and verify that they are significantly greater than 1, as required for
self-consistency. This has already been attempted in several instances,
where it was claimed that the results are consistent with a disk-driven
outflow that originates on scales $\ga 1\,  {\rm AU}$ and carries a
significant fraction of the disk angular momentum (e.g., Ray et al. 2007;
Coffey et al. 2007; Chrysostomou et al. 2008). However, the
interpretation of the measurements has been controversial (e.g., Soker 2005;
Cerqueira et al. 2006) and there have been some conflicting
observational results (e.g., Cabrit et al. 2006), so the issue is not
yet fully settled.

\subsubsection{Critical Points of the Outflow}
\label{subsubsec:critical}

The critical points occur in stationary flows at the locations where
the fluid velocity equals the speed of a backward-propagating
disturbance. Since the disturbances propagate along the
characteristics of the time-dependent equations, the critical points
can be regarded as relics of the initial conditions in a
time-dependent flow (Blandford \& Payne 1982). If one takes the
magnetic field configuration as given and considers the poloidal flow
along $\vec{B}_{\rm p}$, one can derive the location of the critical
points and the values of the critical speeds (i.e., the values of
$v_{\rm p}$ at the critical points) by regarding the Bernoulli
integral as a function of the spatial coordinate along the field line
and of the density (${\cal{H}}={\cal{H}}(s,\varrho)$, where $s$ is
the arc-length of the streamline) and deriving the extrema of
${\cal{H}}$ by setting $\lfrac{\partial {\cal{H}}}{\partial \varrho}
=0$ (which yields the critical speeds) and $\lfrac{\partial
{\cal{H}}}{\partial s} =0$ (which, together with the other relation,
yields the critical point locations). The critical points obtained
from the generalized Bernoulli equation~(\ref{eq:wind12}) correspond
to $v_{\rm p}$ becoming equal to either the slow- or the
fast-magnetosonic wave speeds (e.g., Sakurai 1985). In the full wind
problem, the shape of the field lines must be determined as part of
the solution by solving also the transfield (or Grad-Shafranov)
equation, which involves the force balance {\em across} the flux
surfaces. This equation introduces a critical point corresponding to
$v_{\rm p}=v_{\rm A p}$ (Okamoto 1975). However, the critical points
of the {\em combined Bernoulli and transfield equations} are in
general {\em different} from those obtained when these two equations
are solved separately. The {\em modified} slow, Alfv\'en, and fast
points occur on surfaces that correspond to the so-called {\em
limiting characteristics} (or separatrices; e.g., Tsinganos et
al. 1996; Bogovalov 1997).  The relevance of the modified critical
points has been recognized already in the original radially
self-similar CDW model constructed by Blandford \& Payne (1982). The
modified critical surfaces of the exact, semianalytic wind solutions
obtained in this model are defined by the locations where the
component of the flow velocity that is perpendicular to the directions
of axisymmetry (i.e., $\vec{\hat \phi}$) and self-similarity (i.e.,
the spherical radius vector $\vec{\hat R}$) equals the MHD wave speed
in that direction (the $\vec{\hat \theta}$ direction, using spherical
coordinates $\{R, \theta, \phi\}$). The significance of the modified
fast surface, which in general is located beyond its classical
counterpart, is that the poloidal aceleration of the wind continues
all the way up to it: initially (roughly until the flow reaches the
Alfv\'en surface) the acceleration is primarily centrifugal, but
further out the pressure gradient of the azimuthal field component
comes to dominate.

\paragraph{Example: the slow-magnetosonic critical surface}

This critical surface is relevant to the determination of the mass flux
in a disk-driven wind. There is some uncertainty about this case since
the location of the first critical point is typically very close to the
disk surface (with $|z|$ being $\ll r$), so a priori it is not obvious
that ideal MHD is already a good approximation there. Under nonideal MHD
conditions, all magnetic terms in a disturbance are formally wiped out
by magnetic diffusivity on its backward propagation from ``spatial
infinity,'' and one is left with pure sound waves (e.g., Ferreira \&
Pelletier 1995). Here we discuss the situation where the
critical point is encountered when the charged particles' drift speeds
are already small enough in comparison with the bulk speed to justify
employing ideal MHD. In this case the poloidal flow is parallel to the
poloidal magnetic field (see Eq.~(\ref{eq:wind6})) and the relevant wave
speed is the slow-magnetosonic (sms) one. We consider the standard
(rather than the modified) sms point, and our explicit derivation can
hopefully serve to correct inaccurate statements that have appeared in
some of the previous discussions of this topic in the literature.

In the magnetically dominated region above the disk surface the shape of
the field lines changes on the scale of the spherical radius
$R$. Anticipating that the height $z_{\rm sms}$ of the sms point is $\ll
R$, we approximate the shape of the field line just above the point
$\{r_{\rm s},\, z_{\rm s}\}$ on the disk surface by a straight line
(cf. Sect.~\ref{subsec:centrifugal}):
\begin{equation}
\label{eq:wind16}
r = r_{\rm s} + s \sin{\theta_{\rm s}}\, , \quad z = z_{\rm s} + s
\cos{\theta_{\rm s}}\; ,
\end{equation}
where the angle $\theta_{\rm s}$ gives the field-line inclination at the
disk surface ($\sin{\theta_{\rm s}}=B_{r,{\rm s}}/B_{\rm p,s}$;
$\tan{\theta_{\rm s}} = B_{r,{\rm s}}/B_z$). The Bernoulli
integral (in the form of Eq.~(\ref{eq:wind14})) then becomes, after
substituting $\Phi = -GM_*/(r^2+z^2)^{1/2}$,
\begin{eqnarray}
\label{eq:wind17}
{\cal{H}}(s,\varrho)& =& \frac{k^2B_{\rm p}^2}{2\varrho^2} +
\frac{\Omega_B^2}{2}(r_{\rm s} + s \sin{\theta_{\rm s}})^2\left (
\frac{\Omega}{\Omega_B} \right ) \left (\frac{\Omega}{\Omega_B} - 2\right )
\nonumber \\
& & \mbox{}- \frac{G M_*}{[(r_{\rm s} + s \sin{\theta_{\rm s}})^2 +
(z_{\rm s} + s \cos{\theta_{\rm s}})^2]^{1/2}}
+ C^2 \ln{\left ( \frac{\varrho}{\varrho_{\rm A}}\right )}\; ,
\end{eqnarray}
where $\Omega(s,\varrho)$ is given by combining
\begin{equation}
\label{eq:wind18}
\Omega = \frac{1 - \frac{r_{\rm A}^2\varrho_{\rm A}}{r^2 \varrho}}{1 -
\frac{\varrho_{\rm A}}{\varrho}} \ \Omega_B
\end{equation}
(obtained from Eqs.~(\ref{eq:wind8})--(\ref{eq:wind10})) and
Eq.~(\ref{eq:wind16}).
In the hydrostatic approximation to the disk structure (see
Sect.~\ref{subsec:disk}) $\Omega_B =
\Omega_{\rm K}(r_{\rm s})$. We expect this equality to hold approximately
also for the exact solution, in which $v_z \ne 0$, so that, in particular,
$\Omega_B \approx \Omega$ in the region of interest. 

Setting $\lfrac{\partial {\cal{H}}}{\partial \varrho} = 0$ yields the
speed of an sms wave propagating along $\vec{\hat B}_{\rm p}$:
\begin{equation}
\label{eq:wind19}
v_{\rm sms} = \frac{B_{\rm p,s}}{B_{\rm s}}\ C\; ,
\end{equation}
where $B_{\rm s} = (B_{\rm p,s}^2 + B_{\phi,{\rm s}}^2)^{1/2}$ and we
approximated $(B_{\rm p, sms}/B_{\rm sms})^2$ by $(B_{\rm p, s}/B_{\rm
s})^2$ and assumed $\varrho/\varrho_{\rm A} \gg 1$ and $(\Omega/\Omega_B
- 1)^2 \approx (r_{\rm A}/r)^4(\varrho_{\rm A}/\varrho)^2\ll 1$ in the region
between the top of the disk and the sms surface. Setting also $\partial
H/\partial r = 0$ and approximating $r_{\rm sms} \approx r_{\rm s}$ then
gives the height of this point:
\begin{equation}
\label{eq:wind20}
\frac{z_{\rm sms}}{z_{\rm s}} = \frac{3 \tan^2{\theta_{\rm s}}}{3
\tan^2{\theta_{\rm s}} - 1}\; .
\end{equation}
Equation~(\ref{eq:wind20}) yields a meaningful result only if
$\tan{\theta_{\rm s}} > 1/\sqrt{3}$, i.e., if the field line is inclined
at an angle $>30^{\circ}$ to the $z$ axis. This is the CDW launching
condition in a Keplerian disk (Eq.~(\ref{eq:MHD6})) that was derived in
Sect.~\ref{subsec:centrifugal} using the mechanical analogy to a ``bead
on a rigid wire.'' The relationship between these two derivations
becomes clear when one notes that, in the limit $(\Omega/\Omega_B - 1)^2
\ll 1$, the 2nd term on the right-hand side of Eq.~(\ref{eq:wind17})
becomes equal to the centrifugal potential $-\Omega^2 r^2/2$, so that
the 2nd and 3rd terms (which together dominate the right-hand side of
this equation) are just the effective potential $\Phi_{\rm eff}$ used in
Eq.~(\ref{eq:MHD5}).\footnote{Furthermore, the condition
$\lfrac{\partial {\cal{H}}}{\partial r} = 0$ is equivalent to
$\vec{B}_{\rm p} \cdot \nabla \Phi_{\rm eff} = 0$ in this limit
(Campbell 2002), so the latter relation can also be used to obtain the
location of the sms point in this case (e.g., Ogilvie 1997).} The
correspondence of the sms point to the maximum of $\Phi_{\rm eff}$ was
already noted in Sect.~\ref{subsec:centrifugal}.

The density at the sms point can be related to the density at the disk
surface by evaluating the energy integral~(\ref{eq:wind17}) at both $z_{\rm
s}$ and $z_{\rm sms}$. This yields
\begin{equation}
\label{eq:wind21}
\frac{\varrho_{\rm sms}}{\varrho_{\rm s}} = {\rm exp}{ \left \{ - \frac{1}{2}
- \frac{1}{2}\left ( \frac{z_{\rm s} \Omega_{\rm K}(z_{\rm s})}{C}\right
)^2 \frac{1}{3 \tan^2{\theta_{\rm s}} - 1} \right \}}\; ,
\end{equation}
where we assumed $(v_{\rm p,s}/v_{\rm sms})^2 \ll 1$. The mass flux
injected into the wind from the two sides of the disk is then
\begin{equation}
\label{eq:wind22}
\frac{1}{2\pi r}\frac{\D{\dot{M}_{\rm w}}}{\D{r}} = 2 \varrho_{\rm sms}
v_{\rm sms} \cos{\theta_{\rm s}} = 2 \varrho_{\rm sms} C
\frac{B_z}{B_{\rm s}}
\end{equation}
(e.g., Lovelace et al. 1995).

When $\theta_{\rm s}$ approaches (and
decreases below) $30^{\circ}$ the field-line curvature needs to be taken
into account in the analysis. The height of the sonic point rapidly
increases to $\sim R$, with the potential difference growing to $\sim
GM_*/R\,$: the launching problem becomes essentially that of a thermally driven
spherical wind (e.g., Levinson 2006).

It has been argued (Spruit 1996) that the field-line
inclination increases systematically with radius along the disk surface
and that only in a narrow radial range the conditions are favorable for
driving a wind that both satisfies the CDW launching condition and is
not ``overloaded'' (and hence conceivably highly unstable; cf. Cao \&
Spruit 1994). This is an intriguing suggestion, given the
fact that so far there is no observational evidence for an extended
wind-driving region in protostellar disks (see
Sect.~\ref{subsec:connection}). However, in principle it may be possible
to launch outflows over a large radial range. In particular, as
illustrated by the similarity solution presented in
Sect.~\ref{sec:formation}, if the magnetic flux is advected inward by the
accretion flow, the field-line inclination could be favorable for
launching and need not change strongly along the disk. Furthermore,
evidently the mass loading of stable disk/wind configurations actually
{\em decreases} as $\theta_{\rm s}$ is increased (see
Sect.~\ref{subsec:stability}).

\subsubsection{Exact Wind Solutions}
\label{subsubsec:exact_wind}

As was already noted in Sect.~\ref{subsec:similar}, the radial
self-similarity approach has been used to construct exact global
solutions of CDWs. The basic character of this model is revealed by the
prototypal Blandford \& Payne (1982) solution. The underlying assumption
that all quantities scale as a power law of the spherical radius implies
that all the critical surfaces are conical. Furthermore, all the
relevant speeds (including the fluid, sound, and Alfv\'en) must scale
like the characteristic speed of the problem ($v_{\rm K}$), i.e., as
$R^{-1/2}$. The scaling of the magnetic field amplitude can be inferred
from the vertically integrated thin-disk equations presented in
Sect.~\ref{subsec:core_eqns}, in which we now set $\lfrac{\partial}{\partial
t} \equiv 0$. In particular, if the mass outflow from the disk has only
a negligible effect on the accretion rate then $\dot M_{\rm a} = 2\pi r
|v_r| \Sigma = {\rm const}$ by Eq.~(\ref{eq:form6}) and the angular momentum
conservation equation~(\ref{eq:form8}) in a Keplerian disk can be written as
\begin{equation}
\label{eq:wind23}
\frac{1}{2}\dot M_{\rm a} v_{\rm K} = r^2 |B_z B_{\phi,{\rm s}}|\; .
\end{equation}
Equation~(\ref{eq:wind23}) implies the similarity scaling $B\propto
r^{-5/4}$, from which we infer by dimensional arguments ($v_{\rm A}
\propto B/\sqrt{\varrho}$) that $\varrho \propto r^{-3/2}$ in this
case. It then follows that $\dot M_{\rm w} \propto \ln{r}$
(cf. Eq.~(\ref{eq:wind22})), which is consistent with the underlying
assumption that only a small fraction of the inflowing mass leaves
the disk over each decade in radius. More generally, one can define a mass
{\em ejection index} $\xi>0$ by
\begin{equation}
\label{eq:wind24}
\xi \equiv \frac{\D{\ln{\dot M_{\rm a}}}}{\D{\ln{r}}}
\end{equation}
(e.g., Ferreira \& Pelletier 1995) and deduce
\begin{equation}
\label{eq:wind25}
\frac{\D{\ln{B}}}{\D{\ln{r}}} = \frac{\xi}{2} -  \frac{5}{4}\; , \quad
\frac{\D{\ln{\varrho}}}{\D{\ln{r}}}= \xi -\frac{3}{2}
\end{equation}
(e.g., Contopoulos \& Lovelace 1994), with $\xi\rightarrow
0$ corresponding to the Blandford \& Payne (1982)
solution.\footnote{Interestingly, one can also generalize the Blandford
\& Payne (1982) solution to a class of semianalytic, but {\em
non}--self-similar solutions, defined by the constancy of a certain
function of the magnetic flux that controls the shape of the Alfv\'en
surface (Pelletier \& Pudritz 1992). The Blandford \&
Payne solution then separates wind configurations in which the initial
field-line inclination increases progressively with radius from outflows
that emerge from a bounded region and in which the initial field
inclination decreases with $r$ and the field lines converge into a
cylindrical sheath.} The poloidal electric current scales as $I \propto
rB_\phi \propto r^{(2\xi-1)/4}$. For $\xi > 1/2$ the flow is in the
current-carrying regime, with the poloidal current density being
antiparallel to the magnetic field. In this case the current tends to
zero as the symmetry axis is approached, so such solutions should
provide a good representation of the conditions near the axis of a
highly collimated flow. Conversely, solutions with $\xi < 1/2$
correspond to the return-current regime (in which the poloidal current
density is parallel to the field) and are most suitable at larger
cylindrical distances. The Blandford \& Payne (1982) solution is
sometimes critiqued for having a singular behavior on the
axis.\footnote{It is in fact a double singularity, since on the axis
itself there is an oppositely directed line current that exactly
compensates for the distributed return current.} One should bear in
mind, however, that even though the detailed global current distribution
(which includes both a current-carrying and return-current regimes)
cannot be represented by the simplified self-similar solution, this flaw
is not fundamental. Besides, given that the disk has a finite inner
radius, the issue of the behavior of the flow at $r=0$ is, to a
certain degree, merely academic. Whereas the value of the ejection index
is arbitrary in pure wind models, it becomes an eigenvalue of the
self-similar solution (fixed by the regularity condition at the sonic,
or sms, critical surface; see Sect.~\ref{subsubsec:critical}) when one
considers a ``combined'' disk/wind model (e.g., Li 1996a).

While the magnetic field lines must bend away from the symmetry axis
within the disk in order to satisfy the wind launching
condition~(\ref{eq:MHD6}) at the surface, as soon as the magnetically
dominated region at the base of the wind is reached they start to bend
back toward the axis on account of the magnetic tension force (see
Sect.~\ref{subsec:forces}). Further collimation is achieved in
current-carrying jets by the {\em hoop stress} of the toroidal magnetic
field (the term $-J_z B_\phi/c = - (\lfrac{1}{8\pi r^2})\lfrac{\partial
(r B_\phi)^2}{\partial r}$ in the radial force equation), the analog of
a z-pinch in laboratory plasma experiments. The asymptotic behavior in
general depends on the current distribution: if $I \rightarrow 0$ as $r
\rightarrow \infty $ then the field lines are space-filling paraboloids,
whereas if this limit for the current is finite then the flow is
collimated to cylinders (e.g., Heyvaerts \& Norman 1989).  In practice,
however, self-similar wind solutions typically do not reach the
asymptotic regime but instead self-focus (with streamlines intersecting
on the axis) and terminate at a finite height (e.g., Vlahakis et
al. 2000).

The cold self-similar wind solutions are specified by any two of the
following three parameters:
\begin{eqnarray}
\lambda &\equiv& \frac{l}{v_{\rm K s} \, r_{\rm s}} \quad\quad
{\rm normalized\ specific\ angular\ momentum} \nonumber \\
\kappa &\equiv& k \frac{v_{\rm K s}}{B_{z,{\rm s}}} \quad\quad
{\rm normalized\ mass/magnetic\ flux\ ratio}\\
b_{r,{\rm s}} &\equiv& \frac{B_{r,{\rm s}}}{B_{z,{\rm s}}} \quad\quad
{\rm poloidal\ field\ inclination\ at\ the\ disk\ surface}\nonumber \; ,
\label{eq:wind26}
\end{eqnarray}
where $b_{r,{\rm s}}$ must satisfy the constraint~(\ref{eq:MHD6}). A
viable solution is further characterized by $\lambda > 1$ (typically
$\gg 1$) and $\kappa < 1$ (typically $\ll 1$). A two-parameter choice
yields a solution if the corresponding flow crosses the modified
Alfv\'en critical surface (see Sect.~\ref{subsubsec:critical}); if the
two parameters are $\kappa$ and $\lambda$ then, for any given value of
$\kappa$, this can happen only if $\lambda$ exceeds a minimum value that
approximately satisfies $\kappa \lambda_{\rm min} (\lambda_{\rm min} -
3)^{1/2}=1$.\footnote{This expression differs slightly from
the one that appeared in Blandford \& Payne (1982).}
In Sect.~\ref{subsec:disk} we discuss joining a global wind solution of
this type to a radially localized wind-driving disk solution. In this
case the sonic (or sms) critical-point constraint imposed on the latter
solution yields the value of $\kappa$, and, in turn, the Alfv\'en
critical-point constraint on the wind solution fixes one of the disk
model parameters (for example, by providing the value of $B_{\phi,{\rm
s}}/B_{z,{\rm s}} = - \kappa (\lambda - 1)$ from given values of $\kappa$
and $b_{r,{\rm s}}$).\footnote{Although self-similar CDW solutions in
which the wind passes also through the fast-magnetosonic critical
surface have been obtained, the implications of this additional
constraint to a global disk/wind model remain unclear. Ferreira \& Casse
(2004), for instance, suggested that in order to cross
this surface the outflow must experience significant heating after it
leaves the disk and that the added constraint is related to this
requirement.}

\subsubsection{Observational Implications}
\label{subsubsec:wind_observe}

The exact wind solutions and estimates of the physical conditions around
young stellar objects make it possible to identify a number of physical
characteristics of disk-driven protostellar winds that could have
potentially significant observational implications. While this topic is
not directly within the scope of this book, we nevertheless briefly
describe some of these properties here inasmuch as it may not always be
feasible in practice to distinguish between the observational signatures
of a disk and a wind. Furthermore, invoking a disk-driven wind may
sometimes provide a straightforward explanation to a disk observation
that may otherwise seem puzzling. Besides the references explicitly
listed below, the reader may also consult the review articles by
K\"onigl \& Ruden (1993) and K\"onigl \& Pudritz (2000) on this subject.

\begin{figure}
\centering
\includegraphics[height=8cm]{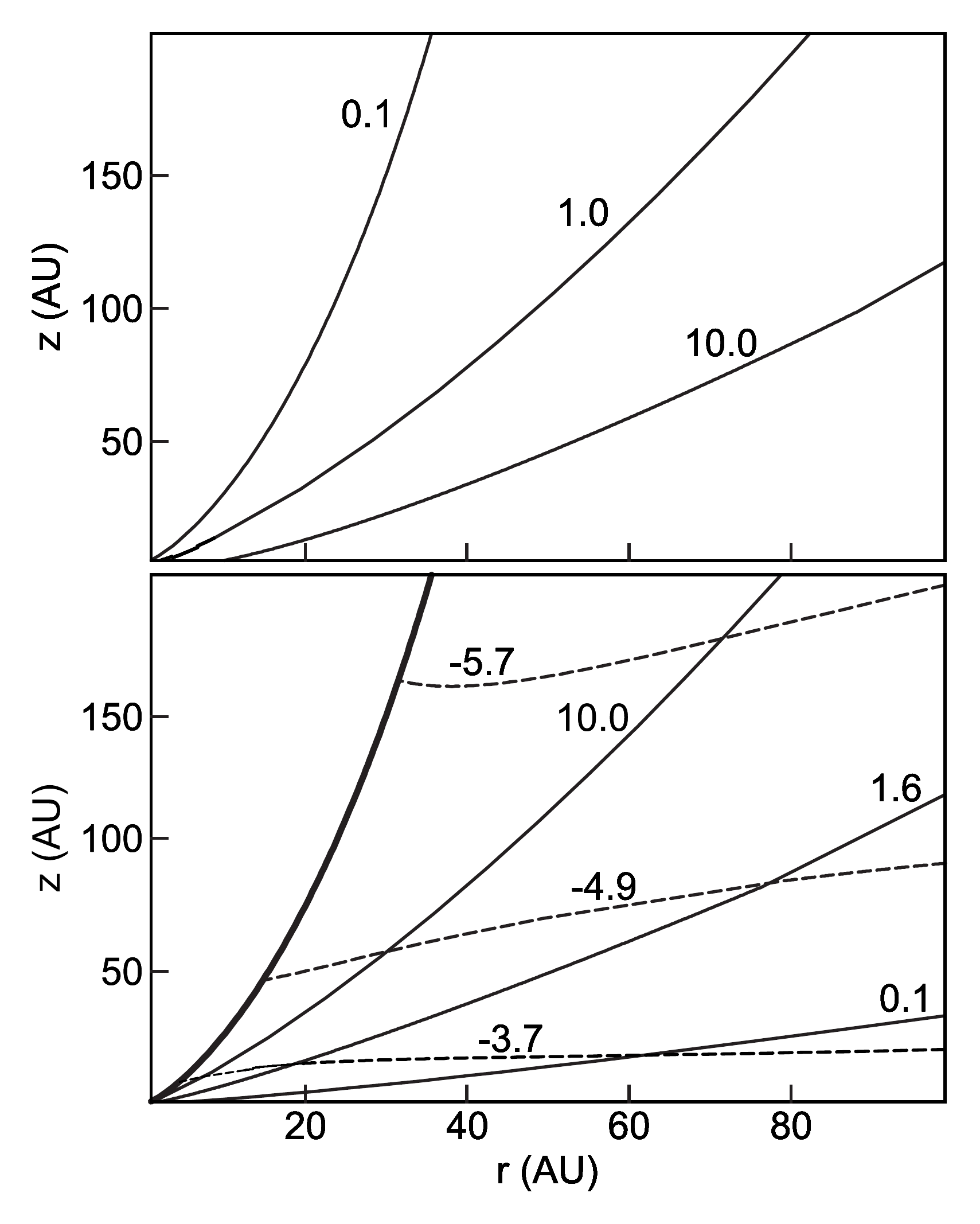}
\caption{Structure of a Blandford-Payne--type self-similar CDW from
a protostellar disk (model C in Safier 1993a). {\bf Top}: Meridional
projections of the streamlines (labeled by the value, in AU, of the
radius where they intersect the disk). {\bf Bottom}: Contours of
$\lfrac{v_{{\rm p},\infty}}{v_{\rm K1}}$ ({\it solid lines}\/),
where $v_{{\rm p},\infty}$ is the asymptotic poloidal speed and
$v_{\rm K1}$ is the Keplerian speed at $1\,$AU, and of
$\log{(\varrho/\varrho_1)}$ ({\it dashed lines}\/), where
$\varrho_1$ is the density at the base of the wind at $1\,$AU. The
heavy line on the left indicates the streamline that originates at
$0.1\,$AU.} \label{fig6_4}
\end{figure}

Centrifugal driving is found to be an efficient acceleration mechanism,
leading to a rapid increase in the poloidal speed, and a correspondingly
strong decrease in the wind density, above the disk surface (see
Fig.~\ref{fig6_4}).\footnote{The model wind presented in
Fig.~\ref{fig6_4} exhibits a comparatively weak collimation. The degree
of collimation depends on the model parameters and is particularly
sensitive to the mass loading of the outflow (e.g., Pudritz et
al. 2007). Protostellar jets are inferred to correspond to more highly
collimated disk winds, in which the enhanced mass loading might be
attributed to the presence of a disk corona (e.g., Dougados et
al. 2004).} This leads to a strong stratification, which, in fact, is
most pronounced in the $\vec{\hat z}$ direction even in the
$R$--self-similar wind models (Safier 1993b). The strong momentum flux
in the wind also implies that an outflow originating from beyond the
{\em dust sublimation radius} could readily {uplift dust from the disk}
(Safier 1993a). This may lead to viewing angle-dependent obscuration and
shielding of the central continuum , whose effect would be distinct from
that of molecular cloud gas undergoing gravitational
collapse.\footnote{In an application to active galactic nuclei, this
effect was invoked by K\"onigl \& Kartje (1994) to account for the
obscuring/absorbing ``molecular torus'' identified in the centers of
Seyfert galaxies.} The stratified dust distribution could intercept a
large portion of the central continuum radiation and reprocess it to the
infrared, which could be relevant to the interpretation of the IR
spectra of low- and intermediate-mass star/disk systems (e.g., K\"onigl
1996; Tambovtseva \& Grinin 2008). Furthermore, scattering by the
uplifted dust could contribute to the observed polarization pattern in
these objects.\footnote{For an application of this model to the
polarization properties of Seyfert galaxies, see Kartje (1995).}

Although, as we already noted, ideal MHD should be an excellent
approximation for modeling the {\em dynamics} of the flow (except
perhaps right above the disk surface, where the nonideal formalism
employed within the disk might still be relevant for determining
the first critical point of the wind), the estimated degree of
ionization at the base of a protostellar disk wind is typically low
enough that energy dissipation induced by ion--neutral drag (ambipolar
diffusion) could play an important role in the wind {\em
thermodynamics}. This is because the volumetric heating rate scales as
$|(\nabla \times \vec{B})\times \vec{B}|^2/\gamma_{\rm i}\varrho_{\rm
i}\varrho \propto 1/\gamma_{\rm i}\varrho_{\rm i}$ (see
Sect.~\ref{subsec:nonideal}), where the distributions of $\vec{B}$ and
$\varrho$ are regarded as being fixed by the flow dynamics, whereas
adiabatic cooling, the most important temperature-lowering mechanism,
remains relatively inefficient in a collimated disk outflow. Under these
circumstances, the temperature might rise rapidly (Safier 1993a),
although its precise terminal level depends on the relevant value of the
collisional drag coefficient $\gamma_{\rm i}$ (Shang et al. 2002).
This heating mechanism could contribute to the observed
forbidden line emission (Safier 1993b; Cabrit et al. 1999;
Garcia et al. 2001a,b) and thermal radio
emission (Martin 1996) from CTTSs, although in both of
these applications an additional source of heating might be needed,
possibly associated with dissipation of weak shocks and turbulence
(e.g., O'Brien et al. 2003; Shang et al. 2004).

On large scales, centrifugally driven outflows assume the structure of a
collimated jet (most noticeable in the density contours) and a
surrounding wide-angle wind (Shu et al. 1995; Li 1996b).  This bears
directly on the general morphology of protostellar sources, and in
particular on the shapes of the (wind/jet)-driven outflow lobes that
form in the surrounding molecular gas.

\subsection{Equilibrium Disk/Wind Models}
\label{subsec:disk}

In contrast with the wind zone, where the dynamics is well described by
an effectively infinite conductivity, the accretion disk in protostellar
systems must be modeled using nonideal MHD. The structure of the
accretion flow thus depends critically on the properties of the
conductivity tensor at each point, which are in turn determined by the
spatial distribution of the degree of ionization (see
Eqs.~(\ref{eq:MHD12})--(\ref{eq:MHD14})). We therefore start this
subsection by reviewing the physical processes that affect the disk
ionization (see also Chap.~II [The Chemical Evolution of Protoplanetary
Disks]), which would allow us to choose the relevant physical parameters
for our models. In reality the accretion flow itself has an influence on
the ionization structure --- through its effect on the disk column
density (see Sect.~\ref{subsec:nonideal}) or on the distribution of dust
grains (see Sect.~\ref{sec:conclude}), for example --- which a truly
self-consistent model must take into account. So far only simpler models
have been constructed, in which this influence is not fully accounted
for and a variety of approximations are employed. We discuss these
models next, describing the basic properties of the derived equilibrium
solutions and distinguishing between ``strongly coupled'' and ``weakly
coupled'' disk configurations.

\subsubsection{Disk Ionization Structure}
\label{subsubsec:ionize}

The dependence of the degree of field--matter coupling on the abundances of the
ionized species in the disk (ions, electrons and charged
dust grains) is a direct outcome of the fact that this coupling is
effected by the collisions of these particles with the much more
abundant neutrals. The degree of ionization is given by the ratio of the
sum of the positively (or, equivalently, negatively) charged particle densities
to the neutral particle density. It is calculated by balancing the
ionization and recombination processes operating in the disk.

Interstellar (and possibly also protostellar) cosmic rays as well as
X-ray and far-UV radiation produced by the magnetically active protostar
(e.g., Glassgold et al. 2000, 2005) are the main potential ionizing
agents in protostellar environments. It is, however, unclear how
effective cosmic-ray ionization really is, since the low-energy
particles most relevant for this purpose may be deflected by magnetized
disk or stellar outflows or by magnetic mirroring near the disk surface,
and they might also be scattered by magnetic turbulence within the
disk. If present, they may be more important than X-rays in the inner disk ($r
\la 1\,$AU), where the surface density is larger than the X-ray
attenuation length. But if cosmic rays are excluded from this region and
ionized particles are not transported there by other means (e.g., Turner
et al. 2007), the gas near the midplane might be ionized only at the low
rate of decay of radioactive elements such as $^{40}{\rm K}$. On the
other hand, free electrons are rapidly lost through recombination
processes that occur both in the gas phase (where electrons recombine
with molecular and metal ions via dissociative and radiative mechanisms,
respectively) and on grain surfaces, and through sticking to dust
particles (e.g., Oppenheimer \& Dalgarno 1974; Nishi et al. 1991). The
ionization equilibrium is, however, very sensitive to the abundance of
metal atoms, as they rapidly remove charges from molecular ions but then
recombine much more slowly with electrons (e.g., Sano et al. 2000;
Fromang et al. 2002).

Dust grains can also significantly affect the degree of field--matter
coupling if they are mixed with the gas. They do so in two ways. First,
they reduce the ionization fraction by absorbing charges from the gas
and by providing additional recombination pathways for ions and
electrons. Second, in high-density regions the grains themselves can
become an important charged species (e.g., Nishi et al. 1991),
leading to a reduction in the magnetic coupling because grains have much
smaller Hall parameters than the much less massive electrons and ions.
We note in this connection that, if dust grains settle to the midplane,
the ionization fraction may become sufficiently large to provide
adequate magnetic coupling even in the inner ($\la 1\,$AU) disk regions
(Wardle 2007). This conclusion could, however, be mitigated in the
presence of turbulence, which might leave a residual population of small
dust grains (carrying a significant fraction of the total grain charge)
suspended in the disk (e.g., Nomura \& Nakagawa 2006; Natta et
al. 2007).

\subsubsection{Exact Disk Solutions}
\label{subsubsec:exact_disk}

\begin{figure}
\centering
\includegraphics[height=7cm]{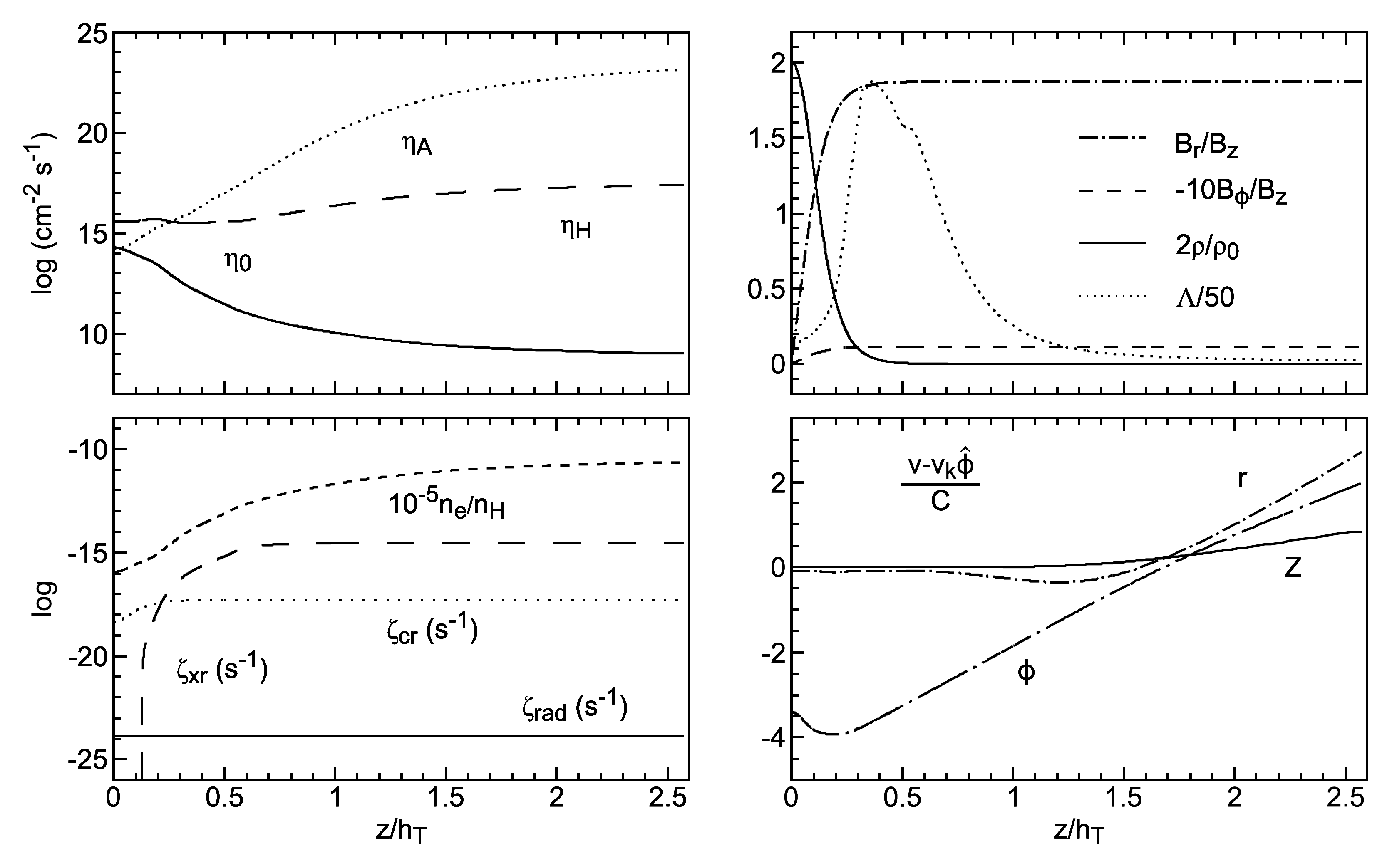}

\caption{Vertical structure of a radially localized wind-driving disk
solution at 1 AU around a Sun-like protostar. The disk is assumed to
have a column density of $\Sigma = 600 \, {\rm g} \, {\rm cm}^{-2}$ and
its model parameters are $a_0 = 0.75$, $C/v_{\rm K} = 0.1$, $\epsilon =
0.1$, and $\epsilon_{\rm B}=0$. The parameters of the matched radially
self-similar CDW are $\kappa = 2.6 \times 10^{-6}$, $\lambda = 4.4
\times 10^3$, and $b_{r,{\rm s}} = 1.6$. ({\bf Left}) {\bf Top}:
Ambipolar, Hall, and Ohm diffusivities. {\bf Bottom}: Ionization rates
by cosmic rays ({\bf cr}), X-rays ({\bf xr}), and radioactivity ({\bf
rad}), and electron fraction $\lfrac{n_{\rm e}}{n_{\rm H}}$. ({\bf
Right}) {\bf Top}: Radial and azimuthal magnetic field components, mass
density, and Elsasser number. {\bf Bottom}: Velocity components. The
mass accretion rate for this model is $7\times 10^{-6}\, M_\odot\, {\rm
yr}^{-1}$, which is consistent with the inferred values for the early
(Class 0/Class I) protostellar accretion phase.}
\label{fig6_5}
\end{figure}

To gain physical insight into the distinguishing properties of
wind-driving disks, we consider a simplified model originally formulated
by Wardle \& K\"onigl (1993). In this model, the entire angular momentum
of the accreted matter is assumed to be transported by a CDW. The disk
is taken to be geometrically thin, vertically isothermal, stellar
gravity-dominated, and threaded by an open magnetic field (possessing
reflection symmetry about the midplane). Under the thin-disk
approximation, the vertical magnetic field component is taken to be
uniform with height in the disk solution. The disk gas is assumed to be
in the ambipolar diffusivity regime, with ions and electrons being the
dominant charge carriers. In the original model, the ion density was
taken to be constant with height (which could be a reasonable
approximation in the outer regions of certain real systems). The main
simplifications involve considering only a {\em radially localized}
($\varDelta{r}\ll r$) disk region and (in accord with the thin-disk
approximation) retaining the $z$ derivatives but neglecting all $r$
derivatives except those of $v_\phi$ (which scales as $r^{-1/2}$) and
$B_r$ (which appears in $\nabla \cdot \vec{B} = 0$). The neglect of the
radial derivatives in the mass conservation equation
(Eq.~(\ref{eq:wind1})) implies that the vertical mass flux $\varrho v_z$
is taken to be uniform with height in the disk solution.\footnote{This
approximation has been critiqued (e.g., Ferreira 1997) for not allowing
$v_z$ to assume negative values within the disk, as it must do in cases
(expected to be typical) in which the disk thickness decreases as the
protostar is approached. This issue can be fully addressed only in the
context of a global disk/wind model. However, it can be expected that
any error introduced by this approximation would be minimized if the
upward mass flux remained small enough for $v_z$ to have only a weak
effect on the behavior of the other variables within the disk. Under the
assumption that $|v_r|$ is of the same order of magnitude as $|v_\phi -
v_{\rm K}|$ one can readily show that the condition for this to hold is
that $v_z/C$ remain $\ll 1$ everywhere within the disk. This can be
checked a posteriori for each derived solution.} This solution is
extended through the first critical point (which is the thermal sonic
point if the flow is still diffusive and the sms point if it is already
in the ideal-MHD regime; see Sect.~\ref{subsubsec:critical}) and then
matched onto a {\em global} Blandford \& Payne (1982) ideal-MHD wind
solution. These simplifications do not compromise the physical essence
of the results. In particular, the qualitative characteristics remain
unchanged when the radially localized disk solution is generalized to a
global configuration in which both the disk and the wind are described
by a single self-similar model that includes both $z$ and $r$
derivatives (e.g., Li 1996a).  Furthermore, solutions with similar
properties are obtained when the disk is in the Hall or Ohm diffusivity
regimes (e.g., Salmeron et al. 2010) and when the full conductivity
tensor is used in conjunction with a realistic ionization profile (see
Fig.~\ref{fig6_5}).

In general, an equilibrium disk solution is specified by the following
parameters:
\begin{enumerate}

\item $a_0 \equiv \lfrac{v_{{\rm A} 0}}{C}$, the midplane ratio of the
Alfv\'en speed (based on the uniform vertical field component) to the
sound speed. This parameter measures the magnetic field
strength.\footnote{Note that $a_0$ is related to the midplane plasma
beta parameter $\beta_0$ through $a_0 = (2 / \gamma \beta_0)^{1/2}$,
where $\gamma$ denotes the adiabatic index of the fluid.}

\item $\lfrac{C}{v_{\rm K}}$, which in a thin isothermal disk is equal to
$\lfrac{h_{\rm T}}{r}$, the ratio of the tidal (i.e., reflecting the
vertical gravitational compression) density scale height to the disk
radius. While this parameter, which measures the geometric thinness of
the disk, does not appear explicitly in the normalized structure
equations, it nevertheless serves to constrain physically viable
solutions (see Eq.~(\ref{eq:disk1}) below).

\item The midplane ratios of the conductivity tensor components:
$[\sigma_{\rm P}/\sigma_\perp]_0$ (or $[\sigma_{\rm H}/\sigma_\perp]_0$)
and $[\sigma_\perp/\sigma_{\rm O}]_0$. They characterize the
conductivity regime of the gas (see Sect.~\ref{subsec:nonideal}). When
there are only two charged species (ions and electrons), one can
equivalently specify the midplane Hall parameters $\beta_{{\rm i} 0}$ and
$\beta_{{\rm e} 0}$. In the inner disk regions the conductivity tensor
components typically vary with height, reflecting the ionization
structure of the disk (see Fig.~\ref{fig6_5}).

\item The midplane Elsasser number $\Lambda_0$ (see Eqs.~(\ref{eq:MHD23})
and~(\ref{eq:MHD25})), which measures the degree of coupling between the
neutrals and the magnetic field.

\item $\epsilon \equiv -\lfrac{v_{r,0}}{C}$, the normalized inward radial
speed at the midplane. Although the value of $\epsilon$ could in
principle be negative (as, in fact, it is in certain viscous disk
models; e.g., Takeuchi \& Lin 2002), it is expected to
remain $>0$ when a large-scale magnetic field dominates the angular
momentum transport.

\item $\epsilon_{\rm B} \equiv -\lfrac{v_{Br,0}}{C}$, the normalized
radial drift speed of the poloidal magnetic field lines (see
Eq.~(\ref{eq:MHD22})). This parameter vanishes in a strictly
steady-state solution but is nonzero if, as expected, the magnetic
field lines drift radially on the ``long'' accretion time scale
$\lfrac{r}{|v_r|}$. It is incorporated into the model through the $z$
component of Eq.~(\ref{eq:MHD20}) (i.e., one sets $\lfrac{\partial
B_z}{\partial t} \ne 0$ but keeps $\lfrac{\partial B_r}{\partial t} =
0$ and $\lfrac{\partial B_\phi}{\partial t} = 0$).\footnote{It is
worth noting, though, that the poloidal components of
Eq.~(\ref{eq:MHD20}) can be written as $\lfrac{\partial \Psi}{\partial
t} = - 2 \pi r B_0 v_{Br,0}$, with $B_z=(\lfrac{1}{2\pi
r})\lfrac{\partial \Psi}{\partial r}$ and $B_r = -(\lfrac{1}{2\pi
r})\lfrac{\partial \Psi}{\partial z}$. Thus, if $B_z$ changes because
of the slow radial diffusion of the flux surfaces, so will also
$B_r$. However, it can be argued (K\"onigl et al. 2010) that $B_r$
(and similarly also $B_\phi$) can in principle change at any given
radial location on the much shorter dynamical time $\lfrac{r}{v_\phi}$
through field-line shearing by the local velocity field. One can
therefore assume that $B_r$ and $B_\phi$ attain their equilibrium
configurations on the ``short'' time scale and neglect, in comparison,
the much slower variations that are expressed by the explicit time
derivate terms. The $B_z$ component is distinct in that it can only
change on the ``long'' radial drift time.}

\end{enumerate}

We already remarked in Sect.~\ref{subsubsec:exact_wind} on how the wind
solution, through the Alfv\'en-surface regularity condition, can be used
to constrain the disk solution when the two are matched. In particular,
if the disk solution is used to fix the wind parameters $\kappa$
(through the sonic/sms critical-point constraint) and $b_{r,{\rm s}}$,
the wind solution yields $B_{\phi,{\rm s}}/B_z$, which in turn can be
used to obtain the value of the disk parameter $\epsilon$. Previous
``combined'' disk/wind models treated $\epsilon_{\rm B}$ as a free
parameter and typically set it equal to zero.\footnote{Wardle \&
K\"onigl (1993) noted, however, that for physical consistency one has to
require $\epsilon_B < \epsilon$. They also demonstrated that the
solution variables (except for $B_{\phi,{\rm s}}$ and $v_{r,0}$) are
insensitive to the value of this parameter, as expected from the fact
that the only modification to the equations introduced by varying
$\epsilon_B$ involves changing the radial velocity of the reference
frame in which the poloidal field lines are stationary.} Under this
approach, $b_{r,{\rm s}}$ was fully determined by the conditions inside
the disk. However, as we noted in Sect.~\ref{subsec:core_eqns} (see also
Ogilvie \& Livio 2001), $b_{r,{\rm s}}$ is in fact also determined by
the conditions outside the disk and can be directly related to the
distribution of $B_z$ along the disk if the force-free field above the
surface can be adequately approximated as being also current free (see
Eq.~(\ref{eq:form15})). In a more general treatment of the disk/wind
problem, this constraint can be used to fix the value of $\epsilon_B$
(see Teitler 2010).

Before turning to the specific solution displayed in Fig.~\ref{fig6_5},
it is instructive to list the dominant magnetic terms (under the
thin-disk approximation) in the neutrals' force equation
(Eq.~(\ref{eq:wind2})) and review their main effects inside the
disk. We have (see Eqs.~(\ref{eq:form3})--(\ref{eq:form5})):
\begin{itemize}
\item {\em radial} component: $\frac{B_z}{4\pi}\frac{\D{B_r}}{\D{z}}$\\
representing the magnetic tension force that acts in opposition to
central gravity;
\item {\em azimuthal} component: $\frac{B_z}{4\pi} \frac{\D{B_\phi}}{\D{z}}$\\
representing the magnetic torque that transfers angular momentum from
the matter to the field;
\item {\em vertical} component: $-\frac{\D}{\D{z}}\frac{B_r^2 +
B_\phi^2}{8\pi}$\\
representing the magnetic squeezing of the disk (which acts in the same
direction as the gravitational tidal force and in opposition to the
thermal pressure-gradient force).
\end{itemize}

\begin{figure}
\centering
\includegraphics[height=6cm]{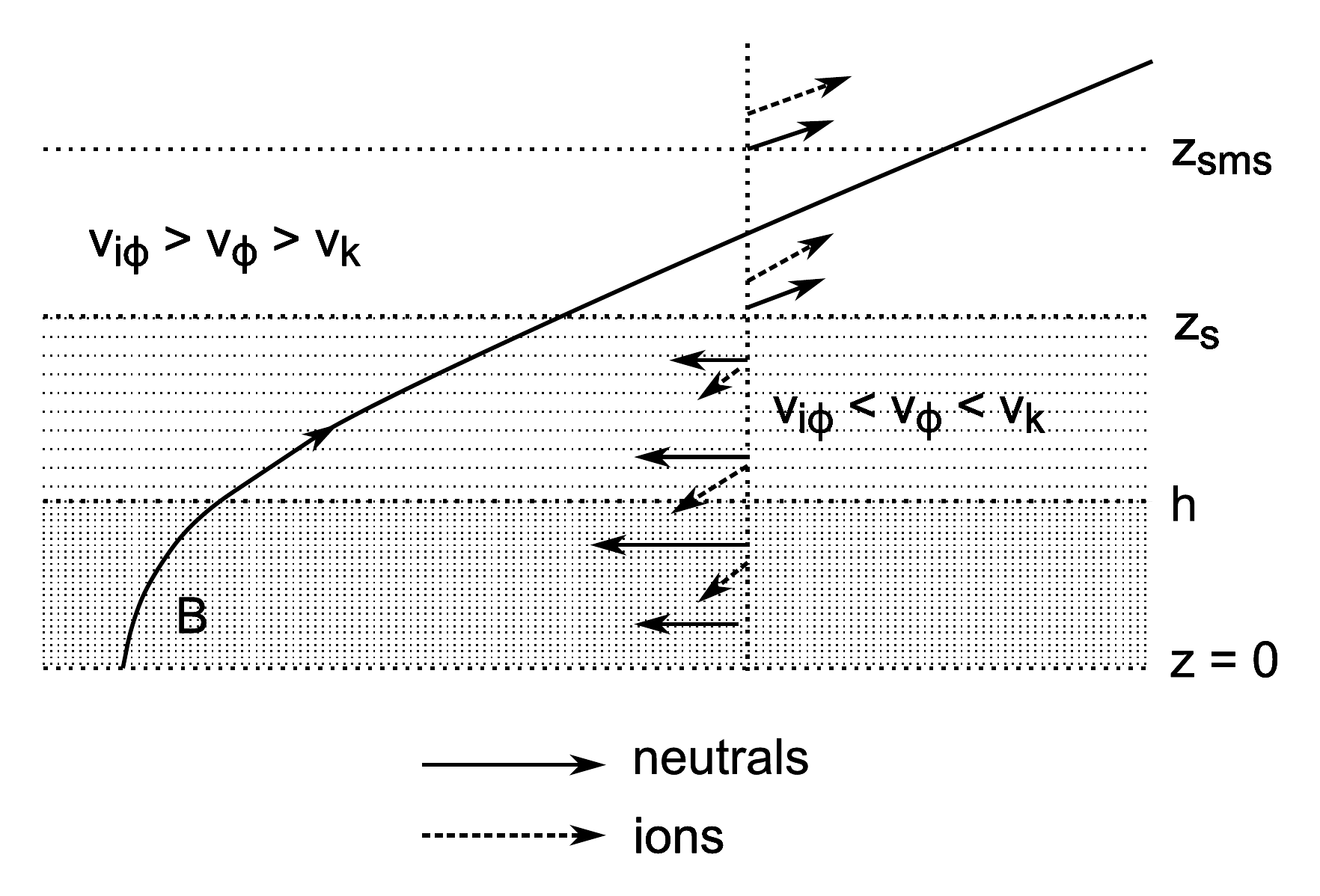}
\caption{Schematic diagram of the vertical structure of an ambipolar
diffusion-dominated disk, showing a representative field line and
the poloidal velocities of the neutral and the ionized gas
components. Note that the poloidal velocity of the ions vanishes at
the midplane ($z=0$), consistent with the assumption that
$\epsilon_B = 0$, and that it is small for both components at the top
of the disk ($z=z_{\rm s}$). The relation between the azimuthal
speeds is also indicated.} \label{fig6_6}
\end{figure}

The solution shown in Fig.~\ref{fig6_5} describes the vertical structure
of a disk that changes from being Hall-dominated near the midplane to
being ambipolar-dominated near the surface. While the qualitative
properties of the solution are not sensitive to the details of the
diffusivity profile, it is heuristically useful to consider a distinct
diffusivity regime. We therefore specialize in the following discussion
to the ``pure ambipolar'' (with $\varrho_{\rm i}(z) = {\rm const}$) case
treated in Wardle \& K\"onigl (1993). As illustrated in
Fig.~\ref{fig6_6} (in which it is assumed that $\epsilon_{\rm B}=0$), the 
disk can be vertically divided into three distinct zones:\footnote{Our 
discussion pertains to disks in which $\Lambda > 1$ throughout their
entire vertical extent, but is also applies to the $\Lambda >1$
wind-driving surface layers of ``weakly coupled'' disks.} a
quasi-hydrostatic region near the midplane, where the bulk of the matter
is concentrated and most of the field-line bending takes place, a
transition zone where the inflow gradually diminishes with height, and
an outflow region that corresponds to the base of the wind. The first
two regions are characterized by a radial inflow and sub-Keplerian
rotation, whereas the gas at the base of the wind flows out with
$v_{\phi} > v_{\rm K}$.

\begin{itemize}
\item The {\em quasi-hydrostatic region} is matter dominated, with the ionized
particles and magnetic field being carried around by the neutral
material. The ions are braked by a magnetic torque, which is transmitted
to the neutral gas through the frictional drag; therefore $v_{{\rm
i}\phi}<v_\phi$ in this region. The neutrals thus lose angular momentum
to the field, and their back reaction leads to a buildup of the
azimuthal field component away from the midplane.  The loss of angular
momentum enables the neutrals to drift toward the center, and in doing
so they exert a radial drag on the field lines. This drag must be
balanced by magnetic tension, so the field lines bend away from the
rotation axis.  This bending builds up the ratio $\lfrac{B_r}{B_z}$,
which needs to exceed $1/\sqrt{3}$ at the disk surface to launch a
centrifugally driven wind. The magnetic tension force, transmitted
through ion--neutral collisions, contributes to the radial support of
the neutral gas and causes it to rotate at sub-Keplerian speeds.
\item The growth of the radial and azimuthal field components on moving
away from the midplane results in a magnetic pressure gradient that
tends to compress the disk. The magnetic energy density comes to
dominate the thermal and gravitational energy densities as the gas
density decreases, marking the beginning of the {\em transition zone}
(at $z \approx h$, where $h$ is the density scale height). The
field above this point is nearly force free and locally straight (see
Sects.~\ref{subsec:forces} and~\ref{subsec:centrifugal}).
\item The field angular velocity $\Omega_B$ is a flux-surface constant
(see Eq.~(\ref{eq:wind8}) and footnote~\ref{fn:Omega_B}; note, however,
that this strictly holds only when $\epsilon_{\rm B}$ is
identically zero). The ion angular velocity $\lfrac{v_{{\rm i}\phi}}{r}$
differs somewhat from $\Omega_B$ but still changes only slightly along
the field. Since the field lines bend away from the symmetry axis, the
cylindrical radius $r$, and hence $v_{{\rm i}\phi}$, increase along any
given field line, whereas $v_\phi$ decreases because of the
near-Keplerian rotation law. Eventually a point is reached where
$(v_{{\rm i}\phi} - v_\phi)$ changes sign. At this point the magnetic
stresses on the neutral gas are small and its angular velocity is almost
exactly Keplerian.\footnote{This result was utilized in the derivation
of the slow-magnetosonic critical surface properties in
Sect.~\ref{subsubsec:critical}.} Above this point the field lines
overtake the neutrals and transfer angular momentum back to the matter,
and the ions start to push the neutrals out in both the radial and the
vertical directions. This region can be regarded as the {\em base of the
wind}, and one can accordingly identify the disk surface $z_{\rm s}$
with the location where $v_\phi$ becomes equal to $v_{\rm K}$. The mass
outflow rate is fixed by the density at the sonic/sms point (marked by
$z_{\rm sms}$ on the diagram).
\end{itemize}

\paragraph{The Hydrostatic Approximation}

The structure equations of the radially localized disk model can be
simplified by setting $v_z \approx 0$. Although this approximation is
most appropriate for the quasi-hydrostatic region, one can nevertheless
extend it to the disk surface to obtain useful algebraic constraints on
viable disk solutions. For the pure ambipolar regime (and again assuming
$\varrho_{\rm i} = {\rm const}$), they are given by
(Wardle \& K\"onigl 1993; K\"onigl 1997)
\begin{equation}
\label{eq:disk1}
(2\Upsilon_0)^{-1/2} \la a_0 \la \sqrt{3}  \la \epsilon\Upsilon_0
\la \lfrac{v_{\rm K}}{2C}\; ,
\end{equation}
where the parameter $\Upsilon_0$ represents the midplane Elsassser
number in the ambipolar limit (see Eq.~(\ref{eq:MHD25})). The four
inequalities in Eq.~(\ref{eq:disk1}) have the following physical meaning
(going from left to right):
\begin{enumerate}
\item The disk remains sub-Keplerian everywhere below its surface.

\item The wind launching condition ($b_{r,{\rm s}}>\lfrac{1}{\sqrt{3}}$)
is satisfied.

\item The top of the disk ($z_{\rm s}$) exceeds a density scale height
($h\approx (\lfrac{a_0}{\epsilon\Upsilon_0})\, h_{\rm T}$), ensuring
that the bulk of the disk material is nearly hydrostatic and that $\dot
M_{\rm w}$ remains $\ll \dot M_{\rm a}$.\footnote{These requirements also place
upper limits on the ratio of the density at the sonic/sms point to the
midplane density (see Eq.~(\ref{eq:wind21}) for the sms case).}

\item The midplane Joule heating rate (see Sect.~\ref{subsec:nonideal}) is less
than the rate of gravitational potential energy release there
($\varrho_0 |v_{r,0}| v_{\rm k}^2/2r$ per unit volume), as the latter
is the ultimate source of energy production in the disk.
\end{enumerate}

The inequalities (1) and (2) together place a lower bound on the
neutral--field coupling parameter $\Upsilon_0$, for which a more detailed
analysis yields the value of $\sim 1$ (see K\"onigl et al. 2010).  The
inequalities (2) and (3), in turn, imply that $\lfrac{h}{h_{\rm T}}< 1$,
i.e., that magnetic squeezing dominates tidal gravity. Analogous
constraints on wind-driving disks can be obtained also in the Hall and
Ohm diffusivity regimes. For example, in the Hall case the parameter
space divides into four sub-regimes with distinct sets of constraints
(K\"onigl et al. 2010; Salmeron et al. 2010). In all of these cases it
is inferred that $\Upsilon_0 > 1$ and $\lfrac{h}{h_{\rm T}}< 1$, which
indicates that these are generic properties of wind-driving disks. In
practice one finds that successful full solutions typically have $a_0
\la 1$ (the midplane magnetic pressure is smaller than the thermal
pressure, but not greatly so) and $\epsilon \la 1$ (the midplane inflow
speed is not much smaller than the speed of sound).

The relations listed in Eq.~(\ref{eq:disk1}) have a few other
interesting implications. For example, one can use the inequality (3)
to argue that wind-driving disks are stable to the fastest growing
linear mode of the MRI, because it implies that the vertical
wavelength of this mode, $\sim \lfrac{v_{\rm A 0}}{\Omega_{\rm K}}$
(see Chap.~V), is larger than the magnetically reduced disk scale
height $h = (\lfrac{a_0}{\epsilon\Upsilon_0})C/\Omega_{\rm K}$. But one
can also argue that the inequality (1) leads to a useful criterion for
the onset of the MRI in the disk. Such a criterion is of interest
since, as we already noted when the Elsasser number was introduced in
Sect.~\ref{subsec:nonideal}, the minimum-coupling condition for the
development of MRI turbulence in a diffusive disk (represented by a
lower bound on $\Lambda$) is evidently similar to that inferred for
driving a disk wind. The question then arises as to whether, in a disk
that is threaded by a large-scale, ordered field, both vertical
(wind-related) and radial (MRI-related) angular momentum transport can
occur at the same radial location; this can alternatively be phrased
as a question about the maximum magnetic field strength for the
operation of the MRI. When the inequality (1) is violated, the
surface layers become super-Keplerian, implying outward streaming
motion that is unphysical in the context of a pure wind-driving
disk. However, as elaborated in Salmeron et al. (2007), such motion
could be associated with the two-channel MRI mode that underlies
MRI-induced turbulence.  If one regards the parameter combination
$2\Upsilon a^2$ that figures in this inequality as a function of
height in the disk rather than being evaluated at the midplane, one
can infer from the fact that it scales as $B_z^2(\lfrac{\varrho_{\rm
i}}{\varrho})$ that it generally increases with $z$. It is thus
conceivable that the inequality (1) (generalized in this manner) is
violated near the midplane but is satisfied closer to the disk
surface. Salmeron et al. (2007) developed disk models in which both
radial turbulent transport and vertical transport associated with the
{\em mean} field take place in the region where the inequality (1) is
violated but only vertical transport (with ultimate deposition of the
removed angular momentum in a disk wind) occurs at greater
heights. They concluded, however, that significant radial overlap
between these two transport mechanisms is unlikely to occur in real
disks.

\paragraph{Weakly Coupled Disks}

In the disk models considered so far, the minimum-coupling condition on
the neutrals, $\Lambda\ga 1$, was satisfied throughout the vertical extent
of the disk. We refer to such disks as being {\em strongly coupled}. As
we have, however, just noted, the parameter values in a real disk could
vary with height, reflecting the vertical stratification of the column
density (which shields the ionizing radiation or cosmic rays) and the
density. In particular, if the disk is in the ambipolar regime near the
surface, in the Hall regime further down, and in the Ohm regime near the
midplane, then $\Lambda$ scales as $\varrho_{\rm i}$,
$\lfrac{\varrho_{\rm i}}{\varrho}$, and $\lfrac{\varrho_{\rm
i}}{\varrho^2}$, respectively, on going from $z=z_{\rm s}$ to $z=0$ (see
Eq.~(\ref{eq:MHD25})). The Elsasser number will thus increase with
height on moving up from $z=0$ as the gas becomes progressively more
ionized and the density decreases. It will generally peak on reaching
the ambipolar regime and will subsequently drop as $\varrho_{\rm i}$ (which
typically scales as $\varrho$ to a power between 0 and 0.5)
decreases. The gas will be weakly coupled in regions where $\Lambda$
remains $\ll 1$. Weakly coupled disks have distinct properties from those
of strongly coupled ones, as first pointed out by Li (1996a).

\begin{figure}
\centering
\includegraphics[height=7cm]{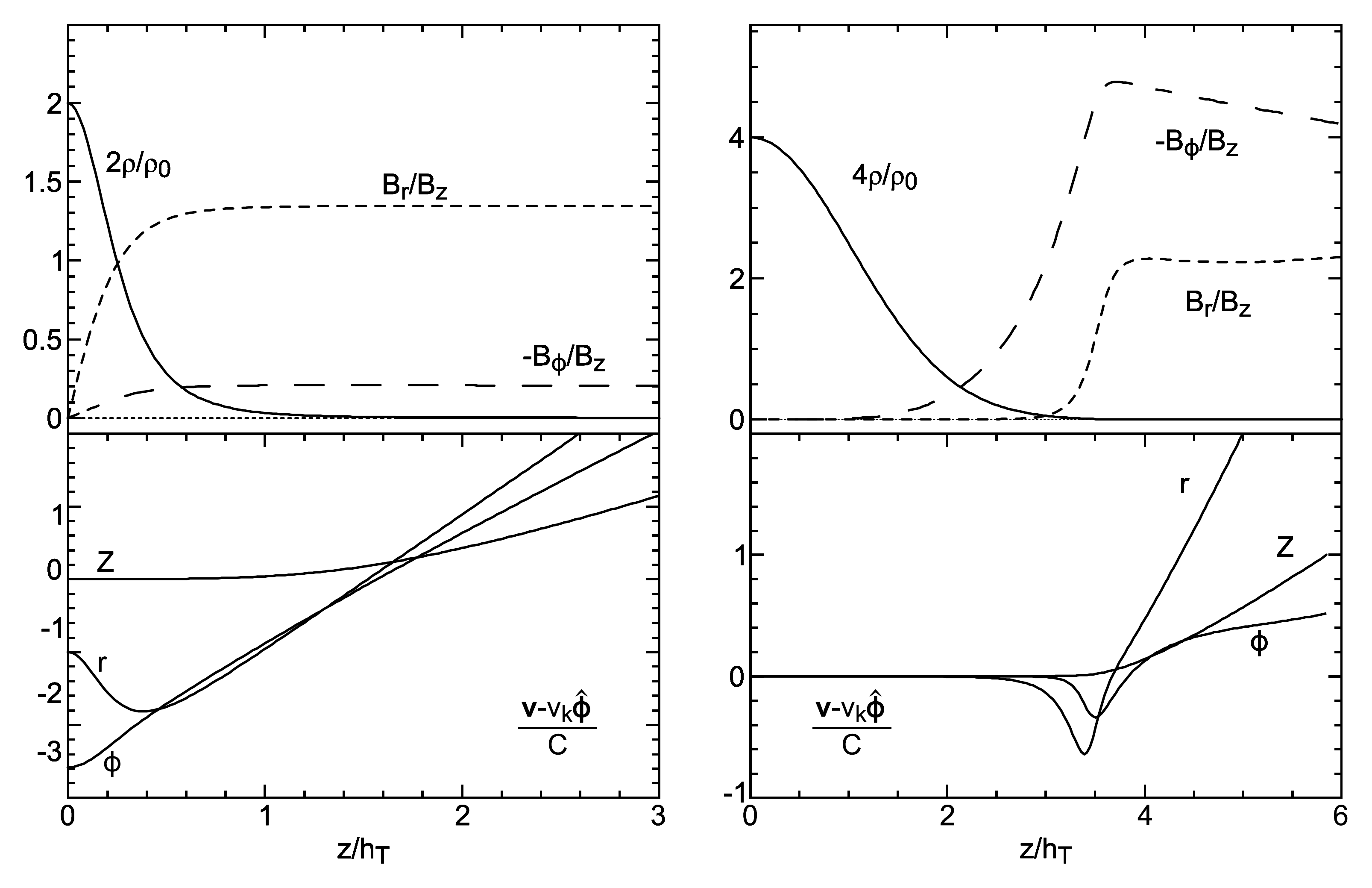}
\caption{Strongly coupled ({\bf left}) vs. \em weakly coupled ({\bf
right}) disk solutions.} \label{fig6_7}
\end{figure}

As an example, consider a protostellar disk that is ionized solely by
cosmic rays (with the ionization rate decreasing exponentially with
depth into the disk with a characteristic attenuation column of $96\,
{\rm g\, cm^{-2}}$) and in which the charge carriers are small, singly
charged grains of equal mass (a reasonable approximation for the
high-density inner regions of a real disk; e.g., Neufeld \& Hollenbach
1994). Near the disk surface the gas is in the ambipolar
regime, but if the disk half-column is $\gg 96\, {\rm g\, cm^{-2}}$, the
degree of ionization near the midplane is very low and the gas there is
in the Ohm regime.\footnote{Owing to the assumed equal mass of the
positive and negative charge carriers, the Hall term in Ohm's Law is
identically zero (see Eq.~(\ref{eq:MHD13})).} Figure~\ref{fig6_7}, taken
from Wardle (1997), depicts two illustrative solutions obtained for
different radii in this model. The solution on the left corresponds to a
large enough radius (and, correspondingly, a sufficiently low disk
column) for the disk to be in the ambipolar regime throughout its
vertical extent. By contrast, the solution on the right depicts a
smaller radius where the column density is large and the disk is weakly
ionized (and in the Ohm regime) near the midplane.

The main differences between these two solutions can be summarized as
follows:

\paragraph{Strongly coupled disks}
\begin{itemize}
\item $v_{{\rm A}0}\la C$ (midplane magnetic pressure comparable to
thermal pressure);
\item $|<v_r>| \sim C$ (mean radial speed comparable to the speed of sound);
\item  $B_{r,{\rm s}} > |B_{\phi,{\rm s}}|$ (with $B_r$
increasing already at $z=0$).
\end{itemize}

\paragraph{Weakly coupled disks}
\begin{itemize}
\item $v_{\rm A 0}\ll C$ (midplane magnetic field is highly subthermal);
\item $|<V_r>| \ll C$ (mean inflow speed is highly
subsonic);\footnote{These solutions can thus evade the short-lifetime
criticism made against strongly coupled disk models, namely, that in the
absence of a vigorous mass supply (which only happens during the early
evolutionary phases of protostars) they would empty out on a relatively
short time scale (e.g., Shu et al. 2008). Recall in this connection that
the disk formation models described in Sect.~\ref{subsec:s_s_soln}
yielded asymptotic ($r\rightarrow 0$) solutions that correspond to
weakly coupled disks. The short-lifetime conundrum could be avoided
altogether even in a strongly coupled disk if angular momentum transport
by a wind only dominated in the innermost disk regions, which is not
inconsistent with existing observational data.}
\item $B_{r,{\rm s}} < |B_{\phi,{\rm s}}|$ (with $B_r$ taking off only
when $\Lambda$ increases above $\sim 1$).\footnote{\label{fn:diffB}
Using the hydrostatic approximation one can derive a differential
equation relating $B_r$ and $B_\phi$, which, assuming $\epsilon_B=0$
and a vanishing Hall term, takes the form $\frac{\D{B_r}}{\D{B_\phi}}
\approx - 2 \Lambda$ at the midplane.}
\end{itemize}

There are two noteworthy features of weakly coupled disk
solutions. First, even though the bulk of the disk volume is nearly
inert, the disk possesses ``active'' surface layers where $\Lambda > 1$,
from which a disk wind can be launched in the presence of a large-scale,
ordered field (or in which MRI-induced turbulence can operate; see
Gammie 1996). Second, in magnetically threaded disks angular momentum is
transported vertically even in regions where $\Lambda$ is still $< 1$
and $B_r \approx 0$. This is because a measurable azimuthal field
component can already exist in these regions ($|\lfrac{B_\phi}{B_r}|$
can be $\gg 1$ when $\Lambda \ll 1$; see footnote~\ref{fn:diffB}) and we
recall that the $z\phi$ stress exerted by the field is $\propto B_z
\lfrac{\D{B_\phi}}{\D{z}}$. As a result, matter can continue to accrete
even in the nominally inert disk regions, which would not be
possible if only a small-scale, disordered field (i.e., MHD turbulence)
were responsible for angular momentum transport (although in practice
turbulent mixing of charges and field might enable accretion in these
regions also in that case; e.g., Fleming \& Stone 2003; Turner et
al. 2007). This could have implications to the ongoing debate about the
nature of ``dead zones'' in protostellar disks (e.g., Pudritz et
al. 2007; see Chap.~V).

\subsection{Stability Considerations}
\label{subsec:stability}

It was noted in Sect.~\ref{subsubsec:exact_disk} that wind-driving disks
should be stable to the most rapidly growing linear mode of the MRI.  On
the other hand, by combining the wind-launching
condition~(\ref{eq:MHD6}) with the condition for the onset of a radial
{\em interchange instability} (e.g., Spruit et al. 1995), one can show
(see K\"onigl \& Wardle 1996) that, to be unstable to radial
interchange, the magnetic term must be comparable to the gravitational
term in the radial momentum equation. This is clearly not the case in
the derived disk solutions, which are characterized by an inherently
subthermal magnetic field (see, e.g., the inequality (2) in
Eq.~(\ref{eq:disk1})). It is thus seen that, when a large-scale magnetic
field is responsibe for the entire angular momentum removal from an
accretion disk through a CDW, it automatically lies in a ``stability
window'' in which it is strong enough not to be affected by the MRI but
not so strong as to be subject to radial interchange.

\begin{figure}
\centering
\includegraphics[height=8cm]{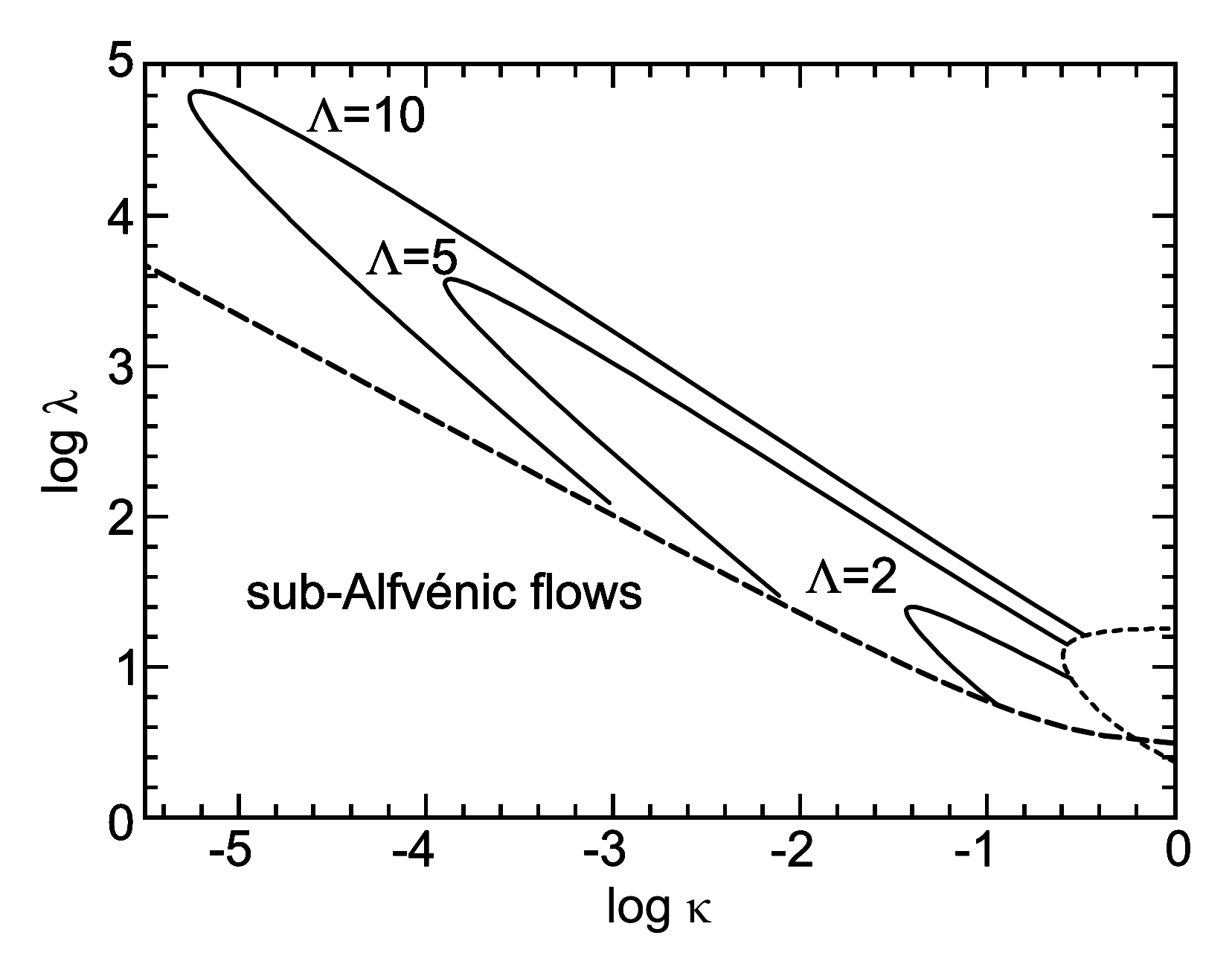}
\caption{Mapping of the wind-driving disk solutions (for several
values of the neutral--field coupling parameter $\Lambda$) onto the
self-similar wind solution space (defined by the values of the
mass-loading parameter $\kappa$ and angular-momentum parameter
$\lambda$). The field-strength parameter $a_0$ increases on moving
counterclockwise along a given curve.} \label{fig6_8}
\end{figure}

There are, of course, other potential instabilities to which such disks
might be susceptible (see K\"onigl \& Wardle 1996 for some examples),
but here we focus on the question of whether there might be an inherent
aspect of the wind-related angular momentum transport that could render
the disk/wind system unstable.\footnote{A distinct, but still relevant,
question is whether the wind by itself may be unstable. Numerical
simulations that treated the disk as providing fixed boundary conditions
for the outflow have indicated that the wind should not be disrupted,
even in the presence of nonaxisymmetric perturbations (e.g., Anderson et
al. 2006; Pudritz et al. 2007).  However, these simulations did not
account for the existence of a feedback between the wind and the disk.}
Based on approximate equilibrium models, Lubow et al. (1994b) and Cao \&
Spruit (2002) suggested that an inherent instability of this sort may,
in fact, exist (see also Campbell 2009). They attributed this
instability to the sensitivity of the outflowing mass flux to changes in
the field-line inclination at the disk surface ($\theta_{\rm s}$),
according to the following feedback loop:
\begin{eqnarray}
|v_r| \ {\rm increases} \ &\Rightarrow& \
\tan{\theta_{\rm s}} \ {\rm increases}\cr
&\Rightarrow& \ {\rm the\ wind\ is \ loaded\ by\
higher\hspace{-0.02in}-\hspace{-0.02in}density\ gas}\cr
&\Rightarrow& \ \dot M_{\rm w} \ ({\rm and \ the \ removed \ angular \
momentum} \ \dot M_{\rm w}\,l) \ {\rm increase}\cr
&\Rightarrow& \ |v_r| \ {\rm increases \ even\ more....}\nonumber
\end{eqnarray}
The issue was reexamined by K\"onigl (2004), who used the disk/wind
model of Wardle \& K\"onigl (1993) and appealed to the fact that the
stability properties generally change at a {\em turning point} of the
equilibrium curve in the solution parameter space. Figure~\ref{fig6_8}
depicts such equilibrium curves, labeled by their vertically constant
Elsasser-number values $\Lambda=\Upsilon$, in the Blandford \& Payne
(1982) $\kappa-\lambda$ wind parameter space. The lower branches of
these curves end on the long dashed curve, below which the outflows
remain sub-Alfv\'enic, whereas the upper branches end on the
short-dashed curve, to the right of which the surface layers of the disk
are super-Keplerian. Any particular solution is determined by the value
of the disk parameter $a_0=(2/ \gamma \beta_0)^{1/2}$, which increases
along each $\Lambda = {\rm const}$ curve from its minimum value on the
``super-Keplerian'' boundary (see the inequality (1) in
Eq.~(\ref{eq:disk1})). It is seen that each curve exhibits a turning
point.\footnote{This feature of the solution curves is also found in the
corresponding plots in the $a_0-\lambda$ plane, where now $\kappa$
varies along each $\Lambda={\rm const}$ curve.} This behavior can be
understood as follows. The vertical hydrostatic equilibrium equation
implies $b_{r,{\rm s}}^2 \approx 2/a_0^2$ (assuming $B_{r,{\rm s}}^2 \gg
B_{\phi,{\rm s}}^2$), and since $|B_\phi|$ evolves with $B_r$ (see
footnote~\ref{fn:diffB}), it follows that $|B_{\phi,{\rm s}}|/B_z$, and
hence the magnitude of the angular momentum that the outflow must carry
away, increases with decreasing $a_0$. Initially, as $a_0$ decreases from
$\sim 1$, $b_{r,{\rm s}}=B_{r,{\rm s}}/B_z$ increases rapidly, and the
corresponding increase in the cylindrical radius of the Alfv\'en surface
(the effective lever arm for the back torque exerted by the outflow on
the disk, which scales as $\lambda^{1/2}$) increases the value of
$\lambda$ and leads to a reduction in the ratio of the mass outflow to
the mass inflow rates (estimated as $\dot M_{\rm w}/\dot M_{\rm a}
\approx 1/[4(\lambda-1)]$). However, as $a_0$ continues to decrease, the
rate of increase of $b_{r,{\rm s}}$ declines while that of $|B_{\phi,{\rm
s}}|/B_z$ increases, and eventually the mass outflow rate must start to
increase (with $\lambda$ going down) to keep up with the angular
momentum removal requirements. The transition between these two modes of
enhanced angular momentum transport: predominantly by the lengthening of
the lever arm (on the lower branch) vs. mainly by a higher mass-loss
rate (on the upper branch) occurs at the turning point of the solution
curve.
One can determine which of the two solution branches is stable and which
is not by considering the competition between radial advection and
diffusion on the magnetic flux evolution (best done in the $a_0-\lambda$
plane), in analogy with thermal stability arguments that analyze the
relative magnitude of heating and cooling on each side of the thermal
equilibrium curve. In this way one finds that the upper branch of the
curves in Fig.~\ref{fig6_8} is unstable: this is the branch along which
an increase in the angular momentum transport is accomplished through a
higher mass outflow rate, precisely the behavior invoked in the
heuristic instability argument reproduced above. Since, however, there
is an alternative way of increasing the angular momentum transport
(namely, through a lengthening of the effective lever arm), there is
also a stable branch of the solution curve. Physically, the reason why a
perturbation that reduces the field inclination to the disk does not
necessarily trigger an instability is that an increase in $b_{r,{\rm
s}}$ also results in greater field-line tension, which tends to oppose
the inward poloidal field bending. Whether a given solution branch is
stable or not is determined by the extent to which this stabilizing
effect can overcome the destabilizing influence of increased angular
momentum removal brought about by the field-line bending.

It is possible to argue (see K\"onigl 2004) that real protostellar
systems likely correspond to the stable branches of the solution
curves. Global numerical simulations could, however, shed more light
on this question, as they have already started to do. For example,
Casse \& Keppens (2002) solved the nonideal MHD equations (assuming
axisymmetry and a polytropic gas) and demonstrated jet launching by
resistive disks. These results are significant on two main counts:
first, they corroborate the basic picture of a magnetically threaded
diffusive disk centrifugally driving a wind that removes the bulk of
its angular momentum, and, second, they indicate that these
configurations are stable inasmuch as they reach a quasi-steady
state. Subsequent simulations have extended this work while confirming
its main conclusions (e.g., Kuwabara et al. 2005; Meliani et al. 2006;
Zanni et al. 2007). Three-dimensional simulations are the natural next
step.

\section{Disk--Star Magnetic Coupling}
\label{sec:disk-star}

We now turn to the best-established example of a large-scale field in
protostars, namely, the stellar dynamo-generated field, and consider the
dynamical interaction between this field and the surrounding accretion
disk. This is a rich topic that has implications to the structure of the
innermost disk region, the stellar rotation and the way mass reaches the
stellar surface, protostellar jets, and the observational signatures of
the region where most of the gravitational potential energy is
liberated. Owing to the nontrivial topology of the stellar (and possibly
also disk-generated) magnetic field and to the complexity of the
field-mediated interaction between the star and the disk even for simple
flux distributions, our understanding is far from complete, with new
progress being driven largely by improved numerical simulations. We
start this section by briefly surveying the relevant observations and
then describe the basic physical processes that underlie this
interaction and the results of some of the recent numerical work
Relevant additional details are provided in Chap.~I [Protoplanetary Disk
Structure and Evolution].

\subsection{Phenomenology}
\label{subsec:interact}

We discuss the observational manifestations of two phenomena in which
a large-scale, ordered magnetic field has been implicated: the
relationship between protostellar rotation and accretion disks, and
magnetic disk truncation and resultant field-channeled accretion. The
theoretical framework that relates these two phenomena is outlined in
the next subsection.

\paragraph{Stellar Rotation and Accretion}

It is now well established (based on $\sim 1700$ measured rotation
periods) that about 50\% of sun-like protostars ($M_* \sim 0.4-1.2\,
M_\odot$) undergo significant rotational braking during their
pre--main-sequence (PMS) contraction to the Zero-Age Main Sequence
(ZAMS), and that the objects whose specific angular momenta decrease with
time are essentially the ones that are already slowly rotating at an
early ($\sim 1\, {\rm My}$) age (e.g., Herbst et al. 2007). A clue
to the braking mechanism is provided by the finding of a clear
correlation between having a comparatively long period ($\ga 2\, {\rm
d}$) and exhibiting an accretion-disk signature (near-IR or mid-IR
excess). This trend is particularly noticeable in higher-mass ($\ga
0.25\, M_\odot$) protostars, but it is also found in very-low-mass stars
and brown dwarfs (in which, however, the braking efficiency is evidently
lower). The connection with disks is supported by the fact that the
inferred maximum stellar braking times ($\sim 5-10\, {\rm Myr}$) are
comparable to the maximum apparent lifetimes of gaseous accretion disks
(which are $\sim 1\, {\rm Myr}$ for $\sim 50\%$ of the stars and $\sim
5\ {\rm Myr}$ for almost all stars). These results imply that the
dominant braking mechanism in protostars is directly tied to active disk
accretion. In the absence of such accretion (or after the disks
disperse), PMS stars nearly conserve specific angular momentum and
therefore spin up as they contract to the ZAMS. As we discuss in the
next subsection, stellar magentic fields likely play a key role in the
braking process.

\paragraph{Disk Truncation and Field-Channeled Accretion}

As was originally inferred in the case of accreting magnetized neutron
stars and white dwarfs, a sufficiently strong protostellar magnetic
field can be expected to {\em truncate the accretion flow}, with the
truncation radius increasing with $B_*$ and decreasing with $\dot M_{\rm
a}$ (see Eqs.~(\ref{eq:couple1}) and~(\ref{eq:couple2}) below). In this
picture, the intercepted matter ``climbs'' onto the field lines and is
{\em magnetically channeled} to some finite stellar latitude. By the
time the matter reaches the stellar surface it is streaming along the
field lines with a near free-fall speed and is therefore stopped in an
{\em accretion shock}. The basic elements of this scenario are supported
by observations of CTTSs (but also of lower-mass brown dwarfs and of
higher-mass Herbig Ae/Be stars, although so far these have been less
well-studied). The main findings have been (see Bouvier et al. 2007):

\begin{itemize}

\item The common occurrence of inverse P Cygni profiles,
with redshifted absorption reaching several hundred ${\rm km\; s^{-1}}$.

\item Observed hydrogen and Na I line profiles that can be adequately
modeled in this picture; the detection in the UV--near-IR spectral range
of predicted statistical correlations between the line fluxes and the
inferred mass accretion rate.

\item Observed spectral energy distributions of optical and UV excesses
that are successfully reproduced by accretion shock models.

\item Periodic visual-flux variations due to ``hot spots'' (interpreted
as nonaxisymmetric accretion shocks on the stellar surface) that are
only observed in actively accreting CTTSs (but {\em not} in the
weak-line T Tauri stars (WTTSs), which lack the signatures of vigorous
accretion). However, in contrast to ``cool spots'' (the analogs of
sunspots), which last for $\sim 10^2-10^3$ rotations, the hot-spot
periodicity only persists for a few rotation periods, indicating a
highly nonsteady configuration. This inference is supported by the
detection of high-amplitude irregular flux variations.

\item Strong near-IR ``veiling'' variability that can be interpreted in
the context of this picture as arising in the interaction region between
the disk and an inclined stellar magnetosphere, with possible
contributions also from the accretion shock and from the
shock-irradiated zone in the inner disk.
\end{itemize}

Zeeman-broadening measurements in a growing number of CTTSs have yielded
an intensity-averaged mean surface magnetic field strength of $\sim
2.5\, {\rm kG}$,\footnote{A similar value has now been inferred also in
a protostar that is at an earlier (Class I) evolutionary phase
(Johns-Krull et al. 2009).} with the field inferred to reach $\sim 4-6\,
{\rm kG}$ in some regions. Furthermore, circular polarization
measurements in lines associated predominantly with the accretion shock
have revealed rotational modulation and demonstrated that the field is
highly organized (with peak value $\sim 2.5\, {\rm kG}$) in the shock
region (covering $<5\%$ of the stellar surface). These results are
consistent with the model predictions. No net polarization has, however,
been found in photospheric absorption lines, implying a complicated
surface field topology (with a global dipole component $\la 0.1\, {\rm
kG}$). A likely physical picture is that there exists a spectrum of
magnetic flux loops of various lengths that extend from the stellar
surface, with a fraction reaching to $\sim 5-20\, R_*$ and intercepting
the disk (see Feigelson et al. 2007). This picture is supported by
analyses of X-ray flares from young stars (e.g., Favata et al. 2005) as
well as by direct radio imaging of the large-scale magnetic structures
(e.g., Loinard et al. 2005) and by spectropolarimetric reconstructions
of the magentospheric topology (e.g., Donati et al. 2007).

Numerical studies of magnetospheric accretion indicate that disk
mass-loaded {\em outflows} (which are often predicted to be nonsteady)
could be produced in the course of the star--disk interaction (see
Sect.~\ref{subsec:coupling_models}). There is now observational evidence
for accretion-induced winds in CTTSs emanating from both the inner disk
and the star (e.g., Edwards et al. 2006). The stellar
component has been inferred to move on radial trajectories and to
undergo full acceleration (up to a few hundred ${\rm km\; s^{-1}}$) in
the stellar vicinity. Although this component may well be launched along
stellar magnetic field lines, the absence of evidence for it in WTTSs
suggests that it, too, is triggered by the interaction with the disk.

\subsection{Disk--Star Coupling Models}
\label{subsec:coupling_models}

\paragraph{Basic Concepts}

Spherical accretion with $\dot M_{\rm a}= 4\pi R^2 \varrho(R) |V_R(R)|$
at the free-fall speed $v_{\rm ff}(R) = (2GM_*/R)^{1/2}$ onto a star of
mass $M_*$ will be stopped by the magnetic stresses of the stellar
magnetic field at a distance where the ram pressure of the flow
($\varrho v_{\rm ff}^2$) becomes comparable to the magnetic pressure of
the stellar field ($B^2/8\pi$). For simplicity, we assume that the field
at that location can be approximated by its dipolar component
(corresponding to a dipole moment $\mu_*$). For accretion in the
equatorial plane, $B(r)=\mu_*/r^3 = B_*R_*^3/r^3$, and one obtains the
nominal {\em accretion Alfv\'en radius}
\begin{equation}
\label{eq:couple1}
r_{\rm A} = \frac{\mu_*^{4/7}}{(2 G M_*)^{1/7}\dot M_{\rm a}^{2/7}}\; .
\end{equation}

Although the radial ram pressure of the flow is relatively small for
disk accretion, one can still define in this case a magnetospheric boundary
radius $r_{\rm m}$ from the requirement that the total material and
magnetic stresses (or energy densities) be comparable, i.e., $B^2/8\pi =
\varrho v_\phi^2 + P \approx \varrho v_\phi^2$. By approximating
$v_\phi \approx v_{\rm K}$, one obtains an expression similar to the
one given by Eq.~(\ref{eq:couple1}). This radius determines
the region within which the stellar magnetic field controls the flow
dynamics and provides a lower bound on the inner disk radius. To estimate
the radius where matter leaves the disk, note that the physical
process of disk truncation requires that the torque exerted by the
stellar field that penetrates the disk be comparable to the rate at which
angular momentum is transported inward by the accretion flow, $- r^2 B_z
B_{\phi,{\rm s}} = \dot M_{\rm a}
\lfrac{\D{(r^2\Omega)}}{\D{r}}$. Taking $|B_{\phi,{\rm s}}| \approx B_z$
(where we use the approximate maximum value of the azimuthal field
component at the disk surface; see Uzdensky et al. 2002a
and cf. Eq.~({\ref{eq:form13})) and setting
$|\lfrac{\D{\Omega}}{\D{r}}| \sim \Omega_{\rm K}/r$ yields an estimate
of the {\em disk truncation radius} $r_{\rm d}$ that is of the order of
$r_{\rm A}$,
\begin{equation}
\label{eq:couple2}
r_{\rm d} = k_{\rm A}\, r_{\rm A}\; .
\end{equation}
Semianalytic models (e.g., Ghosh \& Lamb 1979) and numerical simulations
(e.g., Long et al. 2005) yield $k_{\rm A} \approx 0.5$.

If the star is rotating with angular velocity $\Omega_*$, one can define
the {\em corotation radius} by setting $\Omega_{\rm K}(r) = \Omega_*$,
which gives
\begin{equation}
\label{eq:couple3}
r_{\rm co} = \left (\frac{G M_*}{\Omega_*^2} \right )^{1/3}\; .
\end{equation}
The interaction of the disk with the stellar magnetic field naturally
divides into two qualitatively different regimes depending on the ratio
$\lfrac{r_{\rm d}}{r_{\rm co}}$\/:
\begin{eqnarray}
r_{\rm d} \la r_{\rm co} \quad\quad && \quad\quad {\rm
funnel\hspace{-0.02in}-\hspace{-0.02in}flow\
regime}\cr
r_{\rm d} \ga r_{\rm co}\quad\quad && \quad\quad {\rm propeller\
regime}\, .
\nonumber
\end{eqnarray}
In the funnel-flow regime, the disk angular velocity is higher than that
of the star, and matter can climb onto the stellar magnetic field lines
that thread the disk and reach the stellar surface. By contrast, in the
propeller regime the stellar angular velocity exceeds $\Omega_{\rm
K}(r_{\rm d})$: when disk matter becomes attached to the stellar field
lines its angular momentum increases above the rotational equilibrium
value for that radius and it moves outward. Numerical simulations have
found that, if the disk effective viscosity and magnetic diffusivity are
relatively high, most of the incoming matter is expelled from the
system, both in the form of a wide-angle CDW from the inner regions of
the disk and as a strong, collimated, magnetically dominated outflow
along the open stellar field lines near the axis (e.g., Romanova et
al. 2005; Ustyugova et al. 2006).\footnote{A similar outflow pattern,
comprising a conical disk wind and a higher-velocity, low-density jet
component, was found also in simulations of the funnel-flow regime
(e.g., Romanova et al. 2009).} This mechanism could be relevant to the
initial spindown (on a time scale $< 10^6\, {\rm yr}$) of CTTSs.

\paragraph{Funnel-Flow Regime}

Ghosh \& Lamb (1979) proposed that a ``disk locked'' state in which
the torque exerted by the field lines that thread the disk (as well as
by the material that reaches the star) could keep the star rotating in
equilibrium (neither spinning up nor down). In this picture, the field
lines that connect to the disk within the corotation radius (as well
as the accreted material) tend to spin the star up, whereas the field
lines that connect at $r>r_{\rm co}$ have the opposite effect. This
scenario could potentially explain the relatively low rotation rates
observed in CTTSs (e.g., K\"onigl 1991).\footnote{Naively, one would
expect a star accreting from a rotationally supported disk to rotate
near breakup speed; in reality, CTTSs rotate, on average, at about a
tenth of this speed.}

This picture was, however, challenged on the grounds that the twisting
of the magnetic field lines that thread the disk at $r > r_{\rm co}$
will tend to open them up, thereby reducing the spin-down torque on the
star by more than an order of magnitude compared to the original
calculation (Matt \& Pudritz 2004, 2005). This argument is based on the
fact that stellar field lines that connect to the disk at $r\ne r_{\rm
co}$ are twisted by the differential rotation between their respective
footpoints. Initially, the twisting leads to an increase in $|B_{\phi,{\rm
s}}|$, but then the built-up magnetic stress causes the field lines
to rapidly elongate in a direction making an angle $\sim 60^\circ$ to
the rotation axis (e.g., Lynden-Bell \& Boily 1994). During this phase
$|B_{\phi,{\rm s}}|$ decreases as the field-line twist travels out to the
apex of the elongating field line, where the field is weakest, a process
that can be understood in terms of torque balance along the field line
(e.g., Parker 1979). As the twist angle approaches a certain critical
value ($\sim 4\, {\rm rad}$ for a star linked to a Keplerian disk), the
expansion accelerates and the magnetic field formally reaches a singular
state (a ``finite time singularity''), which in practice means that it
opens up. While the twisting can be countered by magnetic diffusivity in
the disk, this is only possible if the steady-state surface azimuthal
field amplitude $(\lfrac{2\pi r {\cal{S}} \varDelta{\Omega}}{c^2})
B_{z,{\rm s}}$, where ${\cal{S}}$ is the vertically integrated electrical
conductivity (treated as a scalar) and $\varDelta{\Omega}$ is the
differential rotation rate, does not exceed the maximum value of
$|B_{\phi,{\rm s}}|$ in the absence of diffusivity (e.g., Uzdensky et
al. 2002a). The conductivity in the innermost regions of protostellar
disks is probably too high for a steady state to be feasible except in
the immediate vicinity of $r_{\rm co}$ (see also Zweibel et al. 2006),
and in some proposed models (e.g., the X-wind; see Shu et al. 2000) it
has, in fact, been postulated that the stellar field lines can
effectively couple to the disk only in that narrow region.

Numerical simulations have, however, verified that, when the star has a
strong enough field to disrupt the disk at a distance of a few stellar
radii and channel accreting matter along field lines, it can maintain an
equilibrium disk-locking state in which $\Omega_*$ is close to the disk
angular velocity at the truncation radius (Long et al. 2005).
In these simulations it was found that, in equilibrium,
$\lfrac{r_{\rm co}}{r_{\rm d}} \sim 1.2-1.5$, very close to the
prediction of the Ghosh \& Lamb (1979) model. Furthermore, as envisioned
in the latter model, closed magnetic field lines that link the star and
the disk exert the dominant stresses in this interaction. However, in
contrast to the original picture, this linkage does in fact occur
primarily near $r_{\rm co}$. The torque balance is achieved in part
through field-line stretching (which mimics the connection to material
at $r> r_{\rm co}$ in the original model) and by magnetically driven
outflows (which, however, are found to remain comparatively weak in the
simulations). Field-line opening is not a major impediment to this
process, in part because opened field lines tend to reconnect
(especially if the departure from axisymmetry is not large).\footnote{
The idea that twisted field lines that open up could subsequently
reconnect, leading to a repetitive cycle of inflation and reconnection
and resulting in the star--disk linkage being steady only in a
time-averaged sense, was first proposed by Aly \& Kuijpers (1990)
and received support from subsequent investigations (e.g.,
Uzdensky et al. 2002b; Romanova et al. 2002).}

The question of whether (or under what circumstances) the magnetic
transfer of angular momentum to the disk is the most efficient way of
attaining protostellar spin equilibrium is, however, still unresolved.
For example, Matt \& Pudritz (2008) suggested that stellar winds driven
along open magnetic field lines could dominate the braking torque on the
star. These winds are inferred to be powered by the accretion process,
but it remains to be determined how this could happen in practice and
whether the proposed outflows are, in fact, related to the stellar winds
already identified observationally. It is also important to bear in mind
that the star--disk magnetic linkage mechanism may, in reality, be more
complex than the simplified picture outlined above. Some of these
expected complications are already being investigated with the help of
3D MHD codes, including the effects of misalignment between the rotation
and magnetic axes, of an off-centered dipole, and of higher-order
multipole moments (e.g., Romanova et al. 2003, 2004, 2008; Long et al. 2007,
2008; Kulkarni \& Romanova 2009).\footnote{Given the importance of the
question of how matter crosses magnetic field lines in this problem, it
would be helpful to examine these effects using an
explicitly resistive numerical code. So far, however, only the
axisymmetric version of this scenario has been investigated in this way
(e.g., Bessolaz et al. 2008).} Furthermore, the field topology and
the nature of the interaction could be modified if the disk itself
contains a large-scale magnetic field (e.g., Hirose et al. 1997; Miller
\& Stone 2000; Ferreira et al. 2000, 2006; von Rekowski \& Brandenburg
2004, 2006).

\paragraph{Nonsteady accretion}

The magnetic interaction between stars and disks could be variable on a
time scale as short as $\Omega_*^{-1}$. One possibility, which could
naturally give rise to observable ``hot spots,'' is for the system to
lack axisymmetry --- either because of a misalignment between the
magnetic (e.g., dipole) and rotation axes (e.g., Romanova et al. 2004,
2008; Kulkarni \& Romanova 2009)
or because of the intrinsic structure of the magnetic field (e.g.,von
Rekowski \& Brandenburg 2004, 2006). A longer variablity time scale is
implied by the suggestion of Goodson \& Winglee (1999) that the
truncation radius would oscillate on the diffusion time scale of the
field into the disk (resulting in field-line reconnection events and
episodic polar ejections) if the diffusivity at the inner edge of the
disk were relatively low (so $|(v_{B,r}/v_r)-1|< 1$). A different type
of variability is implied by the star/disk dynamo model of von Rekowski
\& Brandenburg (2006). In this model, the magnetic field geometry
changes at irregular time intervals (with magnetic polarity reversals
occurring mostly on time scales of less than a day), and the star/disk
system alternates between magnetically connected and disconnected
states. The model predicts strong outflows, both from the inner disk and
from the stellar surface, but typically only a small fraction of the
disk accretion flow reaches the stellar surface. In fact, there is an
anti-correlation between the stellar magnetic field strength and the
accretion rate, and material that reaches the stellar surface comes in
at a low velocity (von Rekowski \& Piskunov 2006). It remains to be
determined whether these two aspects of the model are consistent with
the polarization data and the evidence for accretion shocks (see
Sect.~\ref{subsec:interact}).

\section{Conclusion}
\label{sec:conclude}

The discussion in this chapter can be summarized as follows:

\begin{itemize}

\item There is strong observational evidence for a disk--wind connection
in protostars. Large-scale, ordered magnetic fields have been implicated
theoretically as the most likely driving mechanism of the observed winds
and jets. The ubiquity of the outflows may be related to the fact that
centrifugally driven winds (CDWs) are a potentially efficient means of
transporting angular momentum from the disk.

\item Ordered magnetic fields could arise in protostellar disks on
account of ({\it i}) advection of interstellar field by the accretion
flow, ({\it ii}) dynamo action in the disk, and ({\it iii}) interaction
with the stellar magnetic field.

\item Semianalytic MHD models have been able to account for the basic
structure of diffusive disks that drive CDWs from their surfaces as well
as for the formation of such systems in the collapse of rotating
molecular cloud cores. Some of these models already incorporate a
realistic disk ionization and conductivity structure. These studies have
established that vertical angular momentum transport by a CDW or through
torsional Alfv\'en waves (magnetic braking) could in principle be the
main angular momentum removal mechanism in protostellar disks and
determined the parameter regime where wind transport can be expected to
dominate radial transport by MRI-induced turbulence. Further progress is
being made by increasingly more elaborate numerical simulations
(involving nonideal MHD codes) that have started to examine the global
properties, time evolution, and dynamical stability of the magnetic
disk/wind system.

\item Robust observational evidence also exists for a magnetic
interaction between CTTS disks and their respective protostars,
including strong indications of a field-channeled flow onto the stellar
surface. This interaction is likely to involve mass ejection and is
thought to be responsible for the comparatively low rotation rates of
CTTSs.  Since the magnetic field geometry in the interaction region is
evidently quite complex and the interaction is likely time dependent,
numerical simulations are an indispensable tool in the study of this
problem.

\end{itemize}

Future advances in this area will probably arise from a combination of
new observational findings, the refinement of current theoretical
approaches, and the incorporation of additional physics into the
models. On the observational side, the main challenge is still to
demonstrate the existence of CDWs in protostars and to determine their
spatial extent (spread out over most of the disk surface or occurring
only near its inner edge, and, if the former, is the launching region
nearly continuous or is it confined to localized patches?). Recent
attempts to measure rotation in the outflows could potentially help to
answer this question, but, as noted in Sect.~\ref{subsec:centrifugal},
the results obtained so far are still inconclusive.

Regarding the further development of theoretical tools, the greatest
impact would likely be produced by numerical simulations that study
vertical angular momentum transport by either a CDW or magnetic
braking with codes that include a realistic conductivity tensor and
that have full 3D and mesh-refinement capabilities. Such simulations
should be able to clarify the relative roles of vertical and radial
angular momentum transport and the possible interplay between them for
relevant combinations of the disk model parameters (see
Sect.~\ref{subsubsec:exact_disk}). An interim step might be to solve
for the evolution of a vertically integrated disk whose properties, at
any radial grid zone, are determined from a vertical integration of a
simplified version of the radially localized, steady-state disk model
described in Sect.~\ref{subsec:disk}. Time-dependent models of this
type could examine the behavior of a magnetically threaded disk after
its mass supply diminishes or stops altogether (corresponding to the
protostellar system evolving into the optically revealed phase), which
has previously been studied only in the context of
``$\alpha$-viscosity'' models. As was already noted above, a
state-of-the-art, 3D, nonideal-MHD code is also crucial for
investigating the stability of disk/wind systems and the various
aspects of the star--disk field-mediated interaction. One could,
however, also benefit from further development of the semianalytic
models, which could include an extension of the
ionization/conductivity scheme, a derivation of self-similar disk/wind
solutions that allow for a radial drift of the poloidal magnetic field
(see Sect.~\ref{subsubsec:exact_disk}), and a calculation of the
predicted observational characteristics of wind-driving disks.

The mass fraction and size distribution of dust grains in the disk have
a strong effect on its ionization and conductivity structure and on the
degree of field--matter coupling (see
Sect.~\ref{subsubsec:ionize}). Existing models incorporate the effect of
dust in a somewhat ad-hoc manner, by adopting an assumed
distribution. In reality, the grain distribution is determined by the
balance of several processes, including grain collisions due to relative
velocities that develop as a result of Brownian motion, differential
vertical-settling speeds, and turbulence, which can lead to either
coagulation or fragmentation. Grains are also subject to a collisional
drag force exerted by the gas and arising from the fact that the gas is
subject to thermal and magnetic forces that do not affect the dust. This
leads to vertical settling as well as to radial migration, directed
either inward or outward depending on whether the gas rotation is sub-
or super-Keplerian, respectively. Furthermore, radial or vertical gas
motions can affect sufficiently small grains through advection. Yet
another effect is evaporation by the ambient radiation field, which
could impact dust located at sufficiently high elevations and small
radii. Some of these effects have already been incorporated into generic
viscous disk models (e.g., Dominik et al. 2007; Brauer et al. 2008), and
one could similarly consider them in the context of a wind-driving disk
model. In view of the fact that the latter model is characterized by a
vertical outflow and by comparatively fast radial inflow speeds, one can
expect to find new types of behavior in this case. In particular, grains
located near the disk surfaces would either settle to the midplane if
they are large enough or be uplifted from the disk if they are
sufficiently small, whereas intermediate-size grains would first leave
the disk and then re-enter at a potentially much larger radius, from
which they could be advected back inward. By including dust dynamics,
one could examine whether the effect of dust on the gas motion (through
its influence on the field--matter coupling) and the effect of gas on
the grain motions (through gas--dust collisions) together place
meaningful constraints on the resulting grain distribution. One could
also investigate whether the predicted radial transport of
intermediate-size grains from small to large radii and their possible
thermal processing outside the disk could be relevant to the
accumulating evidence for an outward transport of crystalline grains in
the protosolar nebula and in other protostellar disks, and whether the
implied dust distribution in the disk and the wind might have distinct
observational signatures that could be tested by spectral and imaging
techniques (see Millan-Gabet et al. 2007,
Sect.~\ref{subsubsec:wind_observe}, and Chaps.~I and~VIII).

Dust particles are thought to be the building blocks of planetesimals
and their distribution in the disk is thus a key ingredient of planet
formation models. In fact, the general properties of a protostellar
disk are evidently relevant to planet formation in light of the
growing evidence that the latter is strongly influenced by physical
processes that occur when the disk is still predominantly gaseous. A
disk threaded by a large-scale, ordered magnetic field could
potentially have unique effects on planet growth and migration. One
such effect is the generation (through magnetic resonances that are
the analogs of Lindblad resonances) of a global torque that may
reduce, or even reverse, the secular inward drift (the so-called Type
I migration) predicted for low-mass planets, which has posed a
conundrum for current theories of planet formation. As was
demonstrated by Terquem (2003) and Fromang et al. (2005), a torque of
this type could be produced if the disk had a comparatively strong
(MRI-stable, but still subthermal) azimuthal field (with a nonzero
vertical average of $B_\phi^2$) that fell off sufficiently fast with
radius ($\propto r^{-1}-r^{-2}$). A poloidal field component could in
principle also contribute to this process (Muto et al. 2008). Given
that a large-scale field with precisely these properties is expected
in wind-driving protostellar disks (see Sects.~\ref{sec:formation}
and~\ref{sec:vertical}), this possibility clearly merits an explicit
investigation in the context of the disk models considered in this
chapter. The influence of the vertical channel of angular momentum
transport and of the overall effect of an ordered, large-scale field
on the disk structure in such systems (e.g., the reduction of the
density scale height by magnetic squeezing) may also be worth
examining in this connection.

\section*{Acknowledgments}
\label{sec:acknowledge}

The authors have greatly benefited from their collaborations over the
years with Glenn Ciolek, Ioannis Contopoulos, Ruben Krasnopolsky,
Christof Litwin, Pedro Safier, Konstantinos Tassis, Seth Teitler, Dmitri
Uzdensky, Nektarios Vlahakis, and, in particular, Mark Wardle on the
topics discussed in this chapter. They are also grateful to Marina
Romanova for helpful correspondence on disk--star magnetic
coupling. Their work was supported in part by NSF grant AST-0908184 (AK)
and Australian Research Council grant DP0342844 (RS).

\printindex
\end{document}